\documentclass[aps,prl,reprint,superscriptaddress]{revtex4-2}

\usepackage[T1]{fontenc}
\usepackage[utf8]{inputenc}
\usepackage{amsmath}
\usepackage{amssymb}
\usepackage{bm}
\usepackage{graphicx}
\usepackage[version=4]{mhchem}
\usepackage{xcolor}
\usepackage{soul}
\sethlcolor{yellow}

\begin{document}

\title{Constrained Nuclear-Electronic Orbital Time-Dependent Density Functional Theory for Excited-State Dynamics with Nuclear Quantum Effects}

\author{Yiwen Wang}
\affiliation{Theoretical Chemistry Institute and Department of Chemistry, University of Wisconsin-Madison, Madison, WI 53706, USA}
\author{Zehua Chen}
\affiliation{Theoretical Chemistry Institute and Department of Chemistry, University of Wisconsin-Madison, Madison, WI 53706, USA}
\author{Xi Xu}
\affiliation{Center for Advanced Materials Research, Beijing Normal University, Zhuhai 519087, China}
\author{Yang Yang}
\thanks{Contact author: yyang222@wisc.edu}
\affiliation{Theoretical Chemistry Institute and Department of Chemistry, University of Wisconsin-Madison, Madison, WI 53706, USA}

\begin{abstract}
Nuclear quantum effects can strongly influence excited-state processes
involving light nuclei, yet their inclusion in practical molecular
dynamics remains challenging. We develop constrained nuclear-electronic
orbital time-dependent density functional theory (CNEO-TDDFT), which
incorporates quantum-nuclear delocalization into full-dimensional
effective excited-state potential energy surfaces while retaining
the formal structure and computational scaling of conventional TDDFT.
Analytic gradients are also developed, enabling efficient on-the-fly excited-state
dynamics. CNEO-TDDFT substantially improves excited-state vibrational
frequencies involving hydrogen motion and reduces the root-mean-square deviation
from experiment from 113~cm$^{-1}$ to 26~cm$^{-1}$. For excited-state
proton transfer in 10-hydroxybenzo{[}$h${]}quinoline (HBQ), CNEO-TDDFT
yields a characteristic transfer time of 14.0~fs, compared with 25.5~fs
from conventional TDDFT, consistent with the experimentally inferred
ultrafast timescale of $12\pm6$~fs. These results establish CNEO-TDDFT
as an accurate and efficient framework for excited-state dynamics
with nuclear quantum effects.
\end{abstract}
\maketitle

\section{Introduction}

Nuclear quantum effects (NQEs) can strongly influence physical and
chemical processes, particularly those involving light nuclei. These
effects often include zero-point effects arising from quantum nuclear delocalization
as well as tunneling, and their importance in ground-state properties
and dynamics has been widely studied.\cite{markland2018nuclear,ceriotti2016nuclear}
NQEs can also play an important role following electronic excitation,
where changes in electronic structure rapidly reshape the underlying
potential energy surface (PES) and can strongly affect hydrogen-bonding
and proton-transfer processes dominated by light-nuclear motion.\cite{horke2016hydrogen,waluk2024nuclear}
Accurate descriptions of such excited-state properties and dynamics
therefore require treatment of NQEs on excited-state PESs, which remains
computationally challenging. Widely used excited-state electronic-structure
methods, such as time-dependent density functional theory (TDDFT),
conventionally treat nuclei classically,\cite{casida2012progress}
whereas combining \textit{ab initio} electronic-structure calculations
with path-integral approaches is generally computationally demanding
for on-the-fly simulations of realistic molecular systems.\cite{markland2018nuclear}
A practical framework that can accurately and efficiently describe
NQEs in excited states is therefore highly desirable.

Multicomponent quantum approaches provide a natural starting point
by treating some or all nuclei quantum mechanically together with
the electrons,\cite{thomas1969protonic,kreibich2001multicomponent,pavosevic2020multicomponent}
and multicomponent extensions to electronically excited states have
also been developed.\cite{yang2018multicomponent,culpitt2019enhancing}
However, when a nucleus is represented by a quantum wave function,
its conventional geometrical coordinate is no longer an independent
variable, and the resulting PES is reduced-dimensional with respect
to the quantum nuclei. On such a reduced-dimensional PES, adiabatic
dynamics breaks down because the entire quantum subsystem,
including the electrons and quantum nuclei, is assumed to respond
instantaneously to the classical nuclear coordinates.
While such an adiabatic separation is justified between electrons and
nuclei by their large mass difference, it is not justified between
quantum and classical nuclei, whose masses are of the same order of magnitude.\cite{chen2026constrained}
Therefore, accurate dynamical simulations with this treatment generally require computationally
expensive coupled electron--nuclear quantum dynamics.\cite{zhao2020nuclear,zhao2020real,zhao2021excited}
The constrained nuclear-electronic orbital (CNEO) framework, recently
developed by our group, addresses this problem by constraining the
position expectation value of each quantum nucleus to a designated
molecular coordinate.\cite{xu2020constrained,chen2026constrained}
The resulting energy is then a function of both the classical nuclear
positions and quantum-nuclear position expectation values, leading
to a full-dimensional effective CNEO PES analogous to that of the
Born--Oppenheimer framework while incorporating nuclear delocalization
and nuclear kinetic-energy contributions. In contrast to the reduced-dimensional
multicomponent PES, the restored full-dimensional effective PES allows adiabatic
dynamics to be performed without unphysically assuming the instantaneous response
of quantum nuclei to classical nuclei. This construction has enabled accurate and
efficient calculations of a range of ground-state properties and dynamics, including
molecular structures, vibrational spectroscopy, proton-transfer dynamics, and
reaction kinetics.\cite{xu2022molecular,zhang2025finite,chen2025constrained}

However, to date, the development and applications of the CNEO framework have
been focused on ground-state properties and dynamics. To extend these capabilities
to electronically excited states, here we develop constrained nuclear-electronic orbital
time-dependent density functional theory (CNEO-TDDFT). In this formulation, linear
response is applied to the CNEO ground-state reference under a frozen-nuclear-orbital
approximation for the electronic excitation response, preserving the
formal structure and computational scaling of conventional linear-response
TDDFT. Because the nuclear position constraints are retained, the
resulting excitation energies remain functions of the same molecular
coordinates that define the CNEO ground-state PES, thereby generating
full-dimensional effective excited-state PESs that incorporate nuclear
quantum effects, particularly quantum-nuclear delocalization effects.
We further develop analytic excitation-energy gradients with respect to
both classical coordinates and quantum nuclear position expectation values, enabling
efficient on-the-fly excited-state molecular dynamics on CNEO-TDDFT PESs.

We assess the resulting excited-state PESs through vibrational analysis
for a series of molecules with experimental excited-state vibrational results available.
CNEO-TDDFT substantially improves predicted excited-state vibrational frequencies
associated with hydrogen motion relative to conventional TDDFT.
We then apply CNEO-TDDFT to excited-state intramolecular proton
transfer in 10-hydroxybenzo{[}h{]}quinoline. Including NQEs produces
markedly faster proton-transfer dynamics than conventional TDDFT,
and brings the simulated ultrafast dynamics into closer agreement
with experiment. These results establish CNEO-TDDFT as an
efficient framework for incorporating NQEs into excited-state potential
energy surfaces and molecular dynamics.

\section{Theory}

In multicomponent quantum theory, some or all nuclei can be described
quantum mechanically together with electrons.\cite{thomas1969protonic,kreibich2001multicomponent,pavosevic2020multicomponent}
A direct consequence is that the positions of these quantum nuclei
no longer appear as independent geometrical variables, so that the
resulting energy surface is a reduced-dimensional PES and adiabatic dynamics on this
surface breaks down. The CNEO framework addresses this challenge
by restoring these geometrical degrees of freedom through constraints
on the position expectation values of the quantum nuclei.\cite{xu2020constrained,xu2020full}
For the $I$-th quantum nucleus,
\begin{equation}
\langle \widehat{\bm{R}}_{I}\rangle =\bm{R}^{\mathrm{geo}}_{I},\label{eq:constraint}
\end{equation}
where $\langle \widehat{\bm{R}}_{I}\rangle $ is the position
expectation value and $\bm{R}^{\mathrm{geo}}_{I}$ is the corresponding
designated molecular position. Because nuclei in most physical and chemical systems of interest
remain sufficiently localized in space with negligible overlap, their
position expectation values can be associated with individual atomic
positions. Under this localized-particle approximation, the quantum
nuclei can be treated as distinguishable particles and represented
by localized nuclear orbitals.

Minimization of the multicomponent total energy subject to Eq.~(\ref{eq:constraint})
yields the CNEO-DFT ground-state Kohn--Sham equations.\cite{xu2020constrained}
The resulting total energy is a function of both the positions of the classical
nuclei and the constrained position expectation values of the quantum
nuclei, thereby defining the full-dimensional effective CNEO-DFT PES.
In contrast to a conventional electronic PES with classical point-charge
nuclei, the nuclear orbitals introduce quantum-nuclear delocalization
and nuclear kinetic-energy contributions. Therefore, the CNEO
PES naturally incorporates NQEs, particularly zero-point
effects associated with nuclear delocalization.\cite{chen2026constrained}

Here, we extend this construction to electronically excited states by formulating
linear-response TDDFT on a CNEO ground-state reference. Following
the Franck--Condon picture, we employ a frozen-nuclear-orbital
approximation for the electronic linear response. Specifically, at a given molecular
geometry, the electronic density responds to the excitation while
the quantum-nuclear orbitals remain at their CNEO ground-state stationary
solutions. The nuclear position constraints therefore naturally remain
unchanged during the electronic response, and the excitation energy
remains a function of the same geometrical variables that define the
CNEO ground-state PES. Note that the frozen-nuclear-orbital approximation
freezes the nuclear orbitals only during the electronic excitation
response at a fixed geometry, while their variation with molecular
geometry is retained through the underlying ground-state CNEO reference
and is also reflected in the excitation-energy analytic nuclear gradient to be discussed later.

With the frozen-nuclear-orbital approximation, the CNEO-TDDFT eigenvalue
equation retains the form of conventional electronic linear-response
TDDFT,\cite{casida2012progress}
\begin{equation}
\begin{pmatrix}A^{\mathrm{CNEO}} & B^{\mathrm{CNEO}}\\
B^{\mathrm{CNEO}} & A^{\mathrm{CNEO}}
\end{pmatrix}\begin{pmatrix}X^{\mathrm{e}}\\
Y^{\mathrm{e}}
\end{pmatrix}=\Omega\begin{pmatrix}I & 0\\
0 & -I
\end{pmatrix}\begin{pmatrix}X^{\mathrm{e}}\\
Y^{\mathrm{e}}
\end{pmatrix},\label{eq:eigenvalue}
\end{equation}
where $\Omega$ is the electronic excitation energy, and $X^{\mathrm{e}}$
and $Y^{\mathrm{e}}$ are the excitation and de-excitation amplitudes,
respectively. Unlike conventional TDDFT, in which the electronic
orbitals are obtained in the electrostatic field of classical point-charge
nuclei, the CNEO-TDDFT response matrices, $A^{\mathrm{CNEO}}$ and
$B^{\mathrm{CNEO}}$, are constructed from CNEO electronic orbitals
and orbital energies, where the electrons interact with spatially
delocalized quantum-nuclear densities. The effective CNEO excited-state
potential energy is therefore
\begin{equation}
E^{\mathrm{CNEO\text{-}TDDFT}}(\bm{R})=E^{\mathrm{CNEO}}_{0}(\bm{R})+\Omega(\bm{R}),\label{eq:excited-state-energy}
\end{equation}
which defines a full-dimensional effective excited-state PES incorporating
NQEs while retaining the electronic response structure of conventional
TDDFT.

To enable efficient on-the-fly dynamics on these excited-state PESs, we further
develop analytic CNEO-TDDFT gradients. The CNEO-TDDFT Lagrangian
is constructed as:
\begin{equation}
\begin{aligned} & L[X,Y,\Omega,C^{\mathrm{e}},Z^{\mathrm{e}},W^{\mathrm{e}},\{C^{\mathrm{n}},Z^{\mathrm{n}},W^{\mathrm{n}},\bm{f}^{\mathrm{n}},\bm{g}^{\mathrm{n}}\}]\\
 & =\langle X,Y|\Lambda|X,Y\rangle-\Omega(\langle X,Y|\Delta|X,Y\rangle-1)\\
 & \quad+\sum_{ia\sigma}Z^{\mathrm{e}}_{ia\sigma}F^{\mathrm{e}}_{ia\sigma}-\sum_{pq\sigma,p\leq q}W^{\mathrm{e}}_{pq\sigma}(S^{\mathrm{e}}_{pq\sigma}-\delta_{pq})\\
 & \quad+\sum_{i'a'I}Z^{\mathrm{n}}_{i'a'I}F^{\mathrm{n}}_{i'a'I}-\sum_{p'q'I,p'\leq q'}W^{\mathrm{n}}_{p'q'I}(S^{\mathrm{n}}_{p'q'I}-\delta_{p'q'})\\
 & \quad+\sum_{I}\bm{g}^{\mathrm{n}}_{I}\cdot(\langle \widehat{\bm{R}}_I \rangle-\bm{R}^{\mathrm{geo}}_{I}).
\end{aligned}
\label{eq:ex-lagrangian}
\end{equation}
Here, $\Lambda$ and $\Delta$ respectively denote the CNEO-TDDFT
response matrix and the metric matrix defined in Eq.~\eqref{eq:eigenvalue}.
$C^{\mathrm{e}}$ and $C^{\mathrm{n}}$ denote the electronic and
nuclear orbital coefficients.
The vectors $\bm{f}^{\mathrm{n}}$
denote the CNEO position-constraint multipliers of the underlying
ground-state reference that enter the nuclear Fock matrix.\cite{xu2020constrained}
The indices $i$, $a$, and $(p,q)$
denote the occupied, virtual, and general electronic orbitals,
respectively, while primed indices denote the corresponding nuclear
orbitals. $I$ labels a distinguishable quantum nucleus, and $\sigma$
labels electronic spin.
$F^{\mathrm{e}}_{ia\sigma}$ and $F^{\mathrm{n}}_{i'a'I}$
denote the occupied-virtual elements of the electronic and nuclear Fock
matrices, respectively, and $S^{\mathrm{e}}_{pq\sigma}$ and $S^{\mathrm{n}}_{p'q'I}$
denote the corresponding orbital overlap matrix elements.  $Z$, $W$,
and $\bm{g}^{\mathrm{n}}$ are Lagrange multipliers associated with
the stationarity conditions described below.

The first two terms in Eq.~\eqref{eq:ex-lagrangian} yield the CNEO-TDDFT
eigenvalue equation and normalization of the excitation amplitudes.
The terms involving $Z^{\mathrm{e}}$ and $Z^{\mathrm{n}}$ enforce
the electronic and nuclear Kohn--Sham stationarity conditions, respectively,
whereas those involving $W^{\mathrm{e}}$ and $W^{\mathrm{n}}$ enforce
the corresponding orbital orthonormality conditions. The final term
involving $\bm{g}^{\mathrm{n}}_{I}$ enforces the nuclear position
constraints. Detailed derivations of the Lagrangian and working equations
for determining $\{Z^{\mathrm{e}},Z^{\mathrm{n}},W^{\mathrm{e}},W^{\mathrm{n}},\bm{g}^{\mathrm{n}}\}$
are provided in the Supplemental Material.~\cite{supplemental}

The excitation-energy gradient can be evaluated at stationarity of
the Lagrangian
\begin{equation}
\frac{\partial\Omega}{\partial\bm{R}^{\mathrm{geo}}}=\left.\frac{\partial L}{\partial\bm{R}^{\mathrm{geo}}}\right|_{\mathrm{stationary}},\label{eq:excitation-gradient}
\end{equation}
and the total excited-state energy gradient is obtained by combining
it with the CNEO ground-state energy gradient.\cite{xu2020full}
Note again that although the nuclear orbitals are frozen in the
vertical electronic linear-response problem, their response to
molecular displacement is still included in the analytic gradient.

The resulting formulation retains the formal computational scaling
of conventional TDDFT while providing forces with respect to both
classical nuclear coordinates and constrained quantum-nuclear position
expectation values.

CNEO-TDDFT excitation energies and analytic gradients were implemented
in a locally modified version of PySCF.\cite{sun2018pyscf,sun2020pyscf,sun2026python,theorychemyang_pyscf}
In all calculations reported here, all hydrogen nuclei were treated
quantum mechanically, while heavier nuclei were treated classically.
Excited-state harmonic frequencies were obtained by diagonalizing
mass-weighted Hessian matrices, which were evaluated by numerical
differentiation of analytic gradients. Further computational details,
including basis sets, density functionals, numerical settings, and
molecular-dynamics protocols, are provided in the Supplemental Material.\cite{supplemental}

\section{Results}

\subsection{Excited-state vibrational analysis}

We first assess the CNEO-TDDFT excited-state PESs through harmonic
vibrational analysis. Vibrational frequencies directly probe the local
curvature of a PES around its minimum and therefore provide a stringent
test of how NQEs modify excited-state nuclear motion. We consider
molecules for which experimental excited-state vibrational frequencies
are available, with particular emphasis on modes involving hydrogen
motion. The benchmark set includes \ce{HCN}, \ce{CH2O}, \ce{C2H2},
\ce{C2H2O2}, \ce{C3H2O}, \ce{C6H6}, and \ce{C6H5OH}, together
with the deuterated isotopologues \ce{C2D2}, \ce{C6H5D}, \ce{C6H4D2},
\ce{C6H3D3}, \ce{C6H2D4}, and \ce{C6D6}.\cite{herzberg1966molecular,hadad1993excited,lischka1986ab,merer2003new,dong1993rotational,brand1963band,christiansen1998coupled,garforth1948excited,bist1967band}
For each system, the first singlet excited-state geometry was optimized
using both conventional TDDFT and CNEO-TDDFT,
and harmonic frequencies were calculated using the respective methods.

\begin{figure}[t]
\centering \includegraphics[width=0.8\columnwidth]{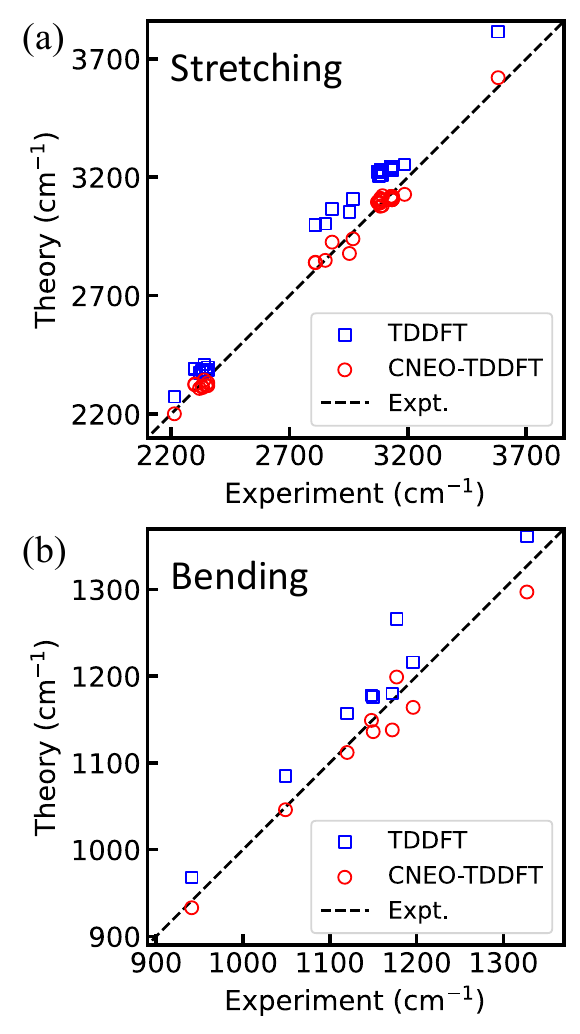} \caption{Excited-state harmonic vibrational frequencies calculated with conventional
TDDFT (blue squares) and CNEO-TDDFT (red circles), compared with experimental reference
values from Refs.~\cite{herzberg1966molecular,hadad1993excited,lischka1986ab,merer2003new,dong1993rotational,brand1963band,christiansen1998coupled,garforth1948excited,bist1967band},
for (a) stretching and (b) bending modes involving hydrogen motion. CNEO-TDDFT
substantially improves agreement with experiment.}
\label{fig:vibrational-frequencies}
\end{figure}

Fig.~\ref{fig:vibrational-frequencies} compares the calculated and
experimental frequencies for modes dominated by hydrogen motion.
Conventional TDDFT systematically overestimates most hydrogen stretching
frequencies, whereas CNEO-TDDFT produces a substantial downshift and
brings them into much closer agreement with experiment. Improvements
are also observed for bending modes involving hydrogen motion. Taken together,
the root-mean-square deviation for the modes in Fig.~\ref{fig:vibrational-frequencies}
decreases from 112.7~cm$^{-1}$ with conventional TDDFT to 26.1~cm$^{-1}$
with CNEO-TDDFT.

For vibrational modes with little contribution from hydrogen motion,
CNEO-TDDFT and conventional TDDFT yield almost identical frequencies,
as shown in Fig.~S1. This similarity is expected in part because only
H and D nuclei are treated quantum mechanically in the present CNEO-TDDFT
calculations. Nevertheless, these modes can still be affected indirectly
through changes in equilibrium geometry and coupling to coordinates
involving hydrogen motion, and no systematic change relative
to conventional TDDFT is observed. Moreover, previous full-quantum
ground-state CNEO calculations, in which all nuclei are treated quantum
mechanically, show much smaller quantum-nuclear corrections for heavier
nuclei because of their more localized nuclear distributions.\cite{xu2021molecular}
Thus, the substantial improvement for modes dominated by hydrogen motion
is achieved while largely preserving the already good TDDFT description
of predominantly heavy-atom vibrations.

\subsection{Excited-state proton transfer dynamics}

We next examine whether the changes in the excited-state PES have
significant consequences for reaction dynamics, which probe regions
far from the equilibrium geometry beyond the local curvature tested
by vibrational frequencies. As a representative process, we consider
excited-state intramolecular proton transfer (ESIPT) in 10-hydroxybenzo{[}h{]}quinoline
(HBQ), which exhibits ultrafast proton-transfer dynamics following
electronic excitation.\cite{martinez1992novel,kim2009coherent,lee2013active,biswas2026tracking}
For CNEO-TDDFT and conventional TDDFT, initial conditions were generated
independently from 15~ps of ground-state $NVT$ sampling at 300~K using
an Andersen thermostat.\cite{andersen1980molecular} Fifty uncorrelated
configurations were selected from each sampling trajectory and subsequently
propagated for 65~fs on the corresponding first singlet excited-state
PES. Proton transfer was monitored using the coordinate
\begin{equation}
\delta_{\mathrm{PT}}=r_{\mathrm{OH}}-r_{\mathrm{NH}},\label{eq:proton-transfer-coordinate}
\end{equation}
where $r_{\mathrm{OH}}$ and $r_{\mathrm{NH}}$ are the distances
of the transferring proton from the donor O and acceptor N atoms,
respectively.

\begin{figure}[t]
\centering \includegraphics[width=1\columnwidth]{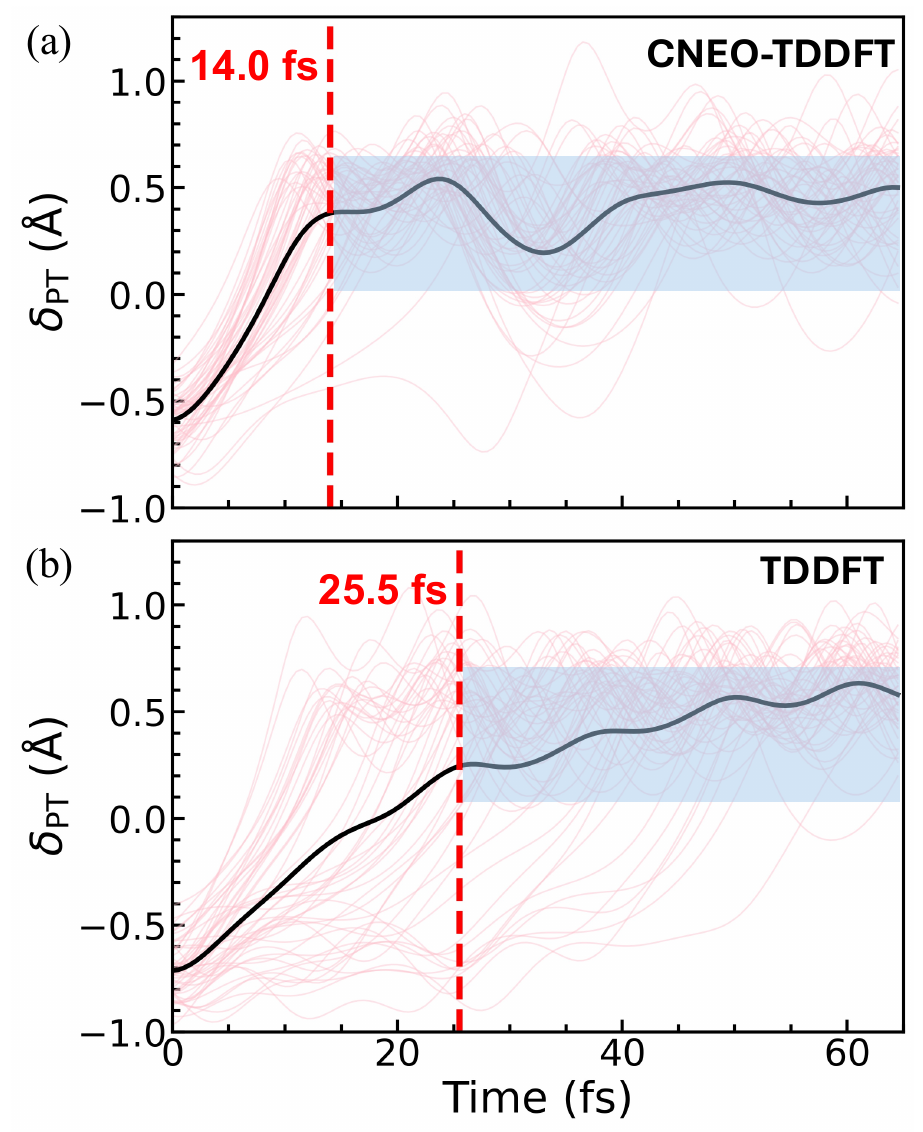} \caption{Proton-transfer coordinate trajectories of HBQ obtained from (a) CNEO-TDDFT
and (b) conventional TDDFT excited-state dynamics. Pink curves show
50 individual trajectories, and black curves show their ensemble averages.
Red dashed lines mark the characteristic transfer times, defined as
the first stationary points of the ensemble-averaged transfer coordinates
after crossing $\delta_{\mathrm{PT}}=0$; these occur at 14.0 fs for
CNEO-TDDFT and 25.5 fs for conventional TDDFT. The shaded regions indicate the subsequent intervals over which
the ensemble-averaged proton-transfer coordinate remains on the product
side.}
\label{fig:proton-transfer-trajectories}
\end{figure}

Fig.~\ref{fig:proton-transfer-trajectories} shows the resulting
proton-transfer trajectories of HBQ obtained from CNEO-TDDFT and conventional
TDDFT. Although the individual trajectories exhibit substantial variation,
their ensemble averages reveal a pronounced difference in the proton-transfer
dynamics. To characterize this difference, we define the characteristic
transfer time as the earliest time after the ensemble-averaged proton transfer
coordinate crosses $\delta_{\mathrm{PT}}=0$ at which its local slope becomes
zero. At this point, the ensemble-averaged proton transfer coordinate reaches
its first stationary point on the product side. This characteristic
time is 14.0~fs with CNEO-TDDFT, compared with 25.5~fs with conventional
TDDFT. As highlighted by the shaded regions in Fig.~\ref{fig:proton-transfer-trajectories},
the ensemble-averaged proton-transfer coordinate subsequently remains
on the product side. Thus, incorporating quantum-nuclear contributions
into the excited-state PES leads to significantly faster proton-transfer
dynamics.

The approximately 14.0-fs characteristic transfer time predicted by
CNEO-TDDFT is close to the experimentally inferred ultrafast ESIPT
timescale of $12\pm6\,\mathrm{fs}$ for HBQ.\cite{lee2013active}
We note that although the simulated structural timescale and
the experimentally extracted spectroscopic time constant are not
identical observables, the acceleration of proton transfer upon
inclusion of nuclear quantum effects is physically well supported:
previous simulations using machine-learning potentials and
path-integral dynamics likewise found that NQEs substantially
accelerate ESIPT in HBQ.\cite{gomez2026excited} However,
unlike approaches that require construction of a separate potential
followed by an additional nuclear-quantum treatment,
CNEO-TDDFT incorporates quantum-nuclear delocalization directly
into the excited-state electronic-structure calculation and enables
efficient on-the-fly dynamics on the resulting effective PES.

\section{Conclusions}

We have developed CNEO-TDDFT as an efficient framework for incorporating
NQEs into full-dimensional effective excited-state potential energy
surfaces and molecular dynamics. By formulating linear-response TDDFT
on a CNEO reference, the method retains the formal structure and computational
scaling of conventional TDDFT while incorporating quantum-nuclear
delocalization and zero-point contributions into the excited-state
PES. Analytic excited-state energy gradients, which include the responses
of the electronic orbitals, quantum-nuclear orbitals, and position
constraints, further enable efficient on-the-fly dynamics on these
quantum-nuclear-corrected surfaces.

Excited-state vibrational benchmarks show that CNEO-TDDFT
substantially improves the prediction of vibrational frequencies
dominated by hydrogen motion. In HBQ, CNEO-TDDFT predicts
substantially faster excited-state proton-transfer dynamics than
conventional TDDFT, with a characteristic timescale in excellent
agreement with experiment. Together, these results demonstrate
that quantum-nuclear delocalization can significantly reshape
the excited-state PES, affecting both local vibrational properties
and reaction dynamics.

CNEO-TDDFT therefore provides a practical route for incorporating
NQEs into excited-state simulations without the computational cost
of a fully quantum treatment of nuclear dynamics. More broadly,
the combination of full-dimensional quantum-nuclear effective
excited-state PESs and analytic forces opens the way to routine
simulations of photoinduced processes in which electronic excitation
drives substantial light-nuclear motion.

\section*{Acknowledgments}

We gratefully acknowledge support from the National Science Foundation
under Grant No.~2238473 for development of the CNEO framework. The
excited-state simulations incorporating nuclear quantum effects were supported
by the U.S. Department of Energy, Office of Science, Office of Advanced
Scientific Computing Research and Office of Basic Energy Sciences,
through the Scientific Discovery through Advanced Computing (SciDAC)
program under the NuQuantEm project at Pacific Northwest National Laboratory
(FWP 85742). Y.W. additionally acknowledges support from a fellowship
from the Molecular Sciences Software Institute under NSF Grant No.~CHE-2136142.
Computational resources were provided by the Center for High Throughput
Computing (CHTC) at the University of Wisconsin--Madison.\cite{chtc}
We thank Mohammad Mostafanejad and Niranjan Govind for helpful discussions.
The authors used ChatGPT (OpenAI) to assist with language editing,
manuscript organization, and refinement of scientific presentation.
All scientific content, interpretations, and conclusions were critically
reviewed and verified by the authors.

\section*{Data Availability}

Data supporting the findings of this study are provided in the main text and the Supplemental Material. The molecular-dynamics trajectory data are available from the corresponding author upon reasonable request.

\nocite{Furche2002,furche2004erratum,apra2020nwchem,wang2016geometric,frisch2016gaussian16,dunning1989cc,kendall1992aug,yu2020nuclearbasis,adamo1999pbe0,hjorth2017atomic,weigend2002fully}

\bibliographystyle{apsrev4-2}
\bibliography{references}

\end{document}

\title{Supplemental Material for ``Constrained Nuclear-Electronic Orbital
Time-Dependent Density Functional Theory for Excited-State Dynamics with
Nuclear Quantum Effects''}
\author{Yiwen Wang}
\affiliation{Theoretical Chemistry Institute and Department of Chemistry,
University of Wisconsin--Madison, Madison, Wisconsin 53706, USA}
\author{Zehua Chen}
\affiliation{Theoretical Chemistry Institute and Department of Chemistry,
University of Wisconsin--Madison, Madison, Wisconsin 53706, USA}
\author{Xi Xu}
\affiliation{Center for Advanced Materials Research, Beijing Normal University,
Zhuhai 519087, China}
\author{Yang Yang}
\email{yyang222@wisc.edu}
\affiliation{Theoretical Chemistry Institute and Department of Chemistry,
University of Wisconsin--Madison, Madison, Wisconsin 53706, USA}
% \date{}
\maketitle

\tableofcontents

\section{Stationary Lagrangian for the CNEO-TDDFT excitation energy}

\label{section:ctd_L}

In this section and the next, we provide the detailed derivation of
the analytic CNEO-TDDFT excitation-energy gradient. The derivation
follows the stationary-Lagrangian strategy established for conventional
TDDFT derivatives~\cite{Furche2002,furche2004erratum}, extended here to account for
the quantum-nuclear orbitals and CNEO position constraints ~\cite{xu2020constrained,xu2020full}.
Because the electronic linear-response problem retains the formal
structure of conventional TDDFT under the frozen-nuclear-orbital approximation,
several purely electronic steps closely parallel those of conventional
TDDFT gradient theory. We nevertheless present the full derivation
to make the additional nuclear and electron--nuclear response contributions
explicit.

Importantly, the frozen-nuclear-orbital approximation applies only
to the electronic excitation response at a fixed molecular geometry.
The electronic and nuclear orbitals of the underlying CNEO reference
still vary with molecular geometry, and these responses must therefore
be included in the analytic excitation-energy gradient.

Throughout the derivation, occupied, virtual, and general electronic
molecular orbitals are indexed by $i,j$, $a,b$, and $p,q$, respectively,
while electronic atomic orbitals are indexed by $\mu,\nu$.
The corresponding primed indices are used for quantum-nuclear molecular orbitals
and atomic orbitals.
The index $I$ labels a distinguishable quantum nucleus, and $\sigma$
denotes electronic spin.

At a fixed molecular geometry and its corresponding converged CNEO
ground-state reference, the CNEO-TDDFT excitation energy is obtained
from the generalized eigenvalue equation given in the main text. For
the purpose of deriving analytic derivatives, it is convenient to
express this eigenvalue problem through the stationary response functional
\begin{equation}
G[X,Y,\Omega]=\langle X,Y|\Lambda|X,Y\rangle-\Omega\left(\langle X,Y|\Delta|X,Y\rangle-1\right),\label{eq:response-functional}
\end{equation}
where
\begin{equation}
\Lambda=\begin{pmatrix}A^{\mathrm{CNEO}} & B^{\mathrm{CNEO}}\\
B^{\mathrm{CNEO}} & A^{\mathrm{CNEO}}
\end{pmatrix},\qquad\Delta=\begin{pmatrix}I & 0\\
0 & -I
\end{pmatrix},
\end{equation}
are the CNEO-TDDFT response matrix and metric, respectively. Here
$\Omega$ acts as the Lagrange multiplier associated with the metric
normalization of the excitation amplitudes and becomes the excitation
energy at the stationary solution.

Stationarity with respect to the excitation amplitudes and $\Omega$
gives
\begin{align}
\frac{\partial G}{\partial\langle X,Y|} & =\left(\Lambda-\Omega\Delta\right)|X,Y\rangle=0,\label{eq:G_stationary_XY}\\
\frac{\partial G}{\partial\Omega} & =-\left(\langle X,Y|\Delta|X,Y\rangle-1\right)=0.\label{eq:G_stationary_Omega}
\end{align}
These conditions recover, respectively, the CNEO-TDDFT generalized
eigenvalue equation and the normalization condition for the excitation
amplitudes.

At the stationary solution,
\begin{align}
G_{0} & \equiv G[X,Y,\Omega]\bigg|_{\{X,Y,\Omega\}_{0}}\\
 & =\langle X,Y|\left(\Lambda-\Omega\Delta\right)|X,Y\rangle+\Omega\\
 & =\Omega,
\end{align}
where the subscript $0$ denotes the stationary solution with respect
to $\{X,Y,\Omega\}$.

We now consider the derivative of the excitation energy with respect
to an external perturbation $\xi$. Because $G$ is stationary with
respect to $X$, $Y$, and $\Omega$, their first-order responses
do not contribute to the derivative of the stationary value of $G$.
Therefore,
\begin{align}
\Omega^{\xi} & =G^{\xi}_{0}=\frac{\partial G}{\partial\xi}\bigg|_{\{X,Y,\Omega\}_{0}}\\
 & =\langle X,Y|\Lambda^{\xi}|X,Y\rangle.\label{eq:G_gradient}
\end{align}

Equation~\ref{eq:G_gradient} eliminates the need to evaluate the
response of the excitation amplitudes, $X^{\xi}$ and $Y^{\xi}$.
However, $\Lambda$ is constructed from the underlying CNEO ground-state
reference, and its derivative $\Lambda^{\xi}$ still contains the
responses of the electronic orbitals, quantum-nuclear orbitals, and
ground-state CNEO constraint multipliers, $C^{\mathrm{e},\xi}$, $C^{\mathrm{n},\xi}$,
and $\bm{f}^{\mathrm{n},\xi}$.

These responses could in principle be obtained by solving the CNEO
coupled-perturbed Kohn--Sham (CNEO-CPKS) equations separately for
each perturbation. For a molecular gradient, however, this would require
a separate response solution for each Cartesian coordinate and would
therefore be computationally inefficient.

To avoid these coordinate-specific response calculations, we augment
the stationary response functional $G$ with the ground-state CNEO
orbital stationarity, orbital orthonormality, and quantum-nuclear
position constraints. The resulting CNEO-TDDFT Lagrangian is constructed
to be stationary not only with respect to the excitation variables
$\{X,Y,\Omega\}$, but also with respect to the underlying CNEO reference
variables $\{C^{\mathrm{e}},C^{\mathrm{n}},\bm{f}^{\mathrm{n}}\}$.
The associated Lagrange multipliers are determined from a coupled
$Z$-vector system that is independent of the perturbation coordinate.
The complete Lagrangian is
\begin{align}
L[X,Y,\Omega,C^{\mathrm{e}},Z^{\mathrm{e}},W^{\mathrm{e}}, & \{C^{\mathrm{n}},Z^{\mathrm{n}},W^{\mathrm{n}},\bm{f}^{\mathrm{n}},\bm{g}^{\mathrm{n}}\}]\nonumber \\
 & =\langle X,Y|\Lambda|X,Y\rangle-\Omega(\langle X,Y|\Delta|X,Y\rangle-1)\nonumber \\
 & +\sum_{ia\sigma}Z^{\mathrm{e}}_{ia\sigma}F^{\mathrm{e}}_{ia\sigma}-\sum_{pq\sigma,p\leq q}W^{\mathrm{e}}_{pq\sigma}(S^{\mathrm{e}}_{pq\sigma}-\delta_{pq})\nonumber \\
 & +\sum_{I}\sum_{i'a'}Z^{\mathrm{n}}_{i'a'I}F^{\mathrm{n}}_{i'a'I}-\sum_{I}\sum_{p'q',p'\leq q'}W^{\mathrm{n}}_{p'q'I}(S^{\mathrm{n}}_{p'q'I}-\delta_{p'q'})\nonumber \\
 & +\sum_{I}\bm{g}^{\mathrm{n}}_{I}\cdot(\langle\widehat{\bm{R}}_{I}\rangle-\bm{R}^{\mathrm{geo}}_{I}).\label{eq:ctd_L}
\end{align}
Here $C^{\mathrm{e}}$ and $C^{\mathrm{n}}$ denote the coefficients
of the electronic and quantum-nuclear orbitals in their respective
AO bases, while $F^{\mathrm{e(n)}}$ and $S^{\mathrm{e(n)}}$ denote
the corresponding Fock and overlap matrices. The vectors $\bm{f}^{\mathrm{n}}_{I}$
are the ground-state CNEO position-constraint multipliers entering
the nuclear Fock matrix~\cite{xu2020constrained}.

The quantities $Z^{\mathrm{e(n)}}$, $W^{\mathrm{e(n)}}$, and $\bm{g}^{\mathrm{n}}_{I}$
are Lagrange multipliers introduced for the analytic derivative formulation.
Variation of $L$ with respect to $Z^{\mathrm{e(n)}}$ enforces the
occupied--virtual orbital-stationarity conditions, variation with
respect to $W^{\mathrm{e(n)}}$ enforces orbital orthonormality, and
variation with respect to $\bm{g}^{\mathrm{n}}_{I}$ enforces the
CNEO position constraints:
\begin{equation}
F^{\mathrm{e}}_{ia\sigma}=0,\ S^{\mathrm{e}}_{pq\sigma}=\delta_{pq},\label{eq:ctd_ks_e}
\end{equation}
\begin{equation}
F^{\mathrm{n}}_{i'a'I}=0,\ S^{\mathrm{n}}_{p'q'I}=\delta_{p'q'},\label{eq:ctd_ks_n}
\end{equation}
\begin{equation}
\langle\widehat{\bm{R}}_{I}\rangle-\bm{R}^{\mathrm{geo}}_{I}=\bm{0}.\label{eq:ctd_ks_r}
\end{equation}
Accordingly,
\begin{equation}
\frac{\partial L}{\partial Z^{\mathrm{e}}_{ia\sigma}}=0,\quad\frac{\partial L}{\partial Z^{\mathrm{n}}_{i'a'I}}=0,\quad\frac{\partial L}{\partial W^{\mathrm{e}}_{pq\sigma}}=0,\quad\frac{\partial L}{\partial W^{\mathrm{n}}_{p'q'I}}=0,\quad\frac{\partial L}{\partial g^{\mathrm{n}}_{I\alpha}}=0,
\end{equation}
recover Eqs.~\ref{eq:ctd_ks_e}--\ref{eq:ctd_ks_r}, where $\alpha\in\{x,y,z\}$ denotes a Cartesian component. In addition,
stationarity with respect to the excitation amplitudes and $\Omega$
recovers the CNEO-TDDFT eigenvalue equation and normalization condition.

The remaining multipliers are determined by requiring stationarity
of $L$ with respect to the underlying CNEO reference variables,
\begin{equation}
\frac{\partial L}{\partial C^{\mathrm{e}}_{\mu p\sigma}}=0,\qquad\frac{\partial L}{\partial C^{\mathrm{n}}_{\mu'p'I}}=0,\qquad\frac{\partial L}{\partial f^{\mathrm{n}}_{I\alpha}}=0.\label{eq:L_reference_stationarity}
\end{equation}
The coupled equations resulting from these
conditions are derived in Sec.~II.

Therefore, the value of $L$ at its stationary point is
\begin{align}
L_{0} & \equiv L[X,Y,\Omega,C^{\mathrm{e}},Z^{\mathrm{e}},W^{\mathrm{e}},\{C^{\mathrm{n}},Z^{\mathrm{n}},W^{\mathrm{n}},\bm{f}^{\mathrm{n}},\bm{g}^{\mathrm{n}}\}]\bigg|_{\{X,Y,\Omega,C^{\mathrm{e}},Z^{\mathrm{e}},W^{\mathrm{e}},\{C^{\mathrm{n}},Z^{\mathrm{n}},W^{\mathrm{n}},\bm{f}^{\mathrm{n}},\bm{g}^{\mathrm{n}}\}\}_{0}}\\
 & =\langle X,Y|\left(\Lambda-\Omega\Delta\right)|X,Y\rangle+\Omega+\sum_{ia\sigma}Z^{\mathrm{e}}_{ia\sigma}\cdot0-\sum_{pq\sigma,p\leq q}W^{\mathrm{e}}_{pq\sigma}\cdot0\nonumber \\
 & +\sum_{i'a'I}Z^{\mathrm{n}}_{i'a'I}\cdot0-\sum_{p'q'I,p'\leq q'}W^{\mathrm{n}}_{p'q'I}\cdot0+\sum_{I}\bm{g}^{\mathrm{n}}_{I}\cdot0\\
 & =\Omega,
\end{align}
where the subscript 0 denotes the complete stationary solution. The
derivative of excitation energy $\Omega$ with respect to the external
perturbation $\xi$ can then be related to the derivative of $L$
at its stationary points
\begin{equation}
\Omega^{\xi}=L^{\xi}_{0}=\frac{\partial L}{\partial\xi}\bigg|_{\{X,Y,\Omega,C^{\mathrm{e}},Z^{\mathrm{e}},W^{\mathrm{e}},\{C^{\mathrm{n}},Z^{\mathrm{n}},W^{\mathrm{n}},\bm{f}^{\mathrm{n}},\bm{g}^{\mathrm{n}}\}\}_{0}}.
\end{equation}

Differentiating Eq.~\ref{eq:ctd_L} and using all stationary conditions
gives
\begin{align}
\Omega^{\xi} & =L^{\xi}_{0}\label{ctd_step1}\\
 & =\langle X,Y|\Lambda^{\xi}|X,Y\rangle+\sum_{ia\sigma}Z^{\mathrm{e}\xi}_{ia\sigma}\cdot0-\sum_{pq\sigma,p\leq q}W^{\mathrm{e}\xi}_{pq\sigma}\cdot0+\sum_{ia\sigma}Z^{\mathrm{e}}_{ia\sigma}F^{\mathrm{e}\xi}_{ia\sigma}-\sum_{pq\sigma,p\leq q}W^{\mathrm{e}}_{pq\sigma}S^{\mathrm{e}\xi}_{pq\sigma}\nonumber \\
 & +\sum_{i'a'I}Z^{\mathrm{n}\xi}_{i'a'I}\cdot0-\sum_{p'q'I,p'\leq q'}W^{\mathrm{n}\xi}_{p'q'I}\cdot0+\sum_{I}\bm{g}^{\mathrm{n}\xi}_{I}\cdot0+\sum_{i'a'I}Z^{\mathrm{n}}_{i'a'I}F^{\mathrm{n}\xi}_{i'a'I}-\sum_{p'q'I,p'\leq q'}W^{\mathrm{n}}_{p'q'I}S^{\mathrm{n}\xi}_{p'q'I}+\sum_{I}\bm{g}^{\text{n}}_{I}\cdot(\langle\widehat{\bm{R}}_{I}\rangle-\bm{R}^{\mathrm{geo}}_{I})^{\xi}\\
 & =\langle X,Y|\Lambda^{\xi}|X,Y\rangle+\sum_{ia\sigma}Z^{\mathrm{e}}_{ia\sigma}F^{\mathrm{e}\xi}_{ia\sigma}-\sum_{pq\sigma,p\leq q}W^{\mathrm{e}}_{pq\sigma}S^{\mathrm{e}\xi}_{pq\sigma}+\sum_{i'a'I}Z^{\mathrm{n}}_{i'a'I}F^{\mathrm{n}\xi}_{i'a'I}-\sum_{p'q'I,p'\leq q'}W^{\mathrm{n}}_{p'q'I}S^{\mathrm{n}\xi}_{p'q'I}+\sum_{I}\bm{g}^{\text{n}}_{I}\cdot(\langle\widehat{\bm{R}}_{I}\rangle-\bm{R}^{\mathrm{geo}}_{I})^{\xi}.\label{eq:ctd_step1_1}
\end{align}
Within the distinguishable-particle CNEO framework, each quantum nucleus
is represented by a single occupied nuclear orbital. We denote the
occupied orbital associated with nucleus $I$ by $i'$. The corresponding
position constraint can therefore be written as
\begin{equation}
\langle\widehat{\bm{R}}_{I}\rangle-\bm{R}^{\mathrm{geo}}_{I}=\left\langle i'\left|\widehat{\bm{R}}_{I}-\bm{R}^{\mathrm{geo}}_{I}\right|i'\right\rangle \equiv\overline{\bm{R}}^{\mathrm{n}}_{i'i'I},
\end{equation}
where
\begin{equation}
\overline{\bm{R}}^{\mathrm{n}}_{p'q'I}\equiv\left\langle p'\left|\widehat{\bm{R}}_{I}-\bm{R}^{\mathrm{geo}}_{I}\right|q'\right\rangle
\end{equation}
denotes the nuclear position-deviation matrix elements. Using this
notation, Eq.~\ref{eq:ctd_step1_1} can be further expressed as
\begin{equation}
\Omega^{\xi}=\langle X,Y|\Lambda^{\xi}|X,Y\rangle+\sum_{ia\sigma}Z^{\mathrm{e}}_{ia\sigma}F^{\mathrm{e}\xi}_{ia\sigma}-\sum_{pq\sigma,p\leq q}W^{\mathrm{e}}_{pq\sigma}S^{\mathrm{e}\xi}_{pq\sigma}+\sum_{i'a'I}Z^{\mathrm{n}}_{i'a'I}F^{\mathrm{n}\xi}_{i'a'I}-\sum_{p'q'I,p'\leq q'}W^{\mathrm{n}}_{p'q'I}S^{\mathrm{n}\xi}_{p'q'I}+\sum_{I}\bm{g}^{\text{n}}_{I}\cdot\overline{\bm{R}}^{\text{n}\xi}_{i'i'I},
\end{equation}

Each total derivative of $\Theta\in\{\Lambda,F^{\mathrm{e(n)}},S^{\mathrm{e(n)}},\overline{\bm{R}}^{\mathrm{n}}_{i'i'I}\}$
with respect to $\xi$ can be decomposed into an explicit derivative
at fixed $\{C^{\mathrm{e}},C^{\mathrm{n}},\bm{f}^{\mathrm{n}}\}$
and implicit contributions arising from the responses of these ground-state
CNEO variables:
\begin{equation}
\Theta^{\xi}=\Theta^{(\xi)}+\sum_{\mu p\sigma}\frac{\partial\Theta}{\partial C^{\mathrm{e}}_{\mu p\sigma}}\frac{\partial C^{\mathrm{e}}_{\mu p\sigma}}{\partial\xi}+\sum_{\mu'p'I}\frac{\partial\Theta}{\partial C^{\mathrm{n}}_{\mu'p'I}}\frac{\partial C^{\mathrm{n}}_{\mu'p'I}}{\partial\xi}+\sum_{I,\alpha=x,y,z}\frac{\partial\Theta}{\partial f^{\mathrm{n}}_{\alpha,I}}\frac{\partial f^{\mathrm{n}}_{\alpha,I}}{\partial\xi}.
\end{equation}
Here the parenthesized superscript $(\xi)$ denotes the explicit derivative
with respect to $\xi$ at fixed $\{C^{\mathrm{e}},C^{\mathrm{n}},\bm{f}^{\mathrm{n}}\}$.
For a molecular-coordinate perturbation, this term contains the explicit
coordinate dependence of the AO basis functions and of the operators
and integrals entering $\Theta$. Substitution into the Lagrangian
derivative gives
\begin{align}
\Omega^{\xi} & =\langle X,Y|\Lambda^{(\xi)}|X,Y\rangle+\sum_{ia\sigma}Z^{\mathrm{e}}_{ia\sigma}F^{\mathrm{e}(\xi)}_{ia\sigma}-\sum_{pq\sigma,p\leq q}W^{\mathrm{e}}_{pq\sigma}S^{\mathrm{e}(\xi)}_{pq\sigma}\nonumber \\
 & +\sum_{i'a'I}Z^{\mathrm{n}}_{i'a'I}F^{\mathrm{n}(\xi)}_{i'a'I}-\sum_{p'q'I,p'\leq q'}W^{\mathrm{n}}_{p'q'I}S^{\mathrm{n}(\xi)}_{p'q'I}+\sum_{I}\bm{g}^{\text{n}}_{I}\cdot\overline{\bm{R}}^{\text{n}(\xi)}_{i'i'I}\nonumber \\
 & +\sum_{\mu p\sigma}\frac{\partial C^{\mathrm{e}}_{\mu p\sigma}}{\partial\xi}\frac{\partial L}{\partial C^{\mathrm{e}}_{\mu p\sigma}}+\sum_{\mu'p'I}\frac{\partial C^{\mathrm{n}}_{\mu'p'I}}{\partial\xi}\frac{\partial L}{\partial C^{\mathrm{n}}_{\mu'p'I}}+\sum_{I,\alpha=x,y,z}\frac{\partial f^{\mathrm{n}}_{\alpha,I}}{\partial\xi}\frac{\partial L}{\partial f^{\mathrm{n}}_{\alpha,I}}.
\end{align}

At full stationarity, the multipliers $Z^{\mathrm{e(n)}}$, $W^{\mathrm{e(n)}}$,
and $\bm{g}^{\mathrm{n}}_{I}$ are determined by the reference-stationarity
conditions in Eq.~\ref{eq:L_reference_stationarity}. Consequently,
all terms proportional to the implicit responses $C^{\mathrm{e},\xi}$,
$C^{\mathrm{n},\xi}$, and $\bm{f}^{\mathrm{n},\xi}$ drop out of
the Lagrangian derivative, leaving
\begin{align}
\Omega^{\xi} & =\langle X,Y|\Lambda^{(\xi)}|X,Y\rangle+\sum_{ia\sigma}Z^{\mathrm{e}}_{ia\sigma}F^{\mathrm{e}(\xi)}_{ia\sigma}-\sum_{pq\sigma,p\leq q}W^{\mathrm{e}}_{pq\sigma}S^{\mathrm{e}(\xi)}_{pq\sigma}+\sum_{i'a'I}Z^{\mathrm{n}}_{i'a'I}F^{\mathrm{n}(\xi)}_{i'a'I}\nonumber \\
 & -\sum_{p'q'I,p'\leq q'}W^{\mathrm{n}}_{p'q'I}S^{\mathrm{n}(\xi)}_{p'q'I}+\sum_{I}\bm{g}^{\text{n}}_{I}\cdot\overline{\bm{R}}^{\text{n}(\xi)}_{i'i'I}.\label{eq:ctd_step2}
\end{align}

We now show that Eq.~\ref{eq:ctd_step2} simplifies further for the
molecular-coordinate derivatives considered here. In the present CNEO
implementation, all nuclear AO basis functions associated with quantum
nucleus $I$ are centered at its designated position $\bm{R}^{\mathrm{geo}}_{I}$
and therefore translate rigidly with that coordinate.

First consider the nuclear AO overlap integral. Because all basis
functions associated with a distinguishable quantum nucleus translate
together with their designated position $\bm{R}^{\mathrm{geo}}_{I}$,

\begin{align}
\langle\mu_{I}'|\nu_{I}'\rangle & =\int\mathrm{d}\bm{R}_I\phi^{\mathrm{n*}}_{\mu'}(\bm{R}_{I}-\bm{R}^{\mathrm{geo}}_{I})\phi^{\mathrm{n}}_{\nu'}(\bm{R}_{I}-\bm{R}^{\mathrm{geo}}_{I})\\
 & =\int\mathrm{d}\bm{x}\phi^{\mathrm{n*}}_{\mu'}(\bm{x})\phi^{\mathrm{n}}_{\nu'}(\bm{x}).
\end{align}
The AO overlap integral therefore has no explicit dependence on $\bm{R}^{\mathrm{geo}}_{I}$,
and its explicit derivative with respect to the molecular coordinates
vanishes. Consequently,
\begin{equation}
S^{\mathrm{n}(\xi)}_{p'q'I}=0.\label{eq:ctd_ovlp}
\end{equation}

Similarly, the AO position-deviation integral is

\begin{align}
\left\langle \mu'_{I}\left|(\widehat{\bm{R}}_{I}-\bm{R}^{\mathrm{geo}}_{I})\right|\nu'_{I}\right\rangle  & =\int\diff\bm{R}_{I}\,\phi^{\mathrm{n}*}_{\mu'}(\bm{R}_{I}-\bm{R}^{\mathrm{geo}}_{I})(\bm{R}_{I}-\bm{R}^{\mathrm{geo}}_{I})\phi^{\mathrm{n}}_{\nu'}(\bm{R}_{I}-\bm{R}^{\mathrm{geo}}_{I})\\
 & =\int\mathrm{d}\bm{x}\phi^{\mathrm{n*}}_{\mu'}(\bm{x})\bm{x}\phi^{\mathrm{n}}_{\nu'}(\bm{x}).
\end{align}
The same change of variables shows that the position-deviation integral
is also invariant under a common translation of the nuclear basis
and its designated position. Therefore, its explicit molecular-coordinate
derivative vanishes:
\begin{equation}
\overline{\bm{R}}^{\text{n}(\xi)}_{i'i'I}=\bm{0}.\label{eq:ctd_pos}
\end{equation}

Eqs.~\ref{eq:ctd_ovlp} and \ref{eq:ctd_pos}
reduce Eq.~\ref{eq:ctd_step2} to the working excitation-energy gradient
\begin{equation}
\Omega^{\xi}=\langle X,Y|\Lambda^{(\xi)}|X,Y\rangle+\sum_{ia\sigma}Z^{\mathrm{e}}_{ia\sigma}F^{\mathrm{e}(\xi)}_{ia\sigma}-\sum_{pq\sigma,p\leq q}W^{\mathrm{e}}_{pq\sigma}S^{\mathrm{e}(\xi)}_{pq\sigma}+\sum_{I}\sum_{i'a'}Z^{\mathrm{n}}_{i'a'I}F^{\mathrm{n}(\xi)}_{i'a'I}.\label{eq:cneo_grad}
\end{equation}
When $\xi$ is a component of either a constrained position $\bm{R}^{\mathrm{geo}}_{I}$
or a classical-nuclear coordinate, Eq.~\ref{eq:cneo_grad} is the
analytic CNEO-TDDFT excitation-energy gradient. With the previously
developed CNEO ground-state analytic gradient,\cite{xu2020full}
the total excited-state gradient is
\begin{equation}
E^{\xi}_{\mathrm{CNEO\text{-}TDDFT}}=E^{\mathrm{CNEO},\xi}_{0}+\Omega^{\xi}.\label{eq:total-excited-gradient}
\end{equation}

\section{Coupled $Z$-vector equations for CNEO-TDDFT}

The remaining task is to determine $Z^{\mathrm{e(n)}}$, $W^{\mathrm{e(n)}}$,
and $\bm{g}^{\mathrm{n}}$ so that Eq.~\ref{eq:L_reference_stationarity}
holds. This section derives all electronic, nuclear, and mixed electron--nuclear
blocks needed for this coupled system.

\subsection{Stationarity with respect to the ground-state position constraint
Lagrange multipliers}

Only the nuclear Fock matrix depends explicitly on $\bm f_I^{\mathrm n}$:
\begin{equation}
F_{r's'I}^{\mathrm{n}}=h_{r's'I}^{\mathrm{n}}-Z_{I}\sum_{i\sigma'}^{\mathrm{occ}}(r's'I|ii\sigma')+V_{r's'I}^{\mathrm{epc}}[\rho^{\mathrm{e}},\rho_{I}^{\mathrm{n}}]+Z_{I}\sum_{j'\ne i',J\neq I}Z_{J}(r's'I|j'j'J)+\bm{f}_{I}^{\mathrm n}\cdot\overline{\bm R}_{r's'I}^{\mathrm n},
\end{equation}
where $h^{\mathrm n}$ is the nuclear core Hamiltonian, $Z_I$ is the
nuclear charge, and $V^{\mathrm{epc}}$ is the electron--proton correlation
potential.  Hence
\begin{align}
\frac{\partial L}{\partial\bm{f}_{I}^{\mathrm n}} & =\sum_{i'a'}Z_{i'a'I}^{\mathrm{n}}\frac{\partial F_{i'a'I}^{\mathrm{n}}}{\partial\bm{f}_{I}^{\mathrm n}}\\
 & =\sum_{i'a'}Z_{i'a'I}^{\mathrm{n}}\overline{\bm R}_{i'a'I}^{\mathrm n}.
\end{align}
Therefore, the third stationarity condition in Eq.~\ref{eq:L_reference_stationarity}
requires
\begin{equation}
\sum_{i'a'}Z^{\mathrm{n}}_{i'a'I}\overline{\bm{R}}^{\mathrm{n}}_{i'a'I}=\bm{0}.\label{eq:constraint_add}
\end{equation}

\subsection{Derivatives of the Fock matrix}

To continue considering the first two terms in Eq.~\ref{eq:L_reference_stationarity},
we first transform them to the following form to simplify the derivation:
\begin{align}
\sum_{\mu}C^{\mathrm{e}}_{\mu q\sigma}\frac{\partial L}{\partial C^{\mathrm{e}}_{\mu p\sigma}} & =0\label{ctd_transform_ce}\\
\sum_{\mu'}C^{\mathrm{n}}_{\mu'q'I}\frac{\partial L}{\partial C^{\mathrm{n}}_{\mu'p'I}} & =0.\label{ctd_transform_cn}
\end{align}
The derivations will be strongly related to the derivatives of nuclear
and electronic Fock matrices with respect to the MO coefficients.
Thus, we want to first derive the derivatives of the Fock matrices.
This section will include the following parts: the derivatives of
electronic Fock matrix with respect to electronic MO coefficients
and nuclear MO coefficients; and the derivatives of nuclear Fock matrix
with respect to electronic MO coefficients and nuclear MO coefficients.

\subsubsection{Electronic Fock matrix--electronic MO coefficients}

\label{subsubsection: fock_ee}

The electronic Fock matrix is
\begin{equation}
F^{\mathrm{e}}_{rs\sigma'}=h^{\mathrm{e}}_{rs\sigma'}+\sum^{\mathrm{occ}}_{i\sigma''}\left[(rs\sigma'|ii\sigma'')-c_{\mathrm{x}}\delta_{\sigma'\sigma''}(ri\sigma'|is\sigma'')\right]+V^{\mathrm{xc}}_{rs\sigma'}[\rho^{\mathrm{e}}]+\sum_{I}V^{\mathrm{epc}}_{rs\sigma'}[\rho^{\mathrm{e}},\rho^{\mathrm{n}}_{I}]-\sum_{Ii'}Z_{I}(rs\sigma'|i'i'I),\label{eq:elec_fock}
\end{equation}
where $h^{\mathrm{e}}$ is the electronic core Hamiltonian matrix,
$c_{\mathrm{x}}$ is the hybrid mixing parameter accounting for the
consideration of Hartree-Fock exact exchange, and $V^{\mathrm{xc}}$
is the exchange-correlation potential described by the selected functional.

To align with the form of Eq.~\ref{ctd_transform_ce}, we express
the derivatives of electronic Fock matrix with respect to electronic
MO coefficients as
\begin{align}
\sum_{\mu}C^{\mathrm{e}}_{\mu q\sigma}\frac{\partial F^{\mathrm{e}}_{rs\sigma'}}{\partial C^{\mathrm{e}}_{\mu p\sigma}} & =\sum_{\mu}C^{\mathrm{e}}_{\mu q\sigma}\frac{\partial}{\partial C^{\mathrm{e}}_{\mu p\sigma}}\bigg\{ h^{\mathrm{e}}_{rs\sigma'}+\sum^{\mathrm{occ}}_{i\sigma''}\left[(rs\sigma'|ii\sigma'')-c_{\mathrm{x}}\delta_{\sigma'\sigma''}(ri\sigma'|is\sigma'')\right]\nonumber \\
 & +V^{\mathrm{xc}}_{rs\sigma'}[\rho^{\mathrm{e}}]+\sum_{I}V^{\mathrm{epc}}_{rs\sigma'}[\rho^{\mathrm{e}},\rho^{\mathrm{n}}_{I}]-\sum_{I}Z_{I}(rs\sigma'|i'i'I)\bigg\}\\
 & =\sum_{\mu}C^{\mathrm{e}}_{\mu q\sigma}\frac{\partial}{\partial C^{\mathrm{e}}_{\mu p\sigma}}\left(\sum_{\nu\lambda}C^{\mathrm{e}}_{\nu r\sigma'}C^{\mathrm{e}}_{\lambda s\sigma'}h^{\mathrm{e}}_{\nu\lambda\sigma'}\right)\nonumber \\
 & +\sum_{\mu}C^{\mathrm{e}}_{\mu q\sigma}\frac{\partial}{\partial C^{\mathrm{e}}_{\mu p\sigma}}\left\{\sum^{\mathrm{occ}}_{i\sigma''}\sum_{\nu\lambda}C^{\mathrm{e}}_{\nu r\sigma'}C^{\mathrm{e}}_{\lambda s\sigma'}\sum_{\eta\gamma}C^{\mathrm{e}}_{\eta i\sigma''}C^{\mathrm{e}}_{\gamma i\sigma''}[(\nu\lambda\sigma'|\eta\gamma\sigma'')-c_{\mathrm{x}}\delta_{\sigma'\sigma''}(\nu\gamma\sigma'|\eta\lambda\sigma')]\right\}\nonumber \\
 & +\sum_{\mu}C^{\mathrm{e}}_{\mu q\sigma}\frac{\partial V^{\mathrm{e},\mathrm{xc}}_{rs\sigma'}}{\partial C^{\mathrm{e}}_{\mu p\sigma}}\nonumber \\
 & +\sum_{I}\sum_{\mu}C^{\mathrm{e}}_{\mu q\sigma}\frac{\partial V^{\mathrm{epc}}_{rs\sigma'}[\rho^{\mathrm{e}},\rho^{\mathrm{n}}_{I}]}{\partial C^{\mathrm{e}}_{\mu p\sigma}}\nonumber \\
 & -\sum_{I}Z_{I}\sum_{\mu}C^{\mathrm{e}}_{\mu q\sigma}\frac{\partial}{\partial C^{\mathrm{e}}_{\mu p\sigma}}\big[\sum_{\nu\lambda}C^{\mathrm{e}}_{\nu r\sigma'}C^{\mathrm{e}}_{\lambda s\sigma'}(\nu\lambda\sigma'|i'i'I)\big].\label{eq:Fock_step1}
\end{align}

To get the final expression, we can start with the derivatives of
$C^{\mathrm{e}}$ with respect to $C^{\mathrm{e}}$:
\begin{equation}
\sum_{\mu}C_{\mu q\sigma}^{\mathrm{e}}\frac{\partial C_{\nu r\sigma'}^{\mathrm{e}}}{\partial C_{\mu p\sigma}^{\mathrm{e}}}=\sum_{\mu}C_{\mu q\sigma}^{\mathrm{e}}\delta_{\mu\nu}\delta_{rp}\delta_{\sigma\sigma'}=C_{\nu q\sigma}^{\mathrm{e}}\delta_{rp}\delta_{\sigma\sigma'}.
\end{equation}
This can give us a general form of the derivatives of two-body integrals:
\begin{align}
 & \sum_{\mu}C_{\mu q\sigma}^{\mathrm{e}}\frac{\partial}{\partial C_{\mu p\sigma}^{\mathrm{e}}}(rs\sigma'|tu\sigma'')\label{eq:two_body_int}\\
 & =\sum_{\mu}C_{\mu q\sigma}^{\mathrm{e}}\frac{\partial}{\partial C_{\mu p\sigma}^{\mathrm{e}}}\sum_{\nu\lambda\eta\gamma}C_{\nu r\sigma'}^{\mathrm{e}}C_{\lambda s\sigma'}^{\mathrm{e}}C_{\eta t\sigma''}^{\mathrm{e}}C_{\gamma u\sigma''}^{\mathrm{e}}(\nu\lambda\sigma'|\eta\gamma\sigma'')\\
 & =\delta_{rp}\delta_{\sigma\sigma'}(qs\sigma'|tu\sigma'')+\delta_{sp}\delta_{\sigma\sigma'}(rq\sigma'|tu\sigma'')+\delta_{tp}\delta_{\sigma\sigma''}(rs\sigma'|qu\sigma'')+\delta_{up}\delta_{\sigma\sigma''}(rs\sigma'|tq\sigma''),
\end{align}
where $r,s,t,u,p,q$ all represent general electronic orbitals. Next,
we consider the derivatives of $V^{\mathrm{xc}}$ with respect to
$C^{\mathrm{e}}$:
\begin{equation}
\sum_{\mu}C_{\mu q\sigma}^{\mathrm{e}}\frac{\partial V_{rs\sigma'}^{\mathrm{xc}}}{\partial C_{\mu p\sigma}^{\mathrm{e}}}=\sum_{\mu}C_{\mu q\sigma}^{\mathrm{e}}\frac{\partial}{\partial C_{\mu p\sigma}^{\mathrm{e}}}\bigg[\int\text{d}\bm{r}\phi_{r\sigma'}^{\mathrm{e}}(\bm{r})\frac{\delta E^{\mathrm{xc}}}{\delta\rho_{\sigma'}^{\mathrm{e}}(\bm{r})}\phi_{s\sigma'}^{\mathrm{e}}(\bm{r})\bigg],\label{Ce_Ce}
\end{equation}
where $\phi^{\mathrm{e}}$ stands for the electronic molecular orbital.
Expand it into AO basis,
\begin{align}
\sum_{\mu}C_{\mu q\sigma}^{\mathrm{e}}\frac{\partial V_{rs\sigma'}^{\mathrm{xc}}}{\partial C_{\mu p\sigma}^{\mathrm{e}}} & =\sum_{\mu}C_{\mu q\sigma}^{\mathrm{e}}\frac{\partial}{\partial C_{\mu p\sigma}^{\mathrm{e}}}\bigg[\int\text{d}\bm{r}\phi_{r\sigma'}^{\mathrm{e}}(\bm{r})\frac{\delta E^{\mathrm{xc}}}{\delta\rho_{\sigma'}^{\mathrm{e}}(\bm{r})}\phi_{s\sigma'}^{\mathrm{e}}(\bm{r})\bigg]\label{vxc_step1}\\
 & =\sum_{\mu}C_{\mu q\sigma}^{\mathrm{e}}\frac{\partial}{\partial C_{\mu p\sigma}^{\mathrm{e}}}\bigg[\sum_{\nu\lambda}C_{\nu r\sigma'}^{\mathrm{e}}C_{\lambda s\sigma'}^{\mathrm{e}}\int\text{d}\bm{r}\chi_{\nu\sigma'}^{\mathrm{e}}(\bm{r})\frac{\delta E^{\mathrm{xc}}}{\delta\rho_{\sigma'}^{\mathrm{e}}(\bm{r})}\chi_{\lambda\sigma'}^{\mathrm{e}}(\bm{r})\bigg],
\end{align}
where $\chi^{\mathrm{e}}$ is the electronic atomic orbital. Utilizing
Eq.~\ref{Ce_Ce}, and being aware of the $C^{\mathrm{e}}$ dependence
in $\rho^{\mathrm{e}}$, we can further express Eq.~\ref{vxc_step1}
as
\begin{align}
 & \sum_{\mu}C^{\mathrm{e}}_{\mu q\sigma}\frac{\partial V^{\mathrm{xc}}_{rs\sigma'}}{\partial C^{\mathrm{e}}_{\mu p\sigma}}\\
 & =\delta_{\sigma\sigma'}\left(V^{\mathrm{xc}}_{qs\sigma}\delta_{rp}+V^{\mathrm{xc}}_{rq\sigma}\delta_{sp}\right)+\sum_{\mu}C^{\mathrm{e}}_{\mu q\sigma}\int\text{d}\bm{r}\text{d}\bm{r}'\phi^{\mathrm{e}}_{r\sigma'}(\bm{r})\sum_{\sigma''}\frac{\delta^{2}E^{\mathrm{xc}}}{\delta\rho^{\mathrm{e}}_{\sigma'}(\bm{r})\delta\rho^{\mathrm{e}}_{\sigma''}(\bm{r}')}\frac{\partial\rho^{\mathrm{e}}_{\sigma''}(\bm{r}')}{\partial C^{\mathrm{e}}_{\mu p\sigma}}\phi^{\mathrm{e}}_{s\sigma'}(\bm{r})\\
 & =\delta_{\sigma\sigma'}\left(V^{\mathrm{xc}}_{qs\sigma}\delta_{rp}+V^{\mathrm{xc}}_{rq\sigma}\delta_{sp}\right)\nonumber \\
 & +\sum_{\mu}C^{\mathrm{e}}_{\mu q\sigma}\int\text{d}\bm{r}\text{d}\bm{r}'\phi^{\mathrm{e}}_{r\sigma'}(\bm{r})\sum_{\sigma''}\frac{\delta^{2}E^{\mathrm{xc}}}{\delta\rho^{\mathrm{e}}_{\sigma'}(\bm{r})\delta\rho^{\mathrm{e}}_{\sigma''}(\bm{r}')}\frac{\partial}{\partial C^{\mathrm{e}}_{\mu p\sigma}}\left[\sum^{\mathrm{occ},\sigma''}_{i}\sum_{\nu\lambda}C^{\mathrm{e}}_{\nu i\sigma''}C^{\mathrm{e}}_{\lambda i\sigma''}\chi^{\mathrm{e}}_{\nu\sigma''}(\bm{r}')\chi^{\mathrm{e}}_{\lambda\sigma''}(\bm{r}')\right]\phi^{\mathrm{e}}_{s\sigma'}(\bm{r})\\
 & =\delta_{\sigma\sigma'}\left(V^{\mathrm{xc}}_{qs\sigma}\delta_{rp}+V^{\mathrm{xc}}_{rq\sigma}\delta_{sp}\right)\nonumber \\
 & +\sum^{\mathrm{occ},\sigma}_{i}\delta_{ip}\int\text{d}\bm{r}\text{d}\bm{r}'\phi^{\mathrm{e}}_{r\sigma'}(\bm{r})\frac{\delta^{2}E^{\mathrm{xc}}}{\delta\rho^{\mathrm{e}}_{\sigma'}(\bm{r})\delta\rho^{\mathrm{e}}_{\sigma}(\bm{r}')}\phi^{\mathrm{e}}_{s\sigma'}(\bm{r})\big[\phi^{\mathrm{e}}_{q\sigma}(\bm{r}')\phi^{\mathrm{e}}_{i\sigma}(\bm{r}')+\phi^{\mathrm{e}}_{i\sigma}(\bm{r}')\phi^{\mathrm{e}}_{q\sigma}(\bm{r}')\big]\\
 & =\delta_{\sigma\sigma'}\left(V^{\mathrm{xc}}_{qs\sigma}\delta_{rp}+V^{\mathrm{xc}}_{rq\sigma}\delta_{sp}\right)+2\sum^{\mathrm{occ},\sigma}_{i}\delta_{ip}f^{\mathrm{xc}}_{rs\sigma'qi\sigma},
\end{align}
where
\begin{equation}
f^{\mathrm{xc}}_{pq\sigma rs\sigma'}=\int\text{d}\bm{r}\text{d}\bm{r}'\phi^{\mathrm{e}}_{p\sigma}(\bm{r})\phi^{\mathrm{e}}_{q\sigma}(\bm{r})\frac{\delta^{2}E^{\mathrm{xc}}}{\delta\rho^{\mathrm{e}}_{\sigma}(\bm{r})\delta\rho^{\mathrm{e}}_{\sigma'}(\bm{r}')}\phi^{\mathrm{e}}_{r\sigma'}(\bm{r}')\phi^{\mathrm{e}}_{s\sigma'}(\bm{r}').
\end{equation}
Similar procedures can be applied to $V^{\mathrm{epc}}$, which will
lead to
\begin{equation}
\sum_{\mu}C^{\mathrm{e}}_{\mu q\sigma}\frac{\partial V^{\mathrm{epc}}_{rs\sigma'}[\rho^{\mathrm{e}},\rho^{\mathrm{n}}_{I}]}{\partial C^{\mathrm{e}}_{\mu p\sigma}}=\delta_{\sigma\sigma'}\left(V^{\mathrm{epc},I}_{qs\sigma}\delta_{rp}+V^{\mathrm{epc},I}_{rq\sigma}\delta_{sp}\right)+2\sum^{\mathrm{occ},\sigma}_{i}\delta_{ip}f^{\mathrm{epc,ee},I}_{rs\sigma'qi\sigma},
\end{equation}
where
\begin{equation}
f^{\mathrm{epc,e(n)e(n),}I}_{pq\sigma(p'q'I),rs\sigma'(r's'I)}=\int\text{d}\bm{r}\text{d}\bm{r}'\phi_{p\sigma(p'I)}(\bm{r})\phi_{q\sigma(q'I)}(\bm{r})\frac{\delta^{2}E^{\mathrm{epc}}[\rho^{\mathrm{e}},\rho^{\mathrm{n}}_{I}]}{\delta\rho^{\mathrm{e(n)}}_{\sigma(I)}(\bm{r})\delta\rho^{\mathrm{e(n)}}_{\sigma'(I)}(\bm{r}')}\phi_{r\sigma'(r'I)}(\bm{r}')\phi_{s\sigma'(s'I)}(\bm{r}'),
\end{equation}
is the vector element of the electron-proton correlation kernel. Combining
all the derivations above, we can express Eq.~\ref{eq:Fock_step1}
as
\begin{align}
 & \sum_{\mu}C_{\mu q\sigma}^{\mathrm{e}}\frac{\partial F_{rs\sigma'}^{\mathrm{e}}}{\partial C_{\mu p\sigma}^{\mathrm{e}}}\nonumber \\
 & =\delta_{\sigma\sigma'}\delta_{rp}h_{qs\sigma}^{\mathrm{e}}+\delta_{\sigma\sigma'}\delta_{sp}h_{rq\sigma}^{\mathrm{e}}\nonumber \\
 & +\sum_{i\sigma''}^{\mathrm{occ}}\bigg\{[(qs\sigma|ii\sigma'')-c_{\mathrm{x}}\delta_{\sigma\sigma''}(qi\sigma|is\sigma)]\delta_{rp}+[(rq\sigma|ii\sigma'')-c_{\mathrm{x}}\delta_{\sigma\sigma''}(ri\sigma|iq\sigma)]\delta_{sp}\bigg\}\delta_{\sigma\sigma'}\nonumber \\
 & +\sum_{i}^{\mathrm{occ},\sigma}\delta_{ip}\bigg\{2(rs\sigma'|qi\sigma)-c_{\mathrm{x}}\delta_{\sigma'\sigma}[(ri\sigma'|qs\sigma')+(rq\sigma'|is\sigma')]\bigg\}\nonumber \\
 & +\delta_{\sigma\sigma'}\left(V_{qs\sigma}^{\mathrm{xc}}\delta_{rp}+V_{rq\sigma}^{\mathrm{xc}}\delta_{sp}\right)+2\sum_{i}^{\mathrm{occ},\sigma}\delta_{ip}f_{rs\sigma'qi\sigma}^{\mathrm{xc}}\nonumber \\
 & +\sum_{I}\left\{ \delta_{\sigma\sigma'}\left(V_{qs\sigma}^{\mathrm{epc},I}\delta_{rp}+V_{rq\sigma}^{\mathrm{epc},I}\delta_{sp}\right)+2\sum_{i}^{\mathrm{occ},\sigma}\delta_{ip}f_{rs\sigma'qi\sigma}^{\mathrm{epc,ee},I}\right\} \nonumber \\
 & -\sum_{i'I}Z_{I}\delta_{\sigma\sigma'}\left((qs\sigma'|i'i'I)\delta_{rp}+(rq\sigma'|i'i'I)\delta_{sp}\right).
\end{align}
Rearranging and using the electronic Fock matrix (Eq.~\ref{eq:elec_fock}),
the above expression can be simplified as
\begin{align}
 & \sum_{\mu}C^{\mathrm{e}}_{\mu q\sigma}\frac{\partial F^{\mathrm{e}}_{rs\sigma'}}{\partial C^{\mathrm{e}}_{\mu p\sigma}}\nonumber \\
 & =\delta_{rp}\delta_{\sigma\sigma'}\bigg\{ h^{\mathrm{e}}_{qs\sigma}+\sum^{\mathrm{occ}}_{i\sigma''}[(qs\sigma|ii\sigma'')-c_{\mathrm{x}}\delta_{\sigma\sigma''}(qi\sigma|is\sigma)]+V^{\mathrm{xc}}_{qs\sigma}+\sum_{I}V^{\mathrm{epc},I}_{qs\sigma}-\sum_{I}Z_{I}(qs\sigma'|i'i'I)\bigg\}\nonumber \\
 & +\delta_{sp}\delta_{\sigma\sigma'}\bigg\{h^{\mathrm{e}}_{rq\sigma}+\sum^{\mathrm{occ}}_{i\sigma''}[(rq\sigma|ii\sigma'')-c_{\mathrm{x}}\delta_{\sigma\sigma''}(ri\sigma|iq\sigma)]+V^{\mathrm{xc}}_{rq\sigma}+\sum_{I}V^{\mathrm{epc},I}_{rq\sigma}-\sum_{I}Z_{I}(rq\sigma'|i'i'I)\bigg\}\nonumber \\
 & +\sum^{\mathrm{occ},\sigma}_{i}\delta_{ip}\bigg\{2(rs\sigma'|qi\sigma)-c_{\mathrm{x}}\delta_{\sigma'\sigma}[(ri\sigma'|qs\sigma')+(rq\sigma'|is\sigma')]\bigg\}\nonumber \\
 & +2\sum^{\mathrm{occ},\sigma}_{i}\delta_{ip}f^{\mathrm{xc}}_{rs\sigma'qi\sigma}+2\sum_{I}\sum^{\mathrm{occ},\sigma}_{i}\delta_{ip}f^{\mathrm{epc,ee},I}_{rs\sigma'qi\sigma}\\
 & =\delta_{rp}\delta_{\sigma\sigma'}F^{\mathrm{e}}_{qs\sigma}+\delta_{sp}\delta_{\sigma\sigma'}F^{\mathrm{e}}_{rq\sigma}\nonumber \\
 & \quad+\sum^{\mathrm{occ},\sigma}_{i}\delta_{ip}\bigg\{2(rs\sigma'|qi\sigma)-c_{\mathrm{x}}\delta_{\sigma'\sigma}[(ri\sigma'|qs\sigma')+(rq\sigma'|is\sigma')]+2f^{\mathrm{xc}}_{rs\sigma'qi\sigma}+2\sum_{I}f^{\mathrm{epc,ee},I}_{rs\sigma'qi\sigma}\bigg\}.\label{eq:ctd_dFe_dCe}
\end{align}

\subsubsection{Electronic Fock matrix--nuclear MO coefficients\label{subsec:Electronic-Fock-matrix-nuclear}}

The electronic Fock matrix only has $C^{\mathrm{n}}$ dependence through
$V^{\mathrm{epc}}$ and the Coulomb interaction with nuclei. Therefore,
\begin{align}
\sum_{\mu'}C_{\mu'q'I}^{\mathrm{n}}\frac{\partial F_{rs\sigma'}^{\mathrm{e}}}{\partial C_{\mu'p'I}^{\mathrm{n}}} & =\sum_{\mu'}C_{\mu'q'I}^{\mathrm{n}}\frac{\partial}{\partial C_{\mu'p'I}^{\mathrm{n}}}\left[\sum_{J}V_{rs\sigma'}^{\mathrm{epc}}[\rho^{\mathrm{e}},\rho_{J}^{\mathrm{n}}]-\sum_{J}Z_{J}(rs\sigma'|j'j'J)\right]\label{eq:ctd_dFe_dCn}\\
 & =2\delta_{i'p'}f_{rs\sigma'i'q'I}^{\mathrm{epc,en},I}-\sum_{\mu'}C_{\mu'q'I}^{\mathrm{n}}\frac{\partial}{\partial C_{\mu'p'I}^{\mathrm{n}}}\left[\sum_{J}\sum_{\nu'\lambda'}C_{\nu'j'J}^{\mathrm{n}}C_{\lambda'j'J}^{\mathrm{n}}Z_{J}(rs\sigma'|\nu'\lambda'J)\right]\\
 & =2\delta_{i'p'}f_{rs\sigma'i'q'I}^{\mathrm{epc,en},I}-Z_{I}(rs\sigma'|q'i'I)\delta_{i'p'}-Z_{I}(rs\sigma'|i'q'I)\delta_{i'p'}\\
 & =2\delta_{i'p'}f_{rs\sigma'i'q'I}^{\mathrm{epc,en},I}-2Z_{I}(rs\sigma'|q'i'I)\delta_{i'p'},
\end{align}
where the first term comes from
\begin{align}
\sum_{\mu'}C^{\mathrm{n}}_{\mu'q'I}\frac{\partial\left(\sum_{J}V^{\mathrm{epc}}_{rs\sigma'}[\rho^{\mathrm{e}},\rho^{\mathrm{n}}_{J}]\right)}{\partial C^{\mathrm{n}}_{\mu'p'I}} & =\sum_{\mu'}C^{\mathrm{n}}_{\mu'q'I}\frac{\partial V^{\mathrm{epc}}_{rs\sigma'}[\rho^{\mathrm{e}},\rho^{\mathrm{n}}_{I}]}{\partial C^{\mathrm{n}}_{\mu'p'I}}\\
 & =\sum_{\mu'}C^{\mathrm{n}}_{\mu'q'I}\int\text{d}\bm{r}\text{d}\bm{r}'\phi_{r\sigma'}(\bm{r})\phi_{s\sigma'}(\bm{r})\frac{\delta^{2}E^{\mathrm{epc}}[\rho^{\mathrm{e}},\rho^{\mathrm{n}}_{I}]}{\delta\rho^{\mathrm{e}}_{\sigma'}\delta\rho^{\mathrm{n}}_{I}}\frac{\partial\rho^{\mathrm{n}}_{I}(\bm{r}')}{\partial C^{\mathrm{n}}_{\mu'p'I}}\\
 & =2\delta_{i'p'}\int\text{d}\bm{r}\text{d}\bm{r}'\phi_{r\sigma'}(\bm{r})\phi_{s\sigma'}(\bm{r})\frac{\delta^{2}E^{\mathrm{epc}}[\rho^{\mathrm{e}},\rho^{\mathrm{n}}_{I}]}{\delta\rho^{\mathrm{e}}_{\sigma'}\delta\rho^{\mathrm{n}}_{I}}\phi_{i'I}(\bm{r}')\phi_{q'I}(\bm{r}')\\
 & =2\delta_{i'p'}f^{\mathrm{epc,en},I}_{rs\sigma'i'q'I}.
\end{align}

\subsubsection{Nuclear Fock matrix--electronic MO coefficients\label{subsec:Nuclear-Fock-matrix-elec}}

The nuclear Fock matrix is
\begin{equation}
F_{r's'I}^{\mathrm{n}}=h_{r's'I}^{\mathrm{n}}-Z_{I}\sum_{i\sigma'}^{\mathrm{occ}}(r's'I|ii\sigma')+V_{r's'I}^{\mathrm{epc}}[\rho^{\mathrm{e}},\rho_{I}^{\mathrm{n}}]+Z_{I}\sum_{j'\ne i',J\neq I}Z_{J}(r's'I|j'j'J)+\bm{f}_{I}^{\mathrm n}\cdot\overline{\bm R}_{r's'I}^{\mathrm n},
\end{equation}
and we can see that only $V_{r's'I}^{\mathrm{epc}}$ and the Coulomb
term between electrons and nuclei depend on $C^{\mathrm{e}}$. Therefore,
\begin{align}
\sum_{\mu}C_{\mu q\sigma}^{\mathrm{e}}\frac{\partial F_{r's'I}^{\mathrm{n}}}{\partial C_{\mu p\sigma}^{\mathrm{e}}} & =\sum_{\mu}C_{\mu q\sigma}^{\mathrm{e}}\frac{\partial}{\partial C_{\mu p\sigma}^{\mathrm{e}}}\left[-Z_{I}\sum_{i\sigma'}^{\mathrm{occ}}(r's'I|ii\sigma')+V_{r's'I}^{\mathrm{epc}}[\rho^{\mathrm{e}},\rho_{I}^{\mathrm{n}}]\right]\label{eq:ctd_dFn_dCe}\\
 & =-Z_{I}\sum_{i\sigma'}^{\mathrm{occ}}\big[(r's'I|qi\sigma')+(r's'I|iq\sigma')\big]\delta_{ip}\delta_{\sigma\sigma'}+2\sum_{i}^{\mathrm{occ},\sigma}\delta_{ip}f_{r's'Iqi\sigma}^{\mathrm{epc,ne},I},
\end{align}
where the last term comes from
\begin{align}
\sum_{\mu}C_{\mu q\sigma}^{\mathrm{e}}\frac{\partial V_{r's'I}^{\mathrm{epc}}[\rho^{\mathrm{e}},\rho_{I}^{\mathrm{n}}]}{\partial C_{\mu p\sigma}^{\mathrm{e}}} & =\sum_{\mu}C_{\mu q\sigma}^{\mathrm{e}}\int\text{d}\bm{r}\text{d}\bm{r}'\phi_{r'}(\bm{r})\sum_{\sigma'}\frac{\delta^{2}E^{\mathrm{epc}}[\rho^{\mathrm{e}},\rho^{\mathrm{n}}]}{\delta\rho^{\mathrm{n}}\delta\rho_{\sigma'}^{\mathrm{e}}}\frac{\partial\rho_{\sigma'}^{\mathrm{e}}(\bm{r}')}{\partial C_{\mu p\sigma}^{\mathrm{e}}}\phi_{s'}(\bm{r})\\
 & =2\sum_{i}^{\mathrm{occ},\sigma}\delta_{ip}f_{r's'Iqi\sigma}^{\mathrm{epc,ne},I}.
\end{align}

\subsubsection{Nuclear Fock matrix--nuclear MO coefficients\label{subsec:Nuclear-Fock-matrix-nuclear}}

Since the quantum nuclei in the CNEO framework are treated as distinguishable
particles, the nuclear Fock matrix contains only one-particle terms and
intercomponent Coulomb interactions, with no exchange interaction between
different quantum nuclei. Therefore, the nuclear--nuclear Fock-matrix
derivatives have a simpler form than their electronic counterparts. Expanding the nuclear Fock in the nuclear AO
basis, and following similar steps in
Section.~\ref{subsubsection: fock_ee}, we can get the nuclear-nuclear
derivative as
\begin{align}
 & \sum_{\mu'}C_{\mu'q'I}^{\mathrm{n}}\frac{\partial F_{r's'K}^{\mathrm{n}}}{\partial C_{\mu'p'I}^{\mathrm{n}}}\nonumber \\
 & =\delta_{r'p'}\delta_{IK}\left[h_{q's'K}^{\mathrm{n}}-Z_{K}\sum_{i\sigma'}(q's'K|ii\sigma')+V_{q's'K}^{\mathrm{epc}}[\rho^{\mathrm{e}},\rho_{K}^{\mathrm{n}}]+Z_{K}\sum_{J\neq K}Z_{J}(q's'K|j'j'J)+\bm{f}_{I}^{\mathrm n}\cdot\overline{\bm R}_{q's'I}^{\mathrm n}\right]\nonumber \\
 & +\delta_{s'p'}\delta_{IK}\left[h_{r'q'K}^{\mathrm{n}}-Z_{K}\sum_{i\sigma'}(r'q'K|ii\sigma')+V_{r'q'K}^{\mathrm{epc}}[\rho^{\mathrm{e}},\rho_{K}^{\mathrm{n}}]+Z_{K}\sum_{J\neq K}Z_{J}(r'q'K|j'j'J)+\bm{f}_{I}^{\mathrm n}\cdot\overline{\bm R}_{r'q'I}^{\mathrm n}\right]\nonumber \\
 & +2f_{r's'Iq'i'I}^{\mathrm{epc,nn},I}\delta_{i'p'}\delta_{IK}+2Z_{K}\sum_{J\neq K}Z_J\delta_{JI}\delta_{j'p'}(r's'K|j'q'J)\\
 & =\delta_{r'p'}\delta_{IK}F_{q's'K}^{\mathrm{n}}+\delta_{s'p'}\delta_{IK}F_{r'q'K}^{\mathrm{n}}+2f_{r's'Iq'i'I}^{\mathrm{epc,nn},I}\delta_{i'p'}\delta_{IK}+2Z_{K}\sum_{J\neq K}Z_J\delta_{JI}\delta_{j'p'}(r's'K|j'q'J).\label{eq:ctd_dFn_dCn}
\end{align}

\subsection{$\Lambda$-related term: $Q$}

Now, with the derivation of Fock matrix derivatives, we will calculate
the derivative of $\langle X,Y|\Lambda|X,Y\rangle$. Following
Ref.~\cite{Furche2002},
we write this term as
\begin{align}
\langle X,Y|\Lambda|X,Y\rangle & =\langle X,Y|\begin{pmatrix}A & B\\
B & A
\end{pmatrix}|X,Y\rangle\label{eq:ctd_lambda}\\
 & =\frac{1}{2}\left[(X+Y)(A+B)(X+Y)+(X-Y)(A-B)(X-Y)\right]\\
 & =\frac{1}{2}\left[\sum_{ia\sigma jb\sigma'}(A+B)_{ia\sigma jb\sigma'}(X+Y)_{ia\sigma}^{*}(X+Y)_{jb\sigma'}+\sum_{ia\sigma jb\sigma'}(A-B)_{ia\sigma jb\sigma'}(X-Y)_{ia\sigma}^{*}(X-Y)_{jb\sigma'}\right].
\end{align}
From the definition of $A,B$ matrices, we have
\begin{align}
(A+B)_{kc\sigma'ld\sigma''}={}&
\delta_{\sigma'\sigma''}\left(F_{cd\sigma'}^{\mathrm{e}}\delta_{kl}
-F_{kl\sigma'}^{\mathrm{e}}\delta_{cd}\right)
+2(kc\sigma'|ld\sigma'') \nonumber\\
&-c_{\mathrm{x}}\delta_{\sigma'\sigma''}\left[(lc\sigma'|kd\sigma')
+(cd\sigma'|kl\sigma')\right]
+2f_{kc\sigma'ld\sigma''}^{\mathrm{xc}}
+2\sum_{I}f_{kc\sigma'ld\sigma''}^{\mathrm{epc,ee},I},
\label{eq:ctd_ApB}
\end{align}
\begin{equation}
(A-B)_{kc\sigma'ld\sigma''}=\delta_{\sigma'\sigma''}\left(F_{cd\sigma'}^{\mathrm{e}}\delta_{kl}-F_{kl\sigma'}^{\mathrm{e}}\delta_{cd}\right)+c_{\mathrm{x}}\delta_{\sigma'\sigma''}\left[(lc\sigma'|kd\sigma')-(cd\sigma'|kl\sigma')\right].\label{eq:ctd_AmB}
\end{equation}
Define the electronic and nuclear source matrices
\begin{equation}
Q_{pq\sigma}^{\mathrm{e}}=\sum_{\mu}C_{\mu q\sigma}^{\mathrm{e}}\frac{\partial}{\partial C_{\mu p\sigma}^{\mathrm{e}}}\langle X,Y|\Lambda|X,Y\rangle,
\end{equation}
\begin{equation}
Q_{p'q'I}^{\mathrm{n}}=\sum_{\mu'}C_{\mu'q'I}^{\mathrm{n}}\frac{\partial}{\partial C_{\mu'p'I}^{\mathrm{n}}}\langle X,Y|\Lambda|X,Y\rangle,
\end{equation}
and we will derive both electronic $Q^{\mathrm{e}}$ and nuclear $Q^{\mathrm{n}}$
in four blocks:
\begin{itemize}
\item occ-occ block: $p=i,q=j$ for $Q^{\mathrm{e}}$ and $p'=i',q'=i'$
for $Q^{\mathrm{n}}$
\item occ-vir block: $p=i,q=a$ for $Q^{\mathrm{e}}$ and $p'=i',q'=a'$
for $Q^{\mathrm{n}}$
\item vir-occ block:$p=a,q=i$ for $Q^{\mathrm{e}}$ and $p'=a',q'=i'$
for $Q^{\mathrm{n}}$
\item vir-vir block: $p=a,q=b$ for $Q^{\mathrm{e}}$and $p'=a',q'=b'$
for $Q^{\mathrm{n}}$
\end{itemize}

\subsubsection{Electronic $Q^{\mathrm{e}}$}

The electronic source matrix includes the exchange--correlation and
electron--proton-correlation contributions to the CNEO response matrix.

\paragraph{occ - occ block ($p=i,q=j$)}

We will first derive the derivatives of each component in $(A+B)$
and $(A-B)$
\begin{align}
\sum_{\mu}C_{\mu j\sigma}^{\mathrm{e}}\frac{\partial F_{kl\sigma'}^{\mathrm{e}}}{\partial C_{\mu i\sigma}^{\mathrm{e}}} & =\delta_{\sigma\sigma'}\delta_{ki}F_{jl\sigma}^{\mathrm{e}}+\delta_{\sigma\sigma'}\delta_{li}F_{kj\sigma}^{\mathrm{e}}\nonumber \\
 & +2(kl\sigma'|ji\sigma)-c_{\mathrm{x}}\delta_{\sigma'\sigma}\big[(ki\sigma'|jl\sigma')+(kj\sigma'|il\sigma')\big]+2f_{kl\sigma'ji\sigma}^{\mathrm{xc}}+2\sum_{I}f_{kl\sigma'ji\sigma}^{\mathrm{epc,ee},I}\\
 & =\delta_{\sigma\sigma'}\delta_{ki}\delta_{jl}\varepsilon_{j\sigma}+\delta_{\sigma\sigma'}\delta_{li}\delta_{kj}\varepsilon_{j\sigma}\nonumber \\
 & +2(kl\sigma'|ji\sigma)-c_{\mathrm{x}}\delta_{\sigma'\sigma}\big[(ki\sigma'|jl\sigma')+(kj\sigma'|il\sigma')\big]+2f_{kl\sigma'ji\sigma}^{\mathrm{xc}}+2\sum_{I}f_{kl\sigma'ji\sigma}^{\mathrm{epc,ee},I}.
\end{align}
Similarly for the derivatives of $F_{cd\sigma'}^{\mathrm{e}}$, we
have
\begin{align}
\sum_{\mu}C_{\mu j\sigma}^{\mathrm{e}}\frac{\partial F_{cd\sigma'}}{\partial C_{\mu i\sigma}^{\mathrm{e}}} & =\delta_{\sigma\sigma'}\delta_{ci}F_{jd\sigma}^{\mathrm{e}}+\delta_{\sigma\sigma'}\delta_{di}F_{cj\sigma}^{\mathrm{e}}\nonumber \\
 & +2(cd\sigma'|ji\sigma)-c_{\mathrm{x}}\delta_{\sigma'\sigma}[(ci\sigma'|jd\sigma')+(cj\sigma'|id\sigma')]+2f_{cd\sigma'ji\sigma}^{\mathrm{xc}}+2\sum_{I}f_{cd\sigma'ji\sigma}^{\mathrm{epc,ee},I}\\
 & =2(cd\sigma'|ji\sigma)-c_{\mathrm{x}}\delta_{\sigma'\sigma}[(ci\sigma'|jd\sigma')+(cj\sigma'|id\sigma')]+2f_{cd\sigma'ji\sigma}^{\mathrm{xc}}+2\sum_{I}f_{cd\sigma'ji\sigma}^{\mathrm{epc,ee},I}.
\end{align}
The derivatives of two-body integrals can all be obtained using Eq.~\ref{eq:two_body_int},
and we additionally need to express the derivatives of $f^{\mathrm{xc}}$
and $f^{\mathrm{epc}}$. The general form of derivatives of $f^{\mathrm{xc}}$
can be derived as:
\begin{align}
\sum_{\mu}C_{\mu q\sigma}^{\mathrm{e}}\frac{\partial f_{rs\sigma'tu\sigma''}^{\mathrm{xc}}}{\partial C_{\mu p\sigma}^{\mathrm{e}}} & =\sum_{\mu}C_{\mu q\sigma}^{\mathrm{e}}\frac{\partial}{\partial C_{\mu p\sigma}^{\mathrm{e}}}\int\mathrm{d}\bm{r}\mathrm{d}\bm{r}'\ \phi_{r\sigma'}^{\mathrm{e}}(\bm{r})\phi_{s\sigma'}^{\mathrm{e}}(\bm{r})\frac{\delta^{2}E_{\mathrm{xc}}}{\delta\rho_{\sigma'}^{\mathrm{e}}(\bm{r})\delta\rho_{\sigma''}^{\mathrm{e}}(\bm{r}')}\phi_{t\sigma''}^{\mathrm{e}}(\bm{r}')\phi_{u\sigma''}^{\mathrm{e}}(\bm{r}')\label{eq:ctd_dxc_de}\\
 & =\delta_{\sigma\sigma'}\delta_{rp}f_{qs\sigma'tu\sigma''}^{\mathrm{xc}}+\delta_{\sigma\sigma'}\delta_{sp}f_{rq\sigma'tu\sigma''}^{\mathrm{xc}}+\delta_{\sigma\sigma''}\delta_{tp}f_{rs\sigma'qu\sigma''}^{\mathrm{xc}}+\delta_{\sigma\sigma''}\delta_{up}f_{rs\sigma'tq\sigma''}^{\mathrm{xc}}\nonumber \\
 & +2\sum_{k}^{\mathrm{occ},\sigma}\delta_{kp}g_{rs\sigma'tu\sigma''qk\sigma}^{\mathrm{xc}},
\end{align}
where
\begin{equation}
g^{\mathrm{xc}}_{rs\sigma tu\sigma'qk\sigma''}=\int\mathrm{d}\bm{r}\mathrm{d}\bm{r}'\mathrm{d}\bm{r}''\ \phi^{\mathrm{e}}_{r\sigma}(\bm{r})\phi^{\mathrm{e}}_{s\sigma}(\bm{r})\frac{\delta^{3}E_{\mathrm{xc}}}{\delta\rho^{\mathrm{e}}_{\sigma}(\bm{r})\delta\rho^{\mathrm{e}}_{\sigma'}(\bm{r}')\delta\rho^{\mathrm{e}}_{\sigma''}(\bm{r}'')}\phi^{\mathrm{e}}_{t\sigma'}(\bm{r}')\phi^{\mathrm{e}}_{u\sigma'}(\bm{r}')\phi^{\mathrm{e}}_{q\sigma''}(\bm{r}'')\phi^{\mathrm{e}}_{k\sigma''}(\bm{r}'').
\end{equation}
Therefore,
\begin{equation}
\sum_{\mu}C_{\mu j\sigma}^{\mathrm{e}}\frac{\partial f_{kc\sigma'ld\sigma''}^{\mathrm{xc}}}{\partial C_{\mu i\sigma}^{\mathrm{e}}}=\delta_{ki}\delta_{\sigma\sigma'}f_{jc\sigma'ld\sigma''}^{\mathrm{xc}}+\delta_{li}\delta_{\sigma\sigma''}f_{kc\sigma'jd\sigma''}^{\mathrm{xc}}+2g_{kc\sigma'ld\sigma''ji\sigma}^{\mathrm{xc}}.
\end{equation}
Similarly, for the epc term, we can have
\begin{equation}
\sum_{\mu}C_{\mu j\sigma}^{\mathrm{e}}\frac{\partial f_{kc\sigma'ld\sigma''}^{\mathrm{epc,ee},I}}{\partial C_{\mu i\sigma}^{\mathrm{e}}}=\delta_{ki}\delta_{\sigma\sigma'}f_{jc\sigma'ld\sigma''}^{\mathrm{epc,ee},I}+\delta_{li}\delta_{\sigma\sigma''}f_{kc\sigma'jd\sigma''}^{\mathrm{epc,ee},I}+2g_{kc\sigma'ld\sigma''ji\sigma}^{\mathrm{epc,eee},I},
\end{equation}
where
\begin{equation}
g_{kc\sigma'ld\sigma''ji\sigma}^{\mathrm{epc,eee},I}=\int\text{d}\bm{r}\text{d}\bm{r}'\text{d}\bm{r}''\frac{\delta^{3}E^{\mathrm{epc},I}[\rho^{\mathrm{e}},\rho_{I}^{\mathrm{n}}]}{\delta\rho_{\sigma}^{\mathrm{e}}(\bm{r})\delta\rho_{\sigma'}^{\mathrm{e}}(\bm{r}')\delta\rho_{\sigma''}^{\mathrm{e}}(\bm{r}'')}\phi_{j\sigma}^{\mathrm{e}}(\bm{r})\phi_{i\sigma}^{\mathrm{e}}(\bm{r})\phi_{k\sigma'}^{\mathrm{e}}(\bm{r}')\phi_{c\sigma'}^{\mathrm{e}}(\bm{r}')\phi_{l\sigma''}^{\mathrm{e}}(\bm{r}'')\phi_{d\sigma''}^{\mathrm{e}}(\bm{r}'').
\end{equation}

Combining all these derivatives, we obtain
\begin{align}
 & \sum_{\mu}C_{\mu j\sigma}^{\mathrm{e}}\frac{\partial(A+B)_{kc\sigma'ld\sigma''}}{\partial C_{\mu i\sigma}^{\mathrm{e}}}\nonumber \\
 & =\delta_{\sigma'\sigma''}\delta_{kl}\left\{ 2(cd\sigma'|ji\sigma)-c_{\mathrm{x}}\delta_{\sigma'\sigma}[(ci\sigma'|jd\sigma')+(cj\sigma'|id\sigma')]+2f_{cd\sigma'ji\sigma}^{\mathrm{xc}}+2\sum_{I}f_{cd\sigma'ji\sigma}^{\mathrm{epc,ee},I}\right\} \nonumber \\
 & -\delta_{\sigma'\sigma''}\delta_{cd}\left\{ \delta_{\sigma\sigma'}\delta_{ki}\delta_{jl}\varepsilon_{j\sigma}+\delta_{\sigma\sigma'}\delta_{li}\delta_{kj}\varepsilon_{j\sigma}+2(kl\sigma'|ji\sigma)-c_{\mathrm{x}}\delta_{\sigma'\sigma}\big[(ki\sigma|jl\sigma)+(kj\sigma|il\sigma)\big]+2f_{kl\sigma'ji\sigma}^{\mathrm{xc}}+2\sum_{I}f_{kl\sigma'ji\sigma}^{\mathrm{epc,ee},I}\right\} \nonumber \\
 & +\delta_{\sigma\sigma'}\delta_{ki}\left\{ 2(jc\sigma|ld\sigma'')+2f_{jc\sigma'ld\sigma''}^{\mathrm{xc}}+2\sum_{I}f_{jc\sigma ld\sigma''}^{\mathrm{epc,ee},I}-c_{\mathrm{x}}\delta_{\sigma'\sigma''}\big[(lc\sigma'|jd\sigma')+(cd\sigma'|jl\sigma')\big]\right\} \nonumber \\
 & +\delta_{\sigma\sigma''}\delta_{li}\left\{ 2(kc\sigma'|jd\sigma)+2f_{kc\sigma'jd\sigma''}^{\mathrm{xc}}+2\sum_{I}f_{kc\sigma'jd\sigma}^{\mathrm{epc,ee},I}-c_{\mathrm{x}}\delta_{\sigma'\sigma''}\big[(jc\sigma'|kd\sigma')+(cd\sigma'|kj\sigma')\big]\right\} \nonumber \\
 & +4g_{kc\sigma'ld\sigma''ji\sigma}^{\mathrm{xc}}+4\sum_{I}g_{kc\sigma'ld\sigma''ji\sigma}^{\mathrm{epc,eee},I},
\end{align}
and
\begin{align}
 & \sum_{\mu}C_{\mu j\sigma}^{\mathrm{e}}\frac{\partial(A-B)_{kc\sigma'ld\sigma''}}{\partial C_{\mu i\sigma}^{\mathrm{e}}}\nonumber \\
 & =\delta_{\sigma'\sigma''}\delta_{kl}\left\{ 2(cd\sigma'|ji\sigma)-c_{\mathrm{x}}\delta_{\sigma'\sigma}[(ci\sigma'|jd\sigma')+(cj\sigma'|id\sigma')]+2f_{cd\sigma'ji\sigma}^{\mathrm{xc}}+2\sum_{I}f_{cd\sigma'ji\sigma}^{\mathrm{epc,ee},I}\right\} \nonumber \\
 & -\delta_{\sigma'\sigma''}\delta_{cd}\left\{ \delta_{\sigma\sigma'}\delta_{ki}\delta_{jl}\varepsilon_{j\sigma}+\delta_{\sigma\sigma'}\delta_{li}\delta_{kj}\varepsilon_{j\sigma}+2(kl\sigma'|ji\sigma)-c_{\mathrm{x}}\delta_{\sigma'\sigma}\big[(ki\sigma|jl\sigma)+(kj\sigma|il\sigma)\big]+2f_{kl\sigma'ji\sigma}^{\mathrm{xc}}+2\sum_{I}f_{kl\sigma'ji\sigma}^{\mathrm{epc,ee},I}\right\} \nonumber \\
 & +c_{\mathrm{x}}\delta_{\sigma'\sigma''}\delta_{\sigma\sigma'}\delta_{ki}\big[(lc\sigma'|jd\sigma')-(cd\sigma'|jl\sigma')\big]\nonumber \\
 & +c_{\mathrm{x}}\delta_{\sigma'\sigma''}\delta_{\sigma\sigma''}\delta_{li}\big[(jc\sigma'|kd\sigma')-(cd\sigma'|kj\sigma')\big].
\end{align}

Now, we can get the expression of $Q_{ij\sigma}^{\mathrm{e}}$ as
\begin{align}
 & Q_{ij\sigma}^{\mathrm{e}}\nonumber \\
 & =\frac{1}{2}\sum_{kc\sigma'ld\sigma''}\bigg[\sum_{\mu}C_{\mu j\sigma}\frac{\partial(A+B)_{kc\sigma'ld\sigma''}}{\partial C_{\mu i\sigma}}(X+Y)_{kc\sigma'}(X+Y)_{ld\sigma''}\nonumber \\
 & \qquad+\sum_{\mu}C_{\mu j\sigma}\frac{\partial(A-B)_{kc\sigma'ld\sigma''}}{\partial C_{\mu i\sigma}}(X-Y)_{kc\sigma'}(X-Y)_{ld\sigma''}\bigg]\\
 & =\frac{1}{2}\sum_{kcd\sigma'}\left[(X+Y)_{kc\sigma'}(X+Y)_{kd\sigma'}+(X-Y)_{kc\sigma'}(X-Y)_{kd\sigma'}\right]\nonumber \\
 & \qquad\times\left[2(cd\sigma'|ji\sigma)-c_{\mathrm{x}}\delta_{\sigma'\sigma}[(ci\sigma'|jd\sigma')+(cj\sigma'|id\sigma')]+2f_{cd\sigma'ji\sigma}^{\mathrm{xc}}+2\sum_{I}f_{cd\sigma'ji\sigma}^{\mathrm{epc,ee},I}\right]\nonumber \\
 & -\sum_{c}\left[(X+Y)_{ic\sigma}(X+Y)_{jc\sigma}+(X-Y)_{ic\sigma}(X-Y)_{jc\sigma}\right]\varepsilon_{j\sigma}\nonumber \\
 & -\frac{1}{2}\sum_{kc\sigma'l}\left[(X+Y)_{kc\sigma'}(X+Y)_{lc\sigma'}+(X-Y)_{kc\sigma'}(X-Y)_{lc\sigma'}\right]\nonumber \\
 & \qquad\times\big[2(kl\sigma'|ji\sigma)-c_{\mathrm{x}}\delta_{\sigma'\sigma}\big[(ki\sigma|jl\sigma)+(kj\sigma|il\sigma)\big]+2f_{kl\sigma'ji\sigma}^{\mathrm{xc}}+2\sum_{I}f_{kl\sigma'ji\sigma}^{\mathrm{epc,ee},I}\big]\nonumber \\
 & +\sum_{kc\sigma'd}(X+Y)_{kc\sigma'}(X+Y)_{id\sigma}\bigg[2(kc\sigma'|jd\sigma)+2f_{kc\sigma'jd\sigma}^{\mathrm{xc}}+2\sum_{I}f_{kc\sigma'jd\sigma}^{\mathrm{epc,ee},I}-c_{\mathrm{x}}\delta_{\sigma'\sigma}\big[(jc\sigma|kd\sigma)+(cd\sigma|kj\sigma)\big]\bigg]\nonumber \\
 & +\sum_{dkc}(X-Y)_{id\sigma}(X-Y)_{kc\sigma}\times c_{\mathrm{x}}\big[(kd\sigma|jc\sigma)-(dc\sigma|jk\sigma)\big]\nonumber \\
 & +2\sum_{kc\sigma'ld\sigma''}(X+Y)_{kc\sigma'}(X+Y)_{ld\sigma''}(g_{kc\sigma'ld\sigma''ji\sigma}^{\mathrm{xc}}+\sum_{I}g_{kc\sigma'ld\sigma''ji\sigma}^{\mathrm{epc,eee},I}).
\end{align}

Defining the unrelaxed difference-density matrix
\begin{align}
T_{ab\sigma} & =\frac{1}{2}\sum_{i}\left[(X+Y)_{ia\sigma}(X+Y)_{ib\sigma}+(X-Y)_{ia\sigma}(X-Y)_{ib\sigma}\right]\label{eq:density T}\\
T_{ij\sigma} & =-\frac{1}{2}\sum_{a}\left[(X+Y)_{ia\sigma}(X+Y)_{ja\sigma}+(X-Y)_{ia\sigma}(X-Y)_{ja\sigma}\right],
\end{align}
the expression of $Q^{\mathrm{e}}_{ij\sigma}$ can thus be simplified
\begin{align}
 & Q^{\mathrm{e}}_{ij\sigma}\nonumber \\
 & =\sum_{cd\sigma'}T_{cd\sigma'}\left[2(cd\sigma'|ji\sigma)-c_{\mathrm{x}}\delta_{\sigma'\sigma}[(ci\sigma'|jd\sigma')+(cj\sigma'|id\sigma')]+2f^{\mathrm{xc}}_{cd\sigma'ji\sigma}+\sum_{I}2f^{\mathrm{epc,ee},I}_{cd\sigma'ji\sigma}\right]\nonumber \\
 & +\sum_{kl\sigma'}T_{kl\sigma'}\left[2(kl\sigma'|ji\sigma)-c_{\mathrm{x}}\delta_{\sigma'\sigma}\big[(ki\sigma|jl\sigma)+(kj\sigma|il\sigma)\big]+2f^{\mathrm{xc}}_{kl\sigma'ji\sigma}+\sum_{I}2f^{\mathrm{epc,ee},I}_{kl\sigma'ji\sigma}\right]\nonumber \\
 & -\sum_{c}\left[(X+Y)_{ic\sigma}(X+Y)_{jc\sigma}+(X-Y)_{ic\sigma}(X-Y)_{jc\sigma}\right]\varepsilon_{j\sigma}\nonumber \\
 & +\sum_{kc\sigma'd}(X+Y)_{kd\sigma'}(X+Y)_{ic\sigma}\bigg[2(kd\sigma'|jc\sigma)+2f^{\mathrm{xc}}_{kd\sigma'jc\sigma}-c_{\mathrm{x}}\delta_{\sigma'\sigma}\big[(jd\sigma|kc\sigma)+(cd\sigma|kj\sigma)\big]+2\sum_{I}f^{\mathrm{epc,ee},I}_{kd\sigma'jc\sigma}\bigg]\nonumber \\
 & +\sum_{dkc\sigma'}\delta_{\sigma\sigma'}(X-Y)_{id\sigma}(X-Y)_{kc\sigma}\times c_{\mathrm{x}}\big[(kd\sigma|jc\sigma)-(dc\sigma|jk\sigma)\big]\nonumber \\
 & +2\sum_{kc\sigma'ld\sigma''}(X+Y)_{kc\sigma'}(X+Y)_{ld\sigma''}\big(g^{\mathrm{xc}}_{kc\sigma'ld\sigma''ji\sigma}+\sum_{I}g^{\mathrm{epc,eee},I}_{kc\sigma'ld\sigma''ji\sigma}\big).
\end{align}

With the linear transformation for arbitrary vectors $V_{rs\sigma'}$
defined as
\begin{equation}
H^{+}_{pq\sigma}[V]=\sum_{rs\sigma'}V_{rs\sigma'}\bigg\{2(pq\sigma|rs\sigma')+2f^{\mathrm{xc}}_{pq\sigma rs\sigma'}-c_{\mathrm{x}}\delta_{\sigma\sigma'}\big[(ps\sigma|rq\sigma)+(pr\sigma|sq\sigma)\big]+2\sum_{I}f^{\mathrm{epc,ee},I}_{pq\sigma rs\sigma'}\bigg\},\label{eq:H_plus_neo-1}
\end{equation}
and utilizing the equation from the expression of $A+B$ and $A-B$:
\begin{equation}
2(kd\sigma'|jc\sigma)+2f_{kd\sigma'jc\sigma}^{\mathrm{xc}}-c_{\mathrm{x}}\delta_{\sigma'\sigma}\big[(jd\sigma|kc\sigma)+(cd\sigma|kj\sigma)\big]+2\sum_{I}f_{kd\sigma'jc\sigma}^{\mathrm{epc,ee},I}=(A+B)_{kd\sigma'jc\sigma}-\delta_{\sigma\sigma'}\delta_{cd}\delta_{jk}(\varepsilon_{c\sigma}-\varepsilon_{j\sigma}),
\end{equation}
\begin{equation}
c_{\mathrm{x}}\delta_{\sigma\sigma'}\big[(kd\sigma|jc\sigma)-(dc\sigma|jk\sigma)\big]=(A-B)_{kd\sigma'jc\sigma}-\delta_{\sigma\sigma'}\delta_{cd}\delta_{kj}(\varepsilon_{c\sigma}-\varepsilon_{j\sigma}),
\end{equation}
we can reformulate $Q_{ij\sigma}^{\mathrm{e}}$ as
\begin{align}
Q_{ij\sigma}^{\mathrm{e}} & =H_{ij\sigma}^{+}[T]-\sum_{c}\left[(X+Y)_{ic\sigma}(X+Y)_{jc\sigma}+(X-Y)_{ic\sigma}(X-Y)_{jc\sigma}\right]\varepsilon_{j\sigma}\nonumber \\
 & +\sum_{kc\sigma'd}(X+Y)_{kd\sigma'}(X+Y)_{ic\sigma}\left[(A+B)_{kd\sigma'jc\sigma}-\delta_{\sigma\sigma'}\delta_{cd}\delta_{jk}(\varepsilon_{c\sigma}-\varepsilon_{j\sigma})\right]\nonumber \\
 & +\sum_{dkc\sigma'}(X-Y)_{id\sigma}(X-Y)_{kc\sigma}\left[(A-B)_{kd\sigma'jc\sigma}-\delta_{\sigma\sigma'}\delta_{cd}\delta_{kj}(\varepsilon_{c\sigma}-\varepsilon_{j\sigma})\right]\nonumber \\
 & +2\sum_{kc\sigma'ld\sigma''}(X+Y)_{kc\sigma'}(X+Y)_{ld\sigma''}\big(g_{kc\sigma'ld\sigma''ji\sigma}^{\mathrm{xc}}+\sum_{I}g_{kc\sigma'ld\sigma''ji\sigma}^{\mathrm{epc,eee},I}\big).
\end{align}
The electron--proton kernel is contained in $H^{+}$, while its third
functional derivative produces the additional $g^{\mathrm{epc}}$
term.

With
\begin{equation}
(A+B)(X+Y)=\Omega(X-Y),
\end{equation}
and
\begin{equation}
(A-B)(X-Y)=\Omega(X+Y),
\end{equation}
we can further simplify the expression as
\begin{align}
Q_{ij\sigma}^{\mathrm{e}} & =H_{ij\sigma}^{+}[T]+2\sum_{kc\sigma'ld\sigma''}(X+Y)_{kc\sigma'}(X+Y)_{ld\sigma''}\big(g_{kc\sigma'ld\sigma''ji\sigma}^{\mathrm{xc}}+\sum_{I}g_{kc\sigma'ld\sigma''ji\sigma}^{\mathrm{epc,eee},I}\big)\nonumber \\
 & -\sum_{c}\left[(X+Y)_{ic\sigma}(X+Y)_{jc\sigma}+(X-Y)_{ic\sigma}(X-Y)_{jc\sigma}\right]\varepsilon_{c\sigma}\nonumber \\
 & +\Omega\sum_{c}\left[(X+Y)_{ic\sigma}(X-Y)_{jc\sigma}+(X-Y)_{ic\sigma}(X+Y)_{jc\sigma}\right].
\end{align}

\paragraph{occ - vir block ($p=i,q=a$)}

With the previously derived Fock matrix derivatives (Eq.\,\ref{eq:ctd_dFe_dCe}),
we have
\begin{align}
\sum_{\mu}C_{\mu a\sigma}^{\mathrm{e}}\frac{\partial F_{cd\sigma'}^{\mathrm{e}}}{\partial C_{\mu i\sigma}^{\mathrm{e}}} & =2(cd\sigma'|ai\sigma)-c_{\mathrm{x}}\delta_{\sigma'\sigma}[(ci\sigma'|ad\sigma')+(ca\sigma'|id\sigma')]+2f_{cd\sigma'ai\sigma}^{\mathrm{xc}}+\sum_{I}2f_{cd\sigma'ai\sigma}^{\mathrm{epc,ee},I},
\end{align}
and
\begin{align}
 & \sum_{\mu}C_{\mu a\sigma}^{\mathrm{e}}\frac{\partial F_{kl\sigma'}^{\mathrm{e}}}{\partial C_{\mu i\sigma}^{\mathrm{e}}}\nonumber \\
 & =\delta_{\sigma\sigma'}\delta_{ki}F_{al\sigma}+\delta_{\sigma\sigma'}\delta_{li}F_{ka\sigma}+2(kl\sigma'|ai\sigma)-c_{\mathrm{x}}\delta_{\sigma'\sigma}\big[(ki\sigma'|al\sigma')+(ka\sigma'|il\sigma')\big]+2f_{kl\sigma'ai\sigma}^{\mathrm{xc}}+2\sum_{I}f_{kl\sigma'ai\sigma}^{\mathrm{epc,ee},I}\\
 & =2(kl\sigma'|ai\sigma)-c_{\mathrm{x}}\delta_{\sigma'\sigma}\big[(ki\sigma|al\sigma)+(ka\sigma|il\sigma)\big]+2f_{kl\sigma'ai\sigma}^{\mathrm{xc}}+2\sum_{I}f_{kl\sigma'ai\sigma}^{\mathrm{epc,ee},I}.
\end{align}
From Eq.\,\ref{eq:ctd_dxc_de}, we can get the derivatives of exchange-correlation
functional as
\begin{equation}
\sum_{\mu}C_{\mu a\sigma}^{\mathrm{e}}\frac{\partial f_{kc\sigma'ld\sigma''}^{\mathrm{xc}}}{\partial C_{\mu i\sigma}^{\mathrm{e}}}=\delta_{ki}\delta_{\sigma\sigma'}f_{ac\sigma'ld\sigma''}^{\mathrm{xc}}+\delta_{li}\delta_{\sigma\sigma''}f_{kc\sigma'ad\sigma''}^{\mathrm{xc}}+2g_{kc\sigma'ld\sigma''ai\sigma}^{\mathrm{xc}},
\end{equation}
and similarly for the epc term
\begin{equation}
\sum_{\mu}C_{\mu a\sigma}^{\mathrm{e}}\frac{\partial f_{kc\sigma'ld\sigma''}^{\mathrm{epc,ee},I}}{\partial C_{\mu i\sigma}^{\mathrm{e}}}=\delta_{ki}\delta_{\sigma\sigma'}f_{ac\sigma'ld\sigma''}^{\mathrm{epc,ee},I}+\delta_{li}\delta_{\sigma\sigma''}f_{kc\sigma'ad\sigma''}^{\mathrm{epc,ee},I}+2g_{kc\sigma'ld\sigma''ai\sigma}^{\mathrm{epc,eee},I}.
\end{equation}

With these derivatives, we can get the derivatives for $A+B$ as
\begin{align}
 & \sum_{\mu}C_{\mu a\sigma}^{\mathrm{e}}\frac{\partial(A+B)_{kc\sigma'ld\sigma''}}{\partial C_{\mu i\sigma}^{\mathrm{e}}}\nonumber \\
 & =\delta_{\sigma'\sigma''}\delta_{kl}\left\{ 2(cd\sigma'|ai\sigma)-c_{\mathrm{x}}\delta_{\sigma'\sigma}[(ci\sigma'|ad\sigma')+(ca\sigma'|id\sigma')]+2f_{cd\sigma'ai\sigma}^{\mathrm{xc}}+2\sum_{I}f_{cd\sigma'ai\sigma}^{\mathrm{epc,ee},I}\right\} \nonumber \\
 & -\delta_{\sigma'\sigma''}\delta_{cd}\left\{ 2(kl\sigma'|ai\sigma)-c_{\mathrm{x}}\delta_{\sigma'\sigma}\big[(ki\sigma|al\sigma)+(ka\sigma|il\sigma)\big]+2f_{kl\sigma'ai\sigma}^{\mathrm{xc}}+2\sum_{I}f_{kl\sigma'ai\sigma}^{\mathrm{epc,ee},I}\right\} \nonumber \\
 & +\delta_{\sigma\sigma'}\delta_{ki}\left\{ 2(ac\sigma|ld\sigma'')+2f_{ac\sigma'ld\sigma''}^{\mathrm{xc}}+2\sum_{I}f_{ac\sigma'ld\sigma''}^{\mathrm{epc,ee},I}-c_{\mathrm{x}}\delta_{\sigma'\sigma''}\big[(lc\sigma'|ad\sigma')+(cd\sigma'|al\sigma')\big]\right\} \nonumber \\
 & +\delta_{\sigma\sigma''}\delta_{li}\left\{ 2(kc\sigma'|ad\sigma)+2f_{kc\sigma'ad\sigma''}^{\mathrm{xc}}+2\sum_{I}f_{kc\sigma'ad\sigma''}^{\mathrm{epc,ee},I}-c_{\mathrm{x}}\delta_{\sigma'\sigma''}\big[(ac\sigma'|kd\sigma')+(cd\sigma'|ka\sigma')\big]\right\} \nonumber \\
 & +4g_{kc\sigma'ld\sigma''ai\sigma}^{\mathrm{xc}}+4\sum_{I}g_{kc\sigma'ld\sigma''ai\sigma}^{\mathrm{epc,eee},I},
\end{align}
and for $A-B$,
\begin{align}
 & \sum_{\mu}C_{\mu a\sigma}^{\mathrm{e}}\frac{\partial(A-B)_{kc\sigma'ld\sigma''}}{\partial C_{\mu i\sigma}^{\mathrm{e}}}\nonumber \\
 & =\delta_{\sigma'\sigma''}\delta_{kl}\left\{ 2(cd\sigma'|ai\sigma)-c_{\mathrm{x}}\delta_{\sigma'\sigma}[(ci\sigma'|ad\sigma')+(ca\sigma'|id\sigma')]+2f_{cd\sigma'ai\sigma}^{\mathrm{xc}}+2\sum_{I}f_{cd\sigma'ai\sigma}^{\mathrm{epc,ee},I}\right\} \nonumber \\
 & -\delta_{\sigma'\sigma''}\delta_{cd}\left\{ 2(kl\sigma'|ai\sigma)-c_{\mathrm{x}}\delta_{\sigma'\sigma}\big[(ki\sigma|al\sigma)+(ka\sigma|il\sigma)\big]+2f_{kl\sigma'ai\sigma}^{\mathrm{xc}}+2\sum_{I}f_{kl\sigma'ai\sigma}^{\mathrm{epc,ee},I}\right\} \nonumber \\
 & +c_{\mathrm{x}}\delta_{\sigma'\sigma''}\delta_{\sigma\sigma'}\delta_{ki}\big[(lc\sigma'|ad\sigma')-(cd\sigma'|al\sigma')\big]\nonumber \\
 & +c_{\mathrm{x}}\delta_{\sigma'\sigma''}\delta_{\sigma\sigma''}\delta_{li}\big[(ac\sigma'|kd\sigma')-(cd\sigma'|ka\sigma')\big].
\end{align}

Combining the previous definition of $H^{+}$ and $T$, the occ-vir
block of electronic $Q$ can be expressed as
\begin{align}
Q_{ia\sigma}^{\mathrm{e}} & =\frac{1}{2}\sum_{kc\sigma'ld\sigma''}\bigg[\sum_{\mu}C_{\mu a\sigma}^{\mathrm{e}}\frac{\partial(A+B)_{kc\sigma'ld\sigma''}}{\partial C_{\mu i\sigma}^{\mathrm{e}}}(X+Y)_{kc\sigma'}(X+Y)_{ld\sigma''}\nonumber \\
 & \qquad+\sum_{\mu}C_{\mu a\sigma}^{\mathrm{e}}\frac{\partial(A-B)_{kc\sigma'ld\sigma''}}{\partial C_{\mu i\sigma}^{\mathrm{e}}}(X-Y)_{kc\sigma'}(X-Y)_{ld\sigma''}\bigg]\\
 & =H_{ia\sigma}^{+}[T]\nonumber \\
 & +\sum_{kc\sigma'd}(X+Y)_{kd\sigma'}(X+Y)_{ic\sigma}\bigg[2(kd\sigma'|ac\sigma)+2f_{kd\sigma'ac\sigma}^{\mathrm{xc}}+2\sum_{I}f_{kd\sigma'ac\sigma}^{\mathrm{epc,ee},I}-c_{\mathrm{x}}\delta_{\sigma'\sigma}\big[(ad\sigma|kc\sigma)+(cd\sigma|ka\sigma)\big]\bigg]\nonumber \\
 & +\sum_{dkc\sigma'}(X-Y)_{id\sigma}(X-Y)_{kc\sigma}\times c_{\mathrm{x}}\delta_{\sigma\sigma'}\big[(kd\sigma|ac\sigma)-(dc\sigma|ak\sigma)\big]\nonumber \\
 & +2\sum_{kc\sigma'ld\sigma''}(X+Y)_{kc\sigma'}(X+Y)_{ld\sigma''}(g_{kc\sigma'ld\sigma''ai\sigma}^{\mathrm{xc}}+\sum_{I}g_{kc\sigma'ld\sigma''ai\sigma}^{\mathrm{epc,eee},I}).
\end{align}
We now define another linear transformation as
\begin{equation}
H_{pq\sigma}^{-}[V]=\sum_{rs\sigma'}c_{\mathrm{x}}\delta_{\sigma\sigma'}[(ps\sigma|rq\sigma)-(pr\sigma|sq\sigma)]V_{rs\sigma'},\label{eq:H_minus}
\end{equation}
which simplifies the expression for $Q_{ia\sigma}^{\mathrm e}$ to
\begin{align}
Q_{ia\sigma}^{\mathrm{e}} & =H_{ia\sigma}^{+}[T]+\sum_{c}(X+Y)_{ic\sigma}H_{ac\sigma}^{+}[X+Y]+\sum_{c}(X-Y)_{ic\sigma}H_{ac\sigma}^{-}[X-Y]\nonumber \\
 & +2\sum_{kc\sigma'ld\sigma''}(X+Y)_{kc\sigma'}(X+Y)_{ld\sigma''}\big(g_{kc\sigma'ld\sigma''ai\sigma}^{\mathrm{xc}}+\sum_{I}g_{kc\sigma'ld\sigma''ai\sigma}^{\mathrm{epc,eee},I}\big).
\end{align}
All electron--proton-correlation contributions are retained through
$H^{+}$ and the explicit third-derivative terms.

\paragraph{vir - occ block ($p=a,q=i$)}

With the previously derived Fock matrix derivatives, we can see that
both the following two derivatives vanish:

\begin{align}
\sum_{\mu}C_{\mu i\sigma}^{\mathrm{e}}\frac{\partial F_{cd\sigma'}^{\mathrm{e}}}{\partial C_{\mu a\sigma}^{\mathrm{e}}} & =\delta_{\sigma\sigma'}\delta_{ac}F_{id\sigma}^{\mathrm{e}}+\delta_{\sigma\sigma'}\delta_{ad}F_{ci\sigma}^{\mathrm{e}}=0,
\end{align}
\begin{align}
\sum_{\mu}C_{\mu i\sigma}^{\mathrm{e}}\frac{\partial F_{kl\sigma'}^{\mathrm{e}}}{\partial C_{\mu a\sigma}^{\mathrm{e}}} & =0,
\end{align}
which lead to the expression for the derivatives of $A+B$ and $A-B$
as
\begin{align}
 & \sum_{\mu}C_{\mu i\sigma}^{\mathrm{e}}\frac{\partial(A+B)_{kc\sigma'ld\sigma''}}{\partial C_{\mu a\sigma}^{\mathrm{e}}}\nonumber \\
 & =\delta_{\sigma\sigma'}\delta_{ac}\left\{ 2(ki\sigma|ld\sigma'')+2f_{ki\sigma'ld\sigma''}^{\mathrm{xc}}+2\sum_{I}f_{ki\sigma ld\sigma''}^{\mathrm{epc,ee},I}-c_{\mathrm{x}}\delta_{\sigma'\sigma''}\big[(li\sigma'|kd\sigma')+(id\sigma'|kl\sigma')\big]\right\} \nonumber \\
 & +\delta_{\sigma\sigma''}\delta_{ad}\left\{ 2(kc\sigma'|li\sigma)+2f_{kc\sigma'li\sigma''}^{\mathrm{xc}}+2\sum_{I}f_{kc\sigma'li\sigma''}^{\mathrm{epc,ee},I}-c_{\mathrm{x}}\delta_{\sigma'\sigma''}\big[(lc\sigma'|ki\sigma')+(ci\sigma'|kl\sigma')\big]\right\},
\end{align}
and
\begin{align}
 & \sum_{\mu}C_{\mu i\sigma}^{\mathrm{e}}\frac{\partial(A-B)_{kc\sigma'ld\sigma''}}{\partial C_{\mu a\sigma}^{\mathrm{e}}}\nonumber \\
 & =c_{\mathrm{x}}\delta_{\sigma'\sigma''}\delta_{\sigma\sigma'}\delta_{ac}\big[(li\sigma'|kd\sigma')-(id\sigma'|kl\sigma')\big]+c_{\mathrm{x}}\delta_{\sigma'\sigma''}\delta_{\sigma\sigma''}\delta_{ad}\big[(lc\sigma'|ki\sigma')-(ci\sigma'|kl\sigma')\big].
\end{align}

With this, $Q_{ai\sigma}^{\mathrm{e}}$ can be expressed as
\begin{align}
 & Q_{ai\sigma}^{\mathrm{e}}\nonumber \\
 & =\frac{1}{2}\sum_{kc\sigma'ld\sigma''}\bigg[\sum_{\mu}C_{\mu i\sigma}\frac{\partial(A+B)_{kc\sigma'ld\sigma''}}{\partial C_{\mu a\sigma}}(X+Y)_{kc\sigma'}(X+Y)_{ld\sigma''}\nonumber \\
 & \qquad+\sum_{\mu}C_{\mu i\sigma}\frac{\partial(A-B)_{kc\sigma'ld\sigma''}}{\partial C_{\mu a\sigma}}(X-Y)_{kc\sigma'}(X-Y)_{ld\sigma''}\bigg]\\
 & =\sum_{kld\sigma''}(X+Y)_{ka\sigma}(X+Y)_{ld\sigma''}\left\{ 2(ki\sigma|ld\sigma'')+2f_{ki\sigma ld\sigma''}^{\mathrm{xc}}-c_{\mathrm{x}}\delta_{\sigma\sigma''}\big[(li\sigma|kd\sigma)+(id\sigma|kl\sigma)\big]+2\sum_{I}f_{ki\sigma ld\sigma''}^{\mathrm{epc,ee},I}\right\} \nonumber \\
 & +\sum_{kcl\sigma''}(X-Y)_{kc\sigma''}(X-Y)_{la\sigma}\times c_{\mathrm{x}}\delta_{\sigma\sigma''}\big[(lc\sigma|ki\sigma)-(ci\sigma|kl\sigma)\big].
\end{align}
Using the definition of linear transformations (Eq.~\ref{eq:H_plus_neo-1}
and \ref{eq:H_minus}), we can get a simplified expression as
\begin{equation}
Q_{ai\sigma}^{\mathrm{e}}=\sum_{k}\left\{ (X+Y)_{ka\sigma}H_{ki\sigma}^{+}[X+Y]+(X-Y)_{ka\sigma}H_{ki\sigma}^{-}[X-Y]\right\},
\end{equation}
where the CNEO definition of $H^{+}$ includes the electron--proton kernel.

\paragraph{vir - vir block ($p=a,q=b$)}

The derivation of $ab$ block is similar to that of $ai$ block. Starting
with Fock matrix derivatives, we have
\begin{equation}
\sum_{\mu}C_{\mu b\sigma}^{\mathrm{e}}\frac{\partial F_{cd\sigma'}^{\mathrm{e}}}{\partial C_{\mu a\sigma}^{\mathrm{e}}}=\delta_{\sigma\sigma'}\varepsilon^{\mathrm{e}}_{b\sigma}
\left(
\delta_{ac}\delta_{bd}
+
\delta_{ad}\delta_{bc}
\right),
\end{equation}
and
\begin{align}
\sum_{\mu}C_{\mu b\sigma}^{\mathrm{e}}\frac{\partial F_{kl\sigma'}^{\mathrm{e}}}{\partial C_{\mu a\sigma}^{\mathrm{e}}} & =0.
\end{align}
This will lead to
\begin{align}
 & Q_{ab\sigma}^{\mathrm{e}}\nonumber \\
 & =\sum_{kld\sigma''}(X+Y)_{ka\sigma}(X+Y)_{ld\sigma''}\bigg\{2(kb\sigma|ld\sigma'')+2f_{kb\sigma ld\sigma''}^{\mathrm{xc}}-c_{\mathrm{x}}\delta_{\sigma\sigma''}\big[(lb\sigma|kd\sigma)+(bd\sigma|kl\sigma)\big]\nonumber \\
 & \qquad+\delta_{kl}\delta_{bd}\delta_{\sigma\sigma''}\varepsilon_{b\sigma}+2\sum_{I}f_{kb\sigma ld\sigma''}^{\mathrm{epc,ee},I}\bigg\}\nonumber \\
 & +\sum_{kdl\sigma''}(X-Y)_{ld\sigma''}(X-Y)_{ka\sigma}\left\{ c_{\mathrm{x}}\delta_{\sigma\sigma''}\big[(kd\sigma|lb\sigma)-(db\sigma|lk\sigma)\big]+\delta_{kl}\delta_{bd}\delta_{\sigma\sigma''}\varepsilon_{b\sigma}\right\}.
\end{align}

Utilizing the expression of $A+B$ and $A-B$, this expression can
be simplified as
\begin{align}
Q_{ab\sigma}^{\mathrm{e}} & =\sum_{kld\sigma''}(X+Y)_{ka\sigma}(X+Y)_{ld\sigma''}\left[(A+B)_{kb\sigma ld\sigma''}+\delta_{\sigma\sigma''}\delta_{bd}\delta_{kl}\varepsilon_{k\sigma}\right]\nonumber \\
 & +\sum_{kld\sigma''}(X-Y)_{ka\sigma}(X-Y)_{ld\sigma''}\left[(A-B)_{kb\sigma ld\sigma''}+\delta_{\sigma\sigma''}\delta_{bd}\delta_{kl}\varepsilon_{k\sigma}\right]\\
 & =\Omega\sum_{k}(X+Y)_{ka\sigma}(X-Y)_{kb\sigma}+\sum_{k}(X+Y)_{ka\sigma}(X+Y)_{kb\sigma}\varepsilon_{k\sigma}\nonumber \\
 & +\Omega\sum_{k}(X-Y)_{ka\sigma}(X+Y)_{kb\sigma}+\sum_{k}(X-Y)_{ka\sigma}(X-Y)_{kb\sigma}\varepsilon_{k\sigma}\\
 & =\sum_{k}\Omega\left[(X+Y)_{ka\sigma}(X-Y)_{kb\sigma}+(X-Y)_{ka\sigma}(X+Y)_{kb\sigma}\right]\nonumber \\
 & +\sum_{k}\varepsilon_{k\sigma}\left[(X+Y)_{ka\sigma}(X+Y)_{kb\sigma}+(X-Y)_{ka\sigma}(X-Y)_{kb\sigma}\right].\label{eq:Q_ab}
\end{align}
This completes the four electronic blocks of $Q^{\mathrm e}$, including all
electron--proton-correlation terms.

\subsubsection{Nuclear $Q^{\mathrm{n}}$}

The nuclear source matrix is
\begin{equation}
Q_{p'q'I}^{\mathrm{n}}=\sum_{\mu'}C_{\mu'q'I}^{\mathrm{n}}\frac{\partial}{\partial C_{\mu'p'I}^{\mathrm{n}}}\langle X,Y|\Lambda|X,Y\rangle.
\end{equation}
From Eqs.\,\ref{eq:ctd_lambda} to \ref{eq:ctd_AmB}, we can see
that for all the terms in $\langle X,Y|\Lambda|X,Y\rangle$, only
$F_{rs\sigma'}^{\mathrm{e}}$ and $f_{kc\sigma'ld\sigma''}^{\mathrm{epc,ee},I}$
have dependence on $\{C_{I}^{\mathrm{n}}\}$. The general form for
the derivatives of $F_{rs\sigma'}^{\mathrm{e}}$ with respect to $C_{I}^{\mathrm{n}}$
has been derived in Sec.~\ref{subsec:Electronic-Fock-matrix-nuclear},
and the general form for the derivatives of $f_{kc\sigma'ld\sigma''}^{\mathrm{epc,ee},I}$
with respect to $C_{I}^{\mathrm{n}}$ can be expressed as
\begin{align}
\sum_{\mu'}C_{\mu'q'I}^{\mathrm{n}}\frac{\partial f_{kc\sigma'ld\sigma''}^{\mathrm{epc,ee},I}}{\partial C_{\mu'p'I}^{\mathrm{n}}} & =\sum_{\mu'}C_{\mu'q'I}^{\mathrm{n}}\frac{\partial}{\partial C_{\mu'p'I}^{\mathrm{n}}}\left(\int\text{d}\bm{r}\text{d}\bm{r}'\phi_{k\sigma'}\phi_{c\sigma'}\frac{\delta^{2}E[\rho^{\mathrm{e}},\rho_{I}^{\mathrm{n}}]}{\delta\rho_{\sigma'}\delta\rho_{\sigma''}}\phi_{l\sigma''}\phi_{d\sigma''}\right)\\
 & =\int\text{d}\bm{r}\text{d}\bm{r}'\phi_{k\sigma'}\phi_{c\sigma'}\frac{\delta^{3}E[\rho^{\mathrm{e}},\rho_{I}^{\mathrm{n}}]}{\delta\rho_{\sigma'}\delta\rho_{\sigma''}\delta\rho_{I}^{\mathrm{n}}}\left(\sum_{\mu'}C_{\mu'q'I}^{\mathrm{n}}\frac{\partial\rho_{I}^{\mathrm{n}}}{\partial C_{\mu'p'I}^{\mathrm{n}}}\right)\phi_{l\sigma''}\phi_{d\sigma''}\\
 & =2\delta_{i'p'}g_{kc\sigma'ld\sigma''i'q'I}^{\mathrm{epc,een},I},\label{eq:ctd_depc_dn}
\end{align}
where
\begin{equation}
g_{kc\sigma ld\sigma'i'q'I}^{\mathrm{epc,een},I}=\int\text{d}\bm{r}\text{d}\bm{r}'\text{d}\bm{r}''\frac{\delta^{3}E^{\mathrm{epc},I}[\rho^{\mathrm{e}},\rho_{I}^{\mathrm{n}}]}{\delta\rho_{\sigma}^{\mathrm{e}}(\bm{r})\delta\rho_{\sigma'}^{\mathrm{e}}(\bm{r}')\delta\rho_{I}^{\mathrm{n}}(\bm{r}'')}\phi_{k\sigma}(\bm{r})\phi_{c\sigma}(\bm{r})\phi_{l\sigma'}(\bm{r}')\phi_{d\sigma'}(\bm{r}')\phi_{i'I}(\bm{r}'')\phi_{q'I}(\bm{r}'').
\end{equation}

With these general form of derivatives, we will derive the expression
of $Q^{\mathrm{n}}$ in four different blocks.

\paragraph{occ - occ block ($p'=i',q'=i'$)}

In this block, the derivatives of $F_{cd\sigma'}^{\mathrm{e}}$ and
$F_{kl\sigma'}^{\mathrm{e}}$ can be expressed as (from Sec.~\ref{subsec:Electronic-Fock-matrix-nuclear})
\begin{equation}
\sum_{\mu'}C_{\mu'i'I}^{\mathrm{n}}\frac{\partial F_{cd\sigma'}^{\mathrm{e}}}{\partial C_{\mu'i'I}^{\mathrm{n}}}=2f_{cd\sigma'i'i'I}^{\mathrm{epc,en},I}-2Z_{I}(cd\sigma'|i'i'I),
\end{equation}
\begin{equation}
\sum_{\mu'}C_{\mu'i'I}^{\mathrm{n}}\frac{\partial F_{kl\sigma'}^{\mathrm{e}}}{\partial C_{\mu'i'I}^{\mathrm{n}}}=2f_{kl\sigma'i'i'I}^{\mathrm{epc,en},I}-2Z_{I}(kl\sigma'|i'i'I),
\end{equation}
and the derivatives of $f_{kc\sigma'ld\sigma''}^{\mathrm{epc,ee},I}$
are
\begin{equation}
\sum_{\mu'}C_{\mu'i'I}^{\mathrm{n}}\frac{\partial f_{kc\sigma'ld\sigma''}^{\mathrm{epc,ee},I}}{\partial C_{\mu'i'I}^{\mathrm{n}}}=2g_{kc\sigma'ld\sigma''i'i'I}^{\mathrm{epc,een},I}.
\end{equation}

Therefore, we have
\begin{align}
 & \sum_{\mu'}C_{\mu'i'I}^{\mathrm{n}}\frac{\partial(A+B)_{kc\sigma'ld\sigma''}}{\partial C_{\mu'i'I}^{\mathrm{n}}}\nonumber \\
 & =\delta_{kl}\delta_{\sigma'\sigma''}\big[2f_{cd\sigma'i'i'I}^{\mathrm{epc,en},I}-2Z_{I}(cd\sigma'|i'i'I)\big]-\delta_{cd}\delta_{\sigma'\sigma''}\big[2f_{kl\sigma'i'i'I}^{\mathrm{epc,en},I}-2Z_{I}(kl\sigma'|i'i'I)\big]\nonumber \\
 & +4g_{kc\sigma'ld\sigma''i'i'I}^{\mathrm{epc,een},I},
\end{align}
and
\begin{align}
 & \sum_{\mu'}C_{\mu'i'I}^{\mathrm{n}}\frac{\partial(A-B)_{kc\sigma'ld\sigma''}}{\partial C_{\mu'i'I}^{\mathrm{n}}}\nonumber \\
 & =\delta_{kl}\delta_{\sigma'\sigma''}\big[2f_{cd\sigma'i'i'I}^{\mathrm{epc,en},I}-2Z_{I}(cd\sigma'|i'i'I)\big]-\delta_{cd}\delta_{\sigma'\sigma''}\big[2f_{kl\sigma'i'i'I}^{\mathrm{epc,en},I}-2Z_{I}(kl\sigma'|i'i'I)\big].
\end{align}
With this, the first block can be derived as
\begin{align}
Q_{i'i'I}^{\mathrm{n}} & =\sum_{kc\sigma'd}\big[2f_{cd\sigma'i'i'I}^{\mathrm{epc,en},I}-2Z_{I}(cd\sigma'|i'i'I)\big]\frac{1}{2}\left[(X+Y)_{kc\sigma'}(X+Y)_{kd\sigma'}+(X-Y)_{kc\sigma'}(X-Y)_{kd\sigma'}\right]\nonumber \\
 & -\sum_{kc\sigma'l}\big[2f_{kl\sigma'i'i'I}^{\mathrm{epc,en},I}-2Z_{I}(kl\sigma'|i'i'I)\big]\frac{1}{2}\left[(X+Y)_{kc\sigma'}(X+Y)_{lc\sigma'}+(X-Y)_{kc\sigma'}(X-Y)_{lc\sigma'}\right]\nonumber \\
 & +\sum_{kc\sigma'ld\sigma''}2g_{kc\sigma'ld\sigma''i'i'I}^{\mathrm{epc,een},I}(X+Y)_{kc\sigma'}(X+Y)_{ld\sigma''}\\
 & =\sum_{c\sigma'd}\big[2f_{cd\sigma'i'i'I}^{\mathrm{epc,en},I}-2Z_{I}(cd\sigma'|i'i'I)\big]T_{cd\sigma'}+\sum_{k\sigma'l}\big[2f_{kl\sigma'i'i'I}^{\mathrm{epc,en},I}-2Z_{I}(kl\sigma'|i'i'I)\big]T_{kl\sigma'}\nonumber \\
 & +\sum_{kc\sigma'ld\sigma''}2g_{kc\sigma'ld\sigma''i'i'I}^{\mathrm{epc,een},I}(X+Y)_{kc\sigma'}(X+Y)_{ld\sigma''}.
\end{align}
To simplify the expression, we define a new linear transformation
as
\begin{equation}
H_{p'q'I}^{\mathrm{ne}}[V]=\sum_{rs\sigma'}\big[-2Z_{I}(rs\sigma'|p'q'I)+2f_{rs\sigma'p'q'I}^{\text{epc,en},I}\big]V_{rs\sigma'},\label{eq:H_ne-1}
\end{equation}
and the expression can be simplified as
\begin{equation}
Q_{i'i'I}^{\mathrm{n}}=H_{i'i'I}^{\mathrm{ne}}[T]+\sum_{kc\sigma'ld\sigma''}2g_{kc\sigma'ld\sigma''i'i'I}^{\mathrm{epc,een},I}(X+Y)_{kc\sigma'}(X+Y)_{ld\sigma''}.
\end{equation}

\paragraph{occ - vir block ($p'=i',q'=a'$)}

In this block, the derivatives of $F_{cd\sigma'}^{\mathrm{e}}$, $F_{kl\sigma'}^{\mathrm{e}}$,
and $f_{kc\sigma'ld\sigma''}^{\mathrm{epc,ee},I}$ can be expressed
as

\begin{equation}
\sum_{\mu'}C_{\mu'a'I}^{\mathrm{n}}\frac{\partial F_{cd\sigma'}^{\mathrm{e}}}{\partial C_{\mu'i'I}^{\mathrm{n}}}=2f_{cd\sigma'i'a'I}^{\mathrm{epc,en},I}-2Z_{I}(cd\sigma'|i'a'I),
\end{equation}
\begin{equation}
\sum_{\mu'}C_{\mu'a'I}^{\mathrm{n}}\frac{\partial F_{kl\sigma'}^{\mathrm{e}}}{\partial C_{\mu'i'I}^{\mathrm{n}}}=2f_{kl\sigma'i'a'I}^{\mathrm{epc,en},I}-2Z_{I}(kl\sigma'|i'a'I),
\end{equation}
and
\begin{equation}
\sum_{\mu'}C_{\mu'a'I}^{\mathrm{n}}\frac{\partial f_{kc\sigma'ld\sigma''}^{\mathrm{epc,ee},I}}{\partial C_{\mu'i'I}^{\mathrm{n}}}=2g_{kc\sigma'ld\sigma''i'a'I}^{\mathrm{epc,een},I}.
\end{equation}
We can see that these derivatives closely resemble those for $Q_{i'i'I}^{\mathrm{n}}$,
except that the orbital indices $i'i'I$ are all replaced by $i'a'I$.
Therefore, the expression of $Q_{i'a'I}^{\mathrm{n}}$ can be analogously
written as
\begin{equation}
Q_{i'a'I}^{\mathrm{n}}=H_{i'a'I}^{\mathrm{ne}}[T]+\sum_{kc\sigma'ld\sigma''}2g_{kc\sigma'ld\sigma''i'a'I}^{\mathrm{epc,een},I}(X+Y)_{kc\sigma'}(X+Y)_{ld\sigma''}.
\end{equation}

\paragraph{vir - occ block ($p'=a',q'=i'$)}

For this block, all the delta functions in the general form of derivatives
for $F_{rs\sigma'}^{\mathrm{e}}$ and $f_{kc\sigma'ld\sigma''}^{\mathrm{epc,ee},I}$
will drop to zero:
\begin{equation}
\sum_{\mu'}C_{\mu'i'I}^{\mathrm{n}}\frac{\partial F_{rs\sigma'}^{\mathrm{e}}}{\partial C_{\mu'a'I}^{\mathrm{n}}}=2\delta_{i'a'}f_{rs\sigma'i'i'I}^{\mathrm{epc,en},I}-2Z_{I}(rs\sigma'|i'i'I)\delta_{i'a'}=0,
\end{equation}
and
\begin{equation}
\sum_{\mu'}C_{\mu'i'I}^{\mathrm{n}}\frac{\partial f_{kc\sigma'ld\sigma''}^{\mathrm{epc,ee},I}}{\partial C_{\mu'a'I}^{\mathrm{n}}}=2\delta_{i'a'}g_{kc\sigma'ld\sigma''i'i'I}^{\mathrm{epc,een},I}=0.
\end{equation}
Consequently, the value of $Q_{a'i'I}^{\mathrm{n}}$ will also drop
to zero
\begin{align}
Q_{a'i'I}^{\mathrm{n}} & =0.
\end{align}

\paragraph{vir - vir block ($p'=a',q'=b'$)}

Similarly as the $a'i'$ block, since $\delta_{i'p'}=\delta_{i'a'}=0$,
all the derivatives will vanish for this block, which leads to
\begin{equation}
Q_{a'b'I}^{\mathrm{n}}=0.
\end{equation}

\subsection{$Z$-related terms}

To solve Eqs.\,\ref{ctd_transform_ce} and \ref{ctd_transform_cn},
we also need to find out the expression for the following four terms
\begin{enumerate}
\item Electron-electron
\begin{equation}
\sum_{kc\sigma'}Z_{kc\sigma'}^{\mathrm{e}}\sum_{\mu}C_{\mu q\sigma}^{\mathrm{e}}\frac{\partial F_{kc\sigma'}^{\mathrm{e}}}{\partial C_{\mu p\sigma}^{\mathrm{e}}};
\end{equation}
\item Electron-nucleus
\begin{equation}
\sum_{kc\sigma'}Z_{kc\sigma'}^{\mathrm{e}}\sum_{\mu'}C_{\mu'q'I}^{\mathrm{n}}\frac{\partial F_{kc\sigma'}^{\mathrm{e}}}{\partial C_{\mu'p'I}^{\mathrm{n}}};
\end{equation}
\item Nucleus-electron
\begin{equation}
\sum_{k'c'K}Z_{k'c'K}^{\mathrm{n}}\sum_{\mu}C_{\mu q\sigma}^{\mathrm{e}}\frac{\partial F_{k'c'K}^{\mathrm{n}}}{\partial C_{\mu p\sigma}^{\mathrm{e}}};
\end{equation}
\item Nucleus-nucleus
\begin{equation}
\sum_{k'c'K}Z_{k'c'K}^{\mathrm{n}}\sum_{\mu'}C_{\mu'q'I}^{\mathrm{n}}\frac{\partial F_{k'c'K}^{\mathrm{n}}}{\partial C_{\mu'p'I}^{\mathrm{n}}}.
\end{equation}
\end{enumerate}

\subsubsection{Electron-electron}

The derivative of the electronic Fock matrix with respect to the electronic
MO coefficients has been derived in Sec.~\ref{subsubsection: fock_ee}.
Using this result, we evaluate the $Z^{\mathrm{e}}$-related contribution
for the four orbital blocks below.

\paragraph{occ - occ block ($p=i,q=j$)}

In this block, the Fock matrix derivatives can be simplified as

\begin{align}
 & \sum_{\mu}C_{\mu j\sigma}^{\mathrm{e}}\frac{\partial F_{kc\sigma'}^{\mathrm{e}}}{\partial C_{\mu i\sigma}^{\mathrm{e}}}\nonumber \\
 & =\delta_{\sigma\sigma'}\delta_{ki}F_{jc\sigma}^{\mathrm{e}}+2(kc\sigma'|ji\sigma)-c_{\mathrm{x}}\delta_{\sigma'\sigma}[(ki\sigma'|jc\sigma')+(kj\sigma'|ic\sigma')]+2f_{kc\sigma'ji\sigma}^{\mathrm{xc}}+\sum_{I}2f_{kc\sigma'ji\sigma}^{\mathrm{epc,ee},I}\\
 & =2(kc\sigma'|ji\sigma)-c_{\mathrm{x}}\delta_{\sigma'\sigma}[(ki\sigma'|jc\sigma')+(kj\sigma'|ic\sigma')]+2f_{kc\sigma'ji\sigma}^{\mathrm{xc}}+\sum_{I}2f_{kc\sigma'ji\sigma}^{\mathrm{epc,ee},I}.
\end{align}
where the KS-DFT condition $F_{jc\sigma}^{\mathrm{e}}=0$ is used.
With the definition of $H^{+}$ in Eq.~\ref{eq:H_plus_neo-1}, we can
get
\begin{align}
 & \sum_{kc\sigma'}Z_{kc\sigma'}^{\mathrm{e}}\sum_{\mu}C_{\mu j\sigma}\frac{\partial F_{kc\sigma'}^{\mathrm{e}}}{\partial C_{\mu i\sigma}^{\mathrm{e}}}\nonumber \\
 & =\sum_{kc\sigma'}Z_{kc\sigma'}^{\mathrm{e}}\bigg\{2(kc\sigma'|ji\sigma)-c_{\mathrm{x}}\delta_{\sigma'\sigma}[(ki\sigma'|jc\sigma')+(kj\sigma'|ic\sigma')]+2f_{kc\sigma'ji\sigma}^{\mathrm{xc}}+\sum_{I}2f_{kc\sigma'ji\sigma}^{\mathrm{epc,ee},I}\bigg\}\\
 & =H_{ij\sigma}^{+}[Z^{\mathrm{e}}].
\end{align}

\paragraph{occ - vir block ($p=i,q=a$)}

For this block, the Fock matrix derivatives are

\begin{align}
\sum_{\mu}C_{\mu a\sigma}^{\mathrm{e}}\frac{\partial F_{kc\sigma'}^{\mathrm{e}}}{\partial C_{\mu i\sigma}^{\mathrm{e}}} & =\delta_{\sigma\sigma'}\delta_{ki}F_{ac\sigma}^{\mathrm{e}}+2(kc\sigma'|ai\sigma)-c_{\mathrm{x}}\delta_{\sigma'\sigma}[(ki\sigma'|ac\sigma')+(ka\sigma'|ic\sigma')]+2f_{kc\sigma'ai\sigma}^{\mathrm{xc}}+\sum_{I}2f_{kc\sigma'ai\sigma}^{\mathrm{epc,ee},I}\\
 & =\delta_{\sigma\sigma'}\delta_{ki}\delta_{ac}\varepsilon_{a\sigma}^{\mathrm{e}}+2(kc\sigma'|ai\sigma)-c_{\mathrm{x}}\delta_{\sigma'\sigma}[(ki\sigma'|ac\sigma')+(ka\sigma'|ic\sigma')]+2f_{kc\sigma'ai\sigma}^{\mathrm{xc}}+\sum_{I}2f_{kc\sigma'ai\sigma}^{\mathrm{epc,ee},I},
\end{align}
where $F_{ac\sigma}^{\mathrm{e}}=\varepsilon_{a\sigma}^{\mathrm{e}}\delta_{ac}$
is used. With the definition of $H^{+}$, we have
\begin{equation}
\sum_{kc\sigma'}Z_{kc\sigma'}^{\mathrm{e}}\sum_{\mu}C_{\mu a\sigma}^{\mathrm{e}}\frac{\partial F_{kc\sigma'}^{\mathrm{e}}}{\partial C_{\mu i\sigma}^{\mathrm{e}}}=H_{ia\sigma}^{+}[Z^{\mathrm{e}}]+\varepsilon_{a\sigma}Z_{ia\sigma}^{\mathrm{e}}.
\end{equation}

\paragraph{vir - occ block ($p=a,q=i$)}

In this block, since $\delta_{ip}=\delta_{ia}=0$, the derivatives
of Fock matrix can be simplified as
\begin{align}
\sum_{\mu}C_{\mu i\sigma}^{\mathrm{e}}\frac{\partial F_{kc\sigma'}^{\mathrm{e}}}{\partial C_{\mu a\sigma}^{\mathrm{e}}} & =\delta_{\sigma\sigma'}\delta_{ca}F_{ki\sigma}^{\mathrm{e}}=\delta_{\sigma\sigma'}\delta_{ca}\delta_{ki}\varepsilon_{i\sigma}^{\mathrm{e}},
\end{align}
and consequently,
\begin{equation}
\sum_{kc\sigma'}Z_{kc\sigma'}^{\mathrm{e}}\sum_{\mu}C_{\mu i\sigma}^{\mathrm{e}}\frac{\partial F_{kc\sigma'}^{\mathrm{e}}}{\partial C_{\mu a\sigma}^{\mathrm{e}}}=\sum_{kc\sigma'}Z_{kc\sigma'}^{\mathrm{e}}\delta_{\sigma\sigma'}\delta_{ca}\delta_{ki}\varepsilon_{i\sigma}^{\mathrm{e}}=\varepsilon_{i\sigma}^{\mathrm{e}}Z_{ia\sigma}^{\mathrm{e}}.\label{eq:Ze_e_ai}
\end{equation}

\paragraph{vir - vir block ($p=a,q=b$)}

Similarly, in this block, we also have $\delta_{ip}=\delta_{ia}=0$,
which gives
\begin{align}
\sum_{\mu}C_{\mu b\sigma}^{\mathrm{e}}\frac{\partial F_{kc\sigma'}^{\mathrm{e}}}{\partial C_{\mu a\sigma}^{\mathrm{e}}} & =\delta_{\sigma\sigma'}\delta_{ca}F_{kb}^{\mathrm{e}}=0,
\end{align}
and
\begin{equation}
\sum_{kc\sigma'}Z_{kc\sigma'}^{\mathrm{e}}\sum_{\mu}C_{\mu b\sigma}^{\mathrm{e}}\frac{\partial F_{kc\sigma'}^{\mathrm{e}}}{\partial C_{\mu a\sigma}^{\mathrm{e}}}=0.
\end{equation}

\subsubsection{Electron-nucleus}

To obtain the expression of
\begin{equation}
\sum_{kc\sigma'}Z_{kc\sigma'}^{\mathrm{e}}\sum_{\mu'}C_{\mu'q'I}^{\mathrm{n}}\frac{\partial F_{kc\sigma'}^{\mathrm{e}}}{\partial C_{\mu'p'I}^{\mathrm{n}}}.
\end{equation}
the derivatives of electronic Fock matrix with respect to nuclear
MO coefficients derived in Sec.~\ref{subsec:Electronic-Fock-matrix-nuclear}
will be utilized:
\begin{equation}
\sum_{\mu'}C_{\mu'q'I}^{\mathrm{n}}\frac{\partial F_{kc\sigma'}^{\mathrm{e}}}{\partial C_{\mu'p'I}^{\mathrm{n}}}=2\delta_{i'p'}f_{kc\sigma'p'q'I}^{\mathrm{epc,en},I}-2Z_{I}(kc\sigma'|q'i'I)\delta_{i'p'}.
\end{equation}

\paragraph{occ - occ block ($p'=i',q'=i'$)}

With the definition of $H^{\mathrm{ne}}$ in Eq.~\ref{eq:H_ne-1},
we can get
\begin{align}
\sum_{kc\sigma'}Z_{kc\sigma'}^{\mathrm{e}}\sum_{\mu'}C_{\mu'i'I}^{\mathrm{n}}\frac{\partial F_{kc\sigma'}^{\mathrm{e}}}{\partial C_{\mu'i'I}^{\mathrm{n}}} & =\sum_{kc\sigma'}Z_{kc\sigma'}^{\mathrm{e}}\left[2f_{kc\sigma'i'i'I}^{\mathrm{epc,en},I}-2Z_{I}(kc\sigma'|i'i'I)\right]\\
 & =H_{i'i'I}^{\mathrm{ne}}[Z^{\mathrm{e}}].
\end{align}

\paragraph{occ - vir block ($p'=i',q'=a'$)}

Similarly, in $i'a'$ block, we have
\begin{align}
\sum_{kc\sigma'}Z_{kc\sigma'}^{\mathrm{e}}\sum_{\mu'}C_{\mu'a'I}^{\mathrm{n}}\frac{\partial F_{kc\sigma'}^{\mathrm{e}}}{\partial C_{\mu'i'I}^{\mathrm{n}}} & =\sum_{kc\sigma'}Z_{kc\sigma'}^{\mathrm{e}}\left[2f_{kc\sigma'i'a'I}^{\mathrm{epc,en},I}-2Z_{I}(kc\sigma'|i'a'I)\right]\\
 & =H_{i'a'I}^{\mathrm{ne}}[Z^{\mathrm{e}}].
\end{align}

\paragraph{vir - occ block ($p'=a',q'=i'$)}

In $a'i'$ block, as $\delta_{i'p'}$ drops to zero, the derivatives
of electronic Fock matrix with respect to nuclear MO coefficients
will vanish,
\begin{equation}
\sum_{\mu'}C_{\mu'q'I}^{\mathrm{n}}\frac{\partial F_{kc\sigma'}^{\mathrm{e}}}{\partial C_{\mu'p'I}^{\mathrm{n}}}=2\delta_{i'a'}f_{kc\sigma'p'q'I}^{\mathrm{epc,en},I}-2Z_{I}(kc\sigma'|q'i'I)\delta_{i'a'}=0,
\end{equation}
leading to
\begin{equation}
\sum_{kc\sigma'}Z_{kc\sigma'}^{\mathrm{e}}\sum_{\mu'}C_{\mu'i'I}^{\mathrm{n}}\frac{\partial F_{kc\sigma'}^{\mathrm{e}}}{\partial C_{\mu'a'I}^{\mathrm{n}}}=0.
\end{equation}

\paragraph{vir - vir block ($p'=a',q'=b'$)}

Similarly, in this block, $\delta_{i'p'}$ also drops to zero, and
thus we have

\begin{equation}
\sum_{kc\sigma'}Z_{kc\sigma'}^{\mathrm{e}}\sum_{\mu'}C_{\mu'b'I}^{\mathrm{n}}\frac{\partial F_{kc\sigma'}^{\mathrm{e}}}{\partial C_{\mu'a'I}^{\mathrm{n}}}=0.
\end{equation}

\subsubsection{Nucleus-electron}

In order to get term
\begin{equation}
\sum_{k'c'K}Z_{k'c'K}^{\mathrm{n}}\sum_{\mu}C_{\mu q\sigma}^{\mathrm{e}}\frac{\partial F_{k'c'K}^{\mathrm{n}}}{\partial C_{\mu p\sigma}^{\mathrm{e}}}.
\end{equation}
the derivatives of nuclear Fock matrix with respect to electronic
MO coefficients derived in Sec.~\ref{subsec:Nuclear-Fock-matrix-elec}
will be utilized:
\begin{equation}
\sum_{\mu}C_{\mu q\sigma}^{\mathrm{e}}\frac{\partial F_{r's'I}^{\mathrm{n}}}{\partial C_{\mu p\sigma}^{\mathrm{e}}}=-Z_{I}\sum_{i\sigma'}^{\mathrm{occ}}\big[(r's'I|qi\sigma')+(r's'I|iq\sigma')\big]\delta_{ip}\delta_{\sigma\sigma'}+2\sum_{i}^{\mathrm{occ},\sigma}\delta_{ip}f_{r's'Iqi\sigma}^{\mathrm{epc,ne},I}.
\end{equation}

\paragraph{occ - occ block ($p=i,q=j$)}

In this block, the $Z^{\mathrm{n}}$ term can be expressed as

\begin{equation}
\sum_{k'c'K}Z_{k'c'K}^{\mathrm{n}}\sum_{\mu}C_{\mu j\sigma}^{\mathrm{e}}\frac{\partial F_{k'c'K}^{\mathrm{n}}}{\partial C_{\mu i\sigma}^{\mathrm{e}}}=\sum_{k'c'K}Z_{k'c'K}^{\mathrm{n}}\big(-2Z_{K}(k'c'K|ji\sigma)+2f_{k'c'Kji\sigma}^{\mathrm{epc,ne},K}\big).
\end{equation}
To simplify the expression, we define a new linear transformation
as
\begin{equation}
H_{pq\sigma}^{\mathrm{en}}[V]=\sum_{r's'I}\big[-2Z_{I}(pq\sigma|r's'I)+2f_{r's'Ipq\sigma}^{\mathrm{epc,ne},I}\big]V_{r's'I}.\label{eq:H_en-1}
\end{equation}
With this, the expression can be simplified as
\begin{equation}
\sum_{k'c'K}Z_{k'c'K}^{\mathrm{n}}\sum_{\mu}C_{\mu j\sigma}^{\mathrm{e}}\frac{\partial F_{k'c'K}^{\mathrm{n}}}{\partial C_{\mu i\sigma}^{\mathrm{e}}}=H_{ij\sigma}^{\mathrm{en}}[Z^{\mathrm{n}}].
\end{equation}

\paragraph{occ - vir block ($p=i,q=a$)}

Similar to the $ij$ block, we can get the expression for $ia$ block
as
\begin{equation}
\sum_{k'c'K}Z_{k'c'K}^{\mathrm{n}}\sum_{\mu}C_{\mu a\sigma}^{\mathrm{e}}\frac{\partial F_{k'c'K}^{\mathrm{n}}}{\partial C_{\mu i\sigma}^{\mathrm{e}}}=H_{ia\sigma}^{\mathrm{en}}[Z^{\mathrm{n}}].
\end{equation}

\paragraph{vir - occ block ($p=a,q=i$)}

For $ai$ block, the term $\delta_{ip}$ will drop to zero, leading
to
\begin{equation}
\sum_{k'c'K}Z_{k'c'K}^{\mathrm{n}}\sum_{\mu}C_{\mu i\sigma}^{\mathrm{e}}\frac{\partial F_{k'c'K}^{\mathrm{n}}}{\partial C_{\mu a\sigma}^{\mathrm{e}}}=0.
\end{equation}

\paragraph{vir - vir block ($p=a,q=b$)}

Similarly, for this block,
\begin{equation}
\sum_{k'c'K}Z_{k'c'K}^{\mathrm{n}}\sum_{\mu}C_{\mu b\sigma}^{\mathrm{e}}\frac{\partial F_{k'c'K}^{\mathrm{n}}}{\partial C_{\mu a\sigma}^{\mathrm{e}}}=0.
\end{equation}

\subsubsection{Nucleus-nucleus}

To obtain
\begin{equation}
\sum_{k'c'K}Z_{k'c'K}^{\mathrm{n}}\sum_{\mu'}C_{\mu'q'I}^{\mathrm{n}}\frac{\partial F_{k'c'K}^{\mathrm{n}}}{\partial C_{\mu'p'I}^{\mathrm{n}}}.
\end{equation}
the derivatives of nuclear Fock matrix with respect to nuclear MO
coefficients derived in Sec.~\ref{subsec:Nuclear-Fock-matrix-nuclear}
will be utilized
\begin{align}
 & \sum_{\mu'}C_{\mu'q'I}^{\mathrm{n}}\frac{\partial F_{k'c'K}^{\mathrm{n}}}{\partial C_{\mu'p'I}^{\mathrm{n}}}\nonumber \\
 & =\delta_{k'p'}\delta_{IK}F_{q'c'K}^{\mathrm{n}}+\delta_{c'p'}\delta_{IK}F_{k'q'K}^{\mathrm{n}}+2f_{k'c'Iq'i'I}^{\mathrm{epc,nn},I}\delta_{i'p'}\delta_{IK}+2Z_{K}Z_{I}(1-\delta_{IK})\delta_{i'p'}(k'c'K|i'q'I).
\end{align}

\paragraph{occ - occ block ($p'=i',q'=i'$)}

In this block, we have

\begin{align}
 & \sum_{k'c'K}Z_{k'c'K}^{\mathrm{n}}\sum_{\mu'}C_{\mu'i'I}^{\mathrm{n}}\frac{\partial F_{k'c'K}^{\mathrm{n}}}{\partial C_{\mu'i'I}^{\mathrm{n}}}\nonumber \\
 & =\sum_{k'c'K}Z_{k'c'K}^{\mathrm{n}}\big[\delta_{k'i'}\delta_{IK}F_{i'c'K}^{\mathrm{n}}+2f_{k'c'Ii'i'I}^{\mathrm{epc,nn},I}\delta_{IK}+2Z_{K}Z_{I}(1-\delta_{IK})(k'c'K|i'i'I)\big]\\
 & =\sum_{k'c'K}Z_{k'c'K}^{\mathrm{n}}[2f_{k'c'Ii'i'I}^{\mathrm{epc,nn},I}\delta_{IK}+2Z_{K}Z_{I}(1-\delta_{IK})(k'c'K|i'i'I)],
\end{align}
Defining another new linear transformation as
\begin{equation}
H_{p'q'I}^{\mathrm{nn}}[V]=\sum_{r's'J}\big[2\delta_{IJ}f_{p'q'Ir's'I}^{\mathrm{epc,nn},I}+2(1-\delta_{IJ})Z_{I}Z_{J}(r's'J|p'q'I)\big]V_{r's'J}.\label{eq:H_nn-1}
\end{equation}
we can get the simplified expression as
\begin{equation}
\sum_{k'c'K}Z_{k'c'K}^{\mathrm{n}}\sum_{\mu'}C_{\mu'i'I}^{\mathrm{n}}\frac{\partial F_{k'c'K}^{\mathrm{n}}}{\partial C_{\mu'i'I}}=H_{i'i'I}^{\mathrm{nn}}[Z^{\mathrm{n}}].
\end{equation}

\paragraph{occ - vir block ($p'=i',q'=a'$)}

With the definition of $H^{\mathrm{nn}}$, we can get
\begin{align}
 & \sum_{k'c'K}Z_{k'c'K}^{\mathrm{n}}\sum_{\mu'}C_{\mu'a'I}^{\mathrm{n}}\frac{\partial F_{k'c'K}^{\mathrm{n}}}{\partial C_{\mu'i'I}^{\mathrm{n}}}\nonumber \\
 & =\sum_{k'c'K}Z_{k'c'K}^{\mathrm{n}}\big[\delta_{k'i'}\delta_{IK}F_{a'c'K}^{\mathrm{n}}+2f_{k'c'Ii'a'I}^{\mathrm{epc,nn},I}\delta_{IK}+2Z_{K}Z_{I}(1-\delta_{IK})(k'c'K|i'a'I)\big]\\
 & =\sum_{k'c'K}Z_{k'c'K}^{\mathrm{n}}\delta_{k'i'}\delta_{IK}\varepsilon_{a'K}^{\mathrm{n}}\delta_{a'c'}+H_{i'a'I}^{\mathrm{nn}}[Z^{\mathrm{n}}]\\
 & =\varepsilon_{a'I}^{\mathrm{n}}Z_{i'a'I}^{\mathrm{n}}+H_{i'a'I}^{\mathrm{nn}}[Z^{\mathrm{n}}].
\end{align}

\paragraph{vir - occ block ($p'=a',q'=i'$)}

For the $a'i'$ block, with $\delta_{i'p'}$ and $\delta_{k'p'}$
dropping to zero, we can get
\begin{align}
\sum_{k'c'K}Z_{k'c'K}^{\mathrm{n}}\sum_{\mu'}C_{\mu'i'I}^{\mathrm{n}}\frac{\partial F_{k'c'K}^{\mathrm{n}}}{\partial C_{\mu'a'I}^{\mathrm{n}}} & =\sum_{k'c'K}Z_{k'c'K}^{\mathrm{n}}\left[\delta_{c'a'}\delta_{IK}F_{k'i'K}^{\mathrm{n}}\right]\\
 & =\sum_{k'c'K}Z_{k'c'K}^{\mathrm{n}}\delta_{c'a'}\delta_{IK}\varepsilon_{k'K}^{\mathrm{n}}\delta_{i'k'}\\
 & =\varepsilon_{i'I}^{\mathrm{n}}Z_{i'a'I}^{\mathrm{n}}.
\end{align}

\paragraph{vir - vir block ($p'=a',q'=b'$)}

For the $a'b'$ block, $\delta_{i'p'}$ and $\delta_{k'p'}$ will
also drop to zero, leading to
\begin{equation}
\sum_{k'c'K}Z_{k'c'K}^{\mathrm{n}}\sum_{\mu'}C_{\mu'b'I}^{\mathrm{n}}\frac{\partial F_{k'c'K}^{\mathrm{n}}}{\partial C_{\mu'a'I}^{\mathrm{n}}}=\sum_{k'c'K}Z_{k'c'K}^{\mathrm{n}}\left[\delta_{c'a'}\delta_{IK}F_{k'b'K}^{\mathrm{n}}\right]=0.
\end{equation}

\subsection{$W$-related terms}

The derivatives of electronic (nuclear) overlap matrix $S^{\mathrm{e(n)}}$
only has dependence on electronic (nuclear) MO coefficients, but not
nuclear (electronic) MO coefficients. Therefore, for the $W$-related
term in $\partial L/\partial C$, we only need to consider the following
two components:
\begin{enumerate}
\item Electron-electron:
\begin{equation}
\sum_{rs\sigma',r\leq s}W_{rs\sigma'}^{\mathrm{e}}\sum_{\mu}C_{\mu q\sigma}^{\mathrm{e}}\frac{\partial S_{rs\sigma'}^{\mathrm{e}}}{\partial C_{\mu p\sigma}^{\mathrm{e}}};\label{eq:ctd_we}
\end{equation}
\item Nucleus-nucleus:
\begin{equation}
\sum_{r's'K,r'\leq s'}W_{r's'K}^{\mathrm{n}}\sum_{\mu'}C_{\mu'q'I}^{\mathrm{n}}\frac{\partial S_{r's'K}^{\mathrm{n}}}{\partial C_{\mu'p'I}^{\mathrm{n}}}.\label{eq:ctd_Wn}
\end{equation}
\end{enumerate}
The general form for the derivatives of the electronic overlap matrix
can be derived as
\begin{align}
\sum_{\mu}C_{\mu q\sigma}^{\mathrm{e}}\frac{\partial S_{rs\sigma'}^{\mathrm{e}}}{\partial C_{\mu p\sigma}^{\mathrm{e}}} & =\sum_{\mu}C_{\mu q\sigma}^{\mathrm{e}}\frac{\partial}{\partial C_{\mu p\sigma}^{\mathrm{e}}}\sum_{\nu\lambda}C_{\nu r\sigma'}^{\mathrm{e}}C_{\lambda s\sigma'}^{\mathrm{e}}S_{\nu\lambda}^{\mathrm{e}}\label{eq:ctd_dSe_dCe}\\
 & =\delta_{\sigma\sigma'}\delta_{rp}S_{qs}^{\mathrm{e}}+\delta_{\sigma\sigma'}\delta_{sp}S_{rq}^{\mathrm{e}}\\
 & =\delta_{\sigma\sigma'}\delta_{rp}\delta_{qs}+\delta_{\sigma\sigma'}\delta_{sp}\delta_{rq}.
\end{align}
Analogously, the general form for the derivatives of the nuclear overlap
matrix can be expressed as
\begin{align}
\sum_{\mu'}C_{\mu'q'I}^{\mathrm{n}}\frac{\partial S_{r's'K}^{\mathrm{n}}}{\partial C_{\mu'p'I}^{\mathrm{n}}} & =\delta_{IK}\delta_{r'p'}\delta_{q's'}+\delta_{IK}\delta_{s'p'}\delta_{r'q'}.\label{eq:ctd_dSn_dCn}
\end{align}

\subsubsection{Electron-electron}

Following the previous derivations, we separate Eq.\,\ref{eq:ctd_we}
in four blocks, and in each block, we will utilize the derivatives
of electronic overlap matrix Eq.\,\ref{eq:ctd_dSe_dCe} to obtain
the expression for $W^{\mathrm{e}}$-related terms.

\paragraph{occ-occ block($p=i,q=j$)}

\begin{align}
\sum_{rs\sigma',r\leq s}W_{rs\sigma'}^{\mathrm{e}}\sum_{\mu}C_{\mu j\sigma}^{\mathrm{e}}\frac{\partial S_{rs\sigma'}^{\mathrm{e}}}{\partial C_{\mu i\sigma}^{\mathrm{e}}} & =\sum_{rs\sigma',r\leq s}W_{rs\sigma'}^{\mathrm{e}}\left(\delta_{\sigma\sigma'}\delta_{ri}\delta_{js}+\delta_{\sigma\sigma'}\delta_{si}\delta_{rj}\right)\label{eq:W_ij}\\
 & =(1+\delta_{ij})W_{ij\sigma}^{\mathrm{e}}\quad\text{if }i\leq j\\
 & \text{or }W_{ji\sigma}^{\mathrm{e}}\quad\text{if }j< i.
\end{align}

\paragraph{occ-vir block($p=i,q=a$)}

\begin{align}
\sum_{rs\sigma',r\leq s}W_{rs\sigma'}^{\mathrm{e}}\sum_{\mu}C_{\mu a\sigma}^{\mathrm{e}}\frac{\partial S_{rs\sigma'}^{\mathrm{e}}}{\partial C_{\mu i\sigma}^{\mathrm{e}}} & =\sum_{rs\sigma',r\leq s}W_{rs\sigma'}^{\mathrm{e}}\left(\delta_{\sigma\sigma'}\delta_{ri}\delta_{as}+\delta_{\sigma\sigma'}\delta_{si}\delta_{ra}\right)\label{eq:W_ia}\\
 & =W_{ia\sigma}^{\mathrm{e}}.
\end{align}

\paragraph{vir-occ block($p=a,q=i$)}

\begin{align}
\sum_{rs\sigma',r\leq s}W_{rs\sigma'}^{\mathrm{e}}\sum_{\mu}C_{\mu i\sigma}^{\mathrm{e}}\frac{\partial S_{rs\sigma'}^{\mathrm{e}}}{\partial C_{\mu a\sigma}^{\mathrm{e}}} & =\sum_{rs\sigma',r\leq s}W_{rs\sigma'}^{\mathrm{e}}\left(\delta_{\sigma\sigma'}\delta_{ra}\delta_{is}+\delta_{\sigma\sigma'}\delta_{sa}\delta_{ri}\right)\label{eq:W_ai}\\
 & =W_{ia\sigma}^{\mathrm{e}}.
\end{align}

\paragraph{vir-vir block($p=a,q=b$)}

\begin{align}
\sum_{rs\sigma',r\leq s}W_{rs\sigma'}^{\mathrm{e}}\sum_{\mu}C_{\mu b\sigma}^{\mathrm{e}}\frac{\partial S_{rs\sigma'}^{\mathrm{e}}}{\partial C_{\mu a\sigma}^{\mathrm{e}}} & =\sum_{rs\sigma',r\leq s}W_{rs\sigma'}^{\mathrm{e}}\left(\delta_{\sigma\sigma'}\delta_{ra}\delta_{bs}+\delta_{\sigma\sigma'}\delta_{sa}\delta_{rb}\right)\label{eq:W_ab}\\
 & =(1+\delta_{ab})W_{ab\sigma}^{\mathrm{e}}\quad\text{if }a\leq b\\
 & \text{or }W_{ba\sigma}^{\mathrm{e}}\quad\text{if }b< a.
\end{align}

\subsubsection{Nucleus-nucleus}

Analogously, we can obtain $W^{\mathrm{n}}$-related terms, Eq.\,\ref{eq:ctd_Wn},
in four blocks with the help of Eq.\,\ref{eq:ctd_dSn_dCn}.

\paragraph{occ-occ block($p'=i',q'=i'$)}

\begin{align}
\sum_{r's'K,r'\leq s'}W_{r's'K}^{\mathrm{n}}\sum_{\mu'}C_{\mu'i'I}^{\mathrm{n}}\frac{\partial S_{r's'K}^{\mathrm{n}}}{\partial C_{\mu'i'I}^{\mathrm{n}}} & =2W_{i'i'I}^{\mathrm{n}}.\label{eq:W_ij-1}
\end{align}

\paragraph{occ-vir block($p'=i',q'=a'$)}

\begin{align}
\sum_{r's'K,r'\leq s'}W_{r's'K}^{\mathrm{n}}\sum_{\mu'}C_{\mu'a'I}^{\mathrm{n}}\frac{\partial S_{r's'K}^{\mathrm{n}}}{\partial C_{\mu'i'I}^{\mathrm{n}}} & =W_{i'a'I}^{\mathrm{n}}.\label{eq:W_ia-1}
\end{align}

\paragraph{vir-occ block($p'=a',q'=i'$)}

\begin{align}
\sum_{r's'K,r'\leq s'}W_{r's'K}^{\mathrm{n}}\sum_{\mu'}C_{\mu'i'I}^{\mathrm{n}}\frac{\partial S_{r's'K}^{\mathrm{n}}}{\partial C_{\mu'a'I}^{\mathrm{n}}} & =W_{i'a'I}^{\mathrm{n}}.\label{eq:W_ai-1}
\end{align}

\paragraph{vir-vir block($p'=a',q'=b'$)}

\begin{align}
\sum_{r's'K,r'\leq s'}W_{r's'K}^{\mathrm{n}}\sum_{\mu'}C_{\mu'b'I}^{\mathrm{n}}\frac{\partial S_{r's'K}^{\mathrm{n}}}{\partial C_{\mu'a'I}^{\mathrm{n}}} & =(1+\delta_{a'b'})W_{a'b'I}^{\mathrm{n}}\quad\text{if }a'\leq b'\label{eq:W_ab-1}\\
 & \text{or }W_{b'a'I}^{\mathrm{n}}\quad\text{if }b'< a'.
\end{align}

\subsection{$\bm{g}$-related term}

The last term to consider is $\sum_{I}\bm{g}_{I}^{\mathrm{n}}\cdot\overline{\bm R}_{i'i'I}^{\mathrm{n}}$
in the Lagrangian. This term depends on the nuclear MO coefficients only
through $\overline{\bm R}_{i'i'I}^{\mathrm{n}}$. The general form for the derivatives
of $\overline{\bm R}_{i'i'I}^{\mathrm{n}}$ with respect to $C^{\mathrm{n}}$
can be expressed as
\begin{align}
\sum_{\mu'}C_{\mu'q'I}^{\mathrm{n}}\frac{\partial\overline{\bm R}_{k'k'K}^{\mathrm n}}{\partial C_{\mu'p'I}^{\mathrm{n}}} & =\delta_{IK}(\delta_{k'p'}\overline{\bm R}_{q'k'K}^{\mathrm n}+\delta_{k'p'}\overline{\bm R}_{k'q'K}^{\mathrm n})\\
 & =2\delta_{IK}\delta_{k'p'}\overline{\bm R}_{k'q'K}^{\mathrm n}.
\end{align}
With this, we can get the expression for the $\bm{g}$-related terms
in the following four blocks.

\paragraph{occ - occ block ($p'=i',q'=i'$)}

\begin{equation}
\sum_{K}\bm{g}_{K}^{\mathrm{n}}\cdot\sum_{\mu'}C_{\mu'i'I}^{\mathrm{n}}\frac{\partial\overline{\bm R}_{k'k'K}^{\mathrm n}}{\partial C_{\mu'i'I}^{\mathrm{n}}}=\sum_{K}\bm{g}_{K}^{\mathrm{n}}\cdot2\delta_{IK}\delta_{k'i'}\overline{\bm R}_{i'i'I}^{\mathrm n}.
\end{equation}
Since at stationary points, the expectation position constraints are
satisfied with $\overline{\bm R}_{i'i'I}^{\mathrm n}=0$, we have
\begin{equation}
\sum_{K}\bm{g}_{K}^{\mathrm{n}}\cdot\sum_{\mu'}C_{\mu'i'I}^{\mathrm{n}}\frac{\partial\overline{\bm R}_{k'k'K}^{\mathrm n}}{\partial C_{\mu'i'I}^{\mathrm{n}}}=0.
\end{equation}

\paragraph{occ - vir block ($p'=i',q'=a'$)}

\begin{align}
\sum_{K}\bm{g}_{K}^{\mathrm{n}}\cdot\sum_{\mu'}C_{\mu'a'I}^{\mathrm{n}}\frac{\partial\overline{\bm R}_{k'k'K}^{\mathrm n}}{\partial C_{\mu'i'I}^{\mathrm{n}}} & =2\sum_{K}\bm{g}_{K}^{\mathrm{n}}\cdot\overline{\bm R}_{k'a'K}^{\mathrm n}\delta_{IK}\delta_{k'i'}\label{eq:cneo_add}\\
 & =2\bm{g}_{I}^{\mathrm{n}}\cdot\overline{\bm R}_{i'a'I}^{\mathrm n}.
\end{align}

For the other two blocks, ($p'=a',q'=i'$) and ($p'=a',q'=b'$), since
$\delta_{k'p'}$ will drop to zero if $p'$ is a virtual orbital,
the $\bm{g}$-related terms will also drop to zero:
\begin{align}
\sum_{K}\bm{g}_{K}^{\mathrm{n}}\cdot\sum_{\mu'}C_{\mu'i'I}^{\mathrm{n}}\frac{\partial\overline{\bm R}_{k'k'K}^{\mathrm n}}{\partial C_{\mu'a'I}^{\mathrm{n}}} & =0\\
\sum_{K}\bm{g}_{K}^{\mathrm{n}}\cdot\sum_{\mu'}C_{\mu'b'I}^{\mathrm{n}}\frac{\partial\overline{\bm R}_{k'k'K}^{\mathrm n}}{\partial C_{\mu'a'I}^{\mathrm{n}}} & =0.
\end{align}

\subsection{Final working equations}

Combining all the derivations above, we can construct the working
equation for solving $Z^{\mathrm{e(n)}}$, $W^{\mathrm{e(n)}}$, and
$\bm{g}^{\mathrm{n}}$ .

\subsubsection{Electronic part}

The electronic equation
\begin{equation}
\sum_{\mu}C_{\mu q\sigma}^{\mathrm{e}}\frac{\partial L}{\partial C_{\mu p\sigma}^{\mathrm{e}}}=0,
\end{equation}
will give us
\begin{align}
 & \sum_{\mu}C_{\mu q\sigma}^{\mathrm{e}}\frac{\partial}{\partial C_{\mu p\sigma}^{\mathrm{e}}}\big(\sum_{rs\sigma',r\leq s}W_{rs\sigma'}^{\mathrm{e}}S_{rs\sigma'}^{\mathrm{e}}\big)\nonumber \\
 & =Q_{pq\sigma}^{\mathrm{e}}+\sum_{kc\sigma'}Z_{kc\sigma'}^{\mathrm{e}}\sum_{\mu}C_{\mu q\sigma}^{\mathrm{e}}\frac{\partial F_{kc\sigma'}^{\mathrm{e}}}{\partial C_{\mu p\sigma}^{\mathrm{e}}}+\sum_{k'c'K}Z_{k'c'K}^{\mathrm{n}}\sum_{\mu}C_{\mu q\sigma}^{\mathrm{e}}\frac{\partial F_{k'c'K}^{\mathrm{n}}}{\partial C_{\mu p\sigma}^{\mathrm{e}}}.
\end{align}
Separating it into four blocks, we can build the following equations:
\begin{equation}
W_{ij\sigma}^{\mathrm{e}}=\frac{1}{1+\delta_{ij}}(Q_{ij\sigma}^{\mathrm{e}}+H_{ij\sigma}^{+}[Z^{\mathrm{e}}]+H_{ij\sigma}^{\mathrm{en}}[Z^{\mathrm{n}}]);\label{eq:e_ij}
\end{equation}
\begin{equation}
W_{ia\sigma}^{\mathrm{e}}=Q_{ia\sigma}^{\mathrm{e}}+H_{ia\sigma}^{+}[Z^{\mathrm{e}}]+\varepsilon_{a\sigma}^{\mathrm{e}}Z_{ia\sigma}^{\mathrm{e}}+H_{ia\sigma}^{\mathrm{en}}[Z^{\mathrm{n}}];\label{eq:e_ia}
\end{equation}
\begin{equation}
W_{ia\sigma}^{\mathrm{e}}=Q_{ai\sigma}^{\mathrm{e}}+\varepsilon_{i\sigma}^{\mathrm{e}}Z_{ia\sigma}^{\mathrm{e}};\label{eq:e_ai}
\end{equation}
\begin{equation}
W_{ab\sigma}^{\mathrm{e}}=\frac{1}{1+\delta_{ab}}Q_{ab\sigma}^{\mathrm{e}}.\label{eq:e_ab}
\end{equation}

\subsubsection{Nuclear part}

Similarly for the nuclear part, from
\begin{equation}
\sum_{\mu'}C_{\mu'q'I}^{\mathrm{n}}\frac{\partial L}{\partial C_{\mu'p'I}^{\mathrm{n}}}=0,
\end{equation}
we can get
\begin{align}
 & \sum_{\mu'}C_{\mu'q'I}^{\mathrm{n}}\frac{\partial}{\partial C_{\mu'p'I}^{\mathrm{n}}}\big(\sum_{r's'K,r'\leq s'}W_{r's'K}^{\mathrm{n}}S_{r's'K}^{\mathrm{n}}\big)\nonumber \\
 & =Q_{p'q'I}^{\mathrm{n}}+\sum_{kc\sigma'}Z_{kc\sigma'}^{\mathrm{e}}\sum_{\mu'}C_{\mu'q'I}^{\mathrm{n}}\frac{\partial F_{kc\sigma'}^{\mathrm{e}}}{\partial C_{\mu'p'I}^{\mathrm{n}}}+\sum_{k'c'K}Z_{k'c'K}^{\mathrm{n}}\sum_{\mu'}C_{\mu'q'I}^{\mathrm{n}}\frac{\partial F_{k'c'K}^{\mathrm{n}}}{\partial C_{\mu'p'I}^{\mathrm{n}}}+\sum_{K}\bm{g}_{K}^{\mathrm{n}}\cdot\sum_{\mu'}C_{\mu'q'I}^{\mathrm{n}}\frac{\partial\overline{\bm R}_{k'k'K}^{\mathrm n}}{\partial C_{\mu'p'I}^{\mathrm{n}}}.
\end{align}
Separating it into four blocks, we can build the following equations:
\begin{equation}
W_{i'i'I}^{\mathrm{n}}=\frac{1}{2}(Q_{i'i'I}^{\mathrm{n}}+H_{i'i'I}^{\mathrm{ne}}[Z^{\mathrm{e}}]+H_{i'i'I}^{\mathrm{nn}}[Z^{\mathrm{n}}]);\label{eq:n_ij}
\end{equation}
\begin{equation}
W_{i'a'I}^{\mathrm{n}}=Q_{i'a'I}^{\mathrm{n}}+H_{i'a'I}^{\mathrm{ne}}[Z^{\mathrm{e}}]+\varepsilon_{a'I}^{\mathrm{n}}Z_{i'a'I}^{\mathrm{n}}+H_{i'a'I}^{\mathrm{nn}}[Z^{\mathrm{n}}]+2\bm{g}_{I}^{\mathrm{n}}\cdot\overline{\bm R}_{i'a'I}^{\mathrm n};\label{eq:n_ia}
\end{equation}
\begin{equation}
W_{i'a'I}^{\mathrm{n}}=Q_{a'i'I}^{\mathrm{n}}+\varepsilon_{i'I}^{\mathrm{n}}Z_{i'a'I}^{\mathrm{n}}=\varepsilon_{i'I}^{\mathrm{n}}Z_{i'a'I}^{\mathrm{n}};\label{eq:n_ai}
\end{equation}
\begin{equation}
W_{a'b'I}^{\mathrm{n}}=\frac{1}{1+\delta_{a'b'}}Q_{a'b'I}^{\mathrm{n}}=0.\label{eq:n_ab}
\end{equation}

\subsubsection{Coupled multicomponent $Z$-vector equations with constraints}

Subtracting Eq.~\ref{eq:e_ai} from Eq.~\ref{eq:e_ia}, subtracting
Eq.~\ref{eq:n_ai} from Eq.~\ref{eq:n_ia}, and appending the constraint
equation~\ref{eq:constraint_add} gives the coupled equations for
$Z^{\mathrm{e}}$, $Z^{\mathrm{n}}$, and $\bm{g}^{\mathrm{n}}$:
\begin{equation}
Q^{\mathrm{e}}_{ia\sigma}-Q^{\mathrm{e}}_{ai\sigma}+H^{+}_{ia\sigma}[Z^{\mathrm{e}}]+(\varepsilon^{\mathrm{e}}_{a\sigma}-\varepsilon^{\mathrm{e}}_{i\sigma})Z^{\mathrm{e}}_{ia\sigma}+H^{\mathrm{en}}_{ia\sigma}[Z^{\mathrm{n}}]=0,\label{eq:zvector-e}
\end{equation}
\begin{equation}
Q^{\mathrm{n}}_{i'a'I}+H^{\mathrm{ne}}_{i'a'I}[Z^{\mathrm{e}}]+(\varepsilon^{\mathrm{n}}_{a'I}-\varepsilon^{\mathrm{n}}_{i'I})Z^{\mathrm{n}}_{i'a'I}+H^{\mathrm{nn}}_{i'a'I}[Z^{\mathrm{n}}]+2\bm{g}^{\mathrm{n}}_{I}\cdot\overline{\bm{R}}^{\mathrm{n}}_{i'a'I}=0,\label{eq:zvector-n}
\end{equation}
\begin{equation}
\sum_{i'a'}Z^{\mathrm{n}}_{i'a'I}\overline{\bm{R}}^{\mathrm{n}}_{i'a'I}=\bm{0}.\label{eq:zvector-constraint}
\end{equation}
These equations constitute the extended coupled $Z$-vector equations
for the multiple CNEO components, along with the corresponding constraint
equations.
Introducing the rescaled position-constraint multipliers
$\widetilde{\bm g}^{\mathrm n}_I=2\bm g^{\mathrm n}_I$,
Eqs.~\ref{eq:zvector-e}--\ref{eq:zvector-constraint}
can be expressed in matrix form as
\begin{equation}
\begin{pmatrix}
\bm A^{\mathrm{ee}} & \bm A^{\mathrm{en}} & 0\\
\bm A^{\mathrm{ne}} & \bm A^{\mathrm{nn}} & \overline{\bm R}^{\mathrm n}\\
0 & (\overline{\bm R}^{\mathrm n})^{\text T} & 0
\end{pmatrix}
\begin{pmatrix}
\bm Z^{\mathrm e}\\
\bm Z^{\mathrm n}\\
\widetilde{\bm g}^{\mathrm n}
\end{pmatrix}
=-
\begin{pmatrix}
\bm q^{\mathrm e}\\
\bm q^{\mathrm n}\\
0
\end{pmatrix},
\end{equation}
where the elements of the source vectors are
\begin{equation}
q^{\mathrm e}_{ia\sigma}
=Q^{\mathrm e}_{ia\sigma}-Q^{\mathrm e}_{ai\sigma},
\qquad
q^{\mathrm n}_{i'a'I}=Q^{\mathrm n}_{i'a'I}.
\end{equation}
Here, $\bm A^{\mathrm{ee}}$, $\bm A^{\mathrm{en}}$,
$\bm A^{\mathrm{ne}}$, and $\bm A^{\mathrm{nn}}$
represent the electron--electron, electron--nuclear,
nuclear--electron, and nuclear--nuclear orbital-response
blocks defined by Eqs.~\ref{eq:zvector-e} and
\ref{eq:zvector-n}, respectively.
The matrix $\overline{\bm R}^{\mathrm n}$ contains the
occupied--virtual nuclear position matrix elements.
All electronic spin channels and quantum-nuclear components
are implicitly included in these blocks and their
corresponding response vectors.
With these definitions, the coefficient matrix is
equivalent to that of the ground-state CNEO-CPKS equations,\cite{xu2021molecular}
whereas the right-hand side is determined by the orbital derivatives
of the CNEO-TDDFT response expression.

After solving Eqs.~\ref{eq:zvector-e}--\ref{eq:zvector-constraint} for $Z^{\mathrm{e}}$ and $Z^{\mathrm{n}}$,
the required $W^{\mathrm{e}}$ and $W^{\mathrm{n}}$ blocks are
determined from Eqs.~\ref{eq:e_ij}--\ref{eq:e_ab} and Eqs.~\ref{eq:n_ij}--\ref{eq:n_ab},
respectively. Substitution of these quantities into the CNEO-TDDFT
excitation energy gradient expression Eq.~\ref{eq:cneo_grad}, followed
by addition of the ground-state CNEO gradient as in Eq.~\ref{eq:total-excited-gradient},
completes the analytic excited-state gradient.

\section{Computational Details}

CNEO-TDDFT excitation energies and analytic excited-state energy gradients
have been implemented in both PySCF~\cite{sun2018pyscf,sun2020pyscf,sun2026python,theorychemyang_pyscf}
and NWChem~\cite{apra2020nwchem}.
The analytic gradients were validated against numerical finite-difference
gradients. Two representative results are shown in Table~\ref{tab:gradient_validation},
covering cases with one and multiple quantum protons.

\begin{table}[htbp]
\centering
\caption{Comparison of analytic and numerical CNEO-TDDFT
excited-state energy gradients for \ce{HF} and \ce{H2O}.
Calculations employed the PBE0 functional, aug-cc-pVDZ electronic
basis set, and PB4D nuclear basis set, with all hydrogen nuclei
treated quantum mechanically.
The SCF energy convergence threshold was set to $10^{-11}$~hartree.
Numerical gradients were obtained using a three-point central
finite-difference scheme with a step size of 0.001~\AA.
$\Delta\mathbf{G}$ denotes the difference between the analytic and
numerical gradients.
MAE, RMSE, and $|\Delta\mathbf{G}|_{\max}$
denote the mean absolute error, root-mean-square error, and maximum absolute
error, respectively, evaluated over all Cartesian gradient components.
All values are in hartree/bohr.}
\label{tab:gradient_validation}
{
\renewcommand{\arraystretch}{1.2}
\setlength{\tabcolsep}{10pt}
\begin{tabular}{cccc}
\hline
Molecule &
MAE &
RMSE &
$|\Delta\mathbf{G}|_{\max}$ \\
\hline
\ce{HF}  & $2.177\times10^{-6}$ & $2.712\times10^{-6}$ & $3.906\times10^{-6}$ \\
\ce{H2O} & $1.604\times10^{-6}$ & $2.146\times10^{-6}$ & $5.073\times10^{-6}$ \\
\hline
\end{tabular}
}
\end{table}

CNEO-TDDFT geometry optimizations were performed using the geomeTRIC
optimizer~\cite{wang2016geometric} interfaced with PySCF, whereas conventional TDDFT
excited-state geometry optimizations and normal-mode analyses
were performed using Gaussian~16.~\cite{frisch2016gaussian16} Ground-state sampling and excited-state molecular dynamics simulations were carried out using the Atomic Simulation Environment (ASE).\cite{hjorth2017atomic}

Unless otherwise specified, calculations employed the PBE0 exchange--correlation functional,~\cite{adamo1999pbe0} the aug-cc-pVDZ electronic basis set,~\cite{dunning1989cc,kendall1992aug} and a numerical integration grid containing 99 radial shells and 590 angular points per shell. The response of the atom-centered numerical integration grid to
nuclear displacements was included in the ground-state CNEO-DFT
gradients for molecular dynamics sampling, but neglected in the CNEO-TDDFT excited-state gradients. In the CNEO calculations, all hydrogen nuclei were treated quantum mechanically. The PB5F nuclear basis
set~\cite{yu2020nuclearbasis} was employed for the excited-state
vibrational calculations. For deuterium nuclei, the proton basis exponents were multiplied by the square root of the deuterium-to-proton nuclear mass ratio. No electron--nucleus correlation functional was included. Conventional TDDFT calculations used the same electronic basis set, exchange--correlation functional, and integration grid.

For the excited-state vibrational calculations, all Coulomb and exact-exchange contributions were evaluated without density fitting. The ground-state CNEO-DFT calculations used an SCF energy convergence threshold of \(10^{-9}\)~hartree and an orbital-gradient threshold of \(10^{-7}\). The CNEO-TDDFT response equations were converged to a residual threshold of \(10^{-5}\). Excited-state geometry optimizations used energy-change and maximum-gradient thresholds of \(10^{-6}\). CNEO-TDDFT excited-state Hessians were evaluated by central finite differences of the analytic gradients with a displacement of \(0.001\)~\AA, and harmonic vibrational frequencies were obtained by diagonalizing the corresponding mass-weighted Hessians. Conventional TDDFT calculations employed the default Gaussian~16
SCF, TDDFT, and geometry-optimization convergence criteria,
including a root-mean-square (RMS) density-matrix convergence
threshold of $10^{-8}$ for the SCF calculations and a
response-vector convergence threshold of $10^{-6}$ for the
TDDFT response equations.

For the excited-state intramolecular proton-transfer simulations, the PB4D nuclear basis set was used in the CNEO-TDDFT calculations. Initial configurations and velocities were sampled from ground-state molecular dynamics at 300~K using an Andersen thermostat and a time step of 0.5~fs. Following equilibration, 50 sets of nuclear positions and velocities were selected to initialize the excited-state trajectories. Each trajectory was propagated for 65~fs in the microcanonical (NVE) ensemble using a time step of 0.5~fs. Density fitting with the aug-cc-pVDZ-JKFIT auxiliary basis set~\cite{weigend2002fully} was used for the electron--electron Coulomb and exact-exchange contributions during both ground-state sampling and excited-state dynamics.

\section{Excited-State Vibrations with Minimal Hydrogen Involvement}

\begin{figure}[htbp!!]
    \centering
    \includegraphics[width=0.4\textwidth]{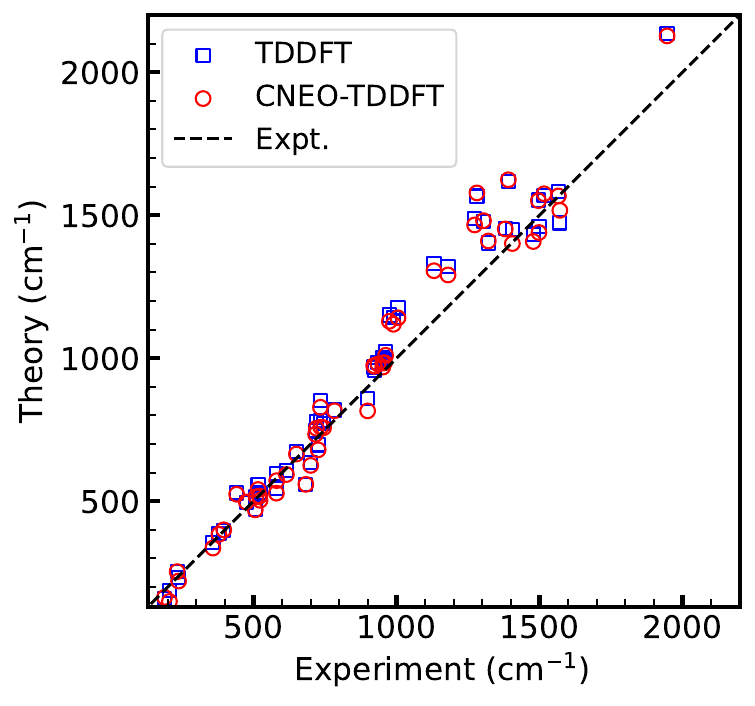}
    \caption{Calculated excited-state vibrational frequencies versus
    experimental reference values for modes with minimal hydrogen
    involvement. Results from conventional TDDFT (blue squares) and CNEO-TDDFT
    (red circles) largely overlap, indicating comparable accuracy for these modes.
    The dashed line denotes experimental references.}
    \label{fig:SI_figure1}
\end{figure}

In addition to the modes involving substantial hydrogen motion discussed in the main text, Fig.~\ref{fig:SI_figure1} compares calculated frequencies with experimental reference values for modes dominated by heavier-atom motion. Conventional TDDFT and CNEO-TDDFT yield nearly identical frequencies. This similarity is expected in part because only H and D nuclei are treated quantum mechanically in the present CNEO-TDDFT calculations. Nevertheless, these modes can still be affected indirectly through changes in equilibrium geometry and coupling to light-nuclear motion. For the modes examined here, no systematic change in agreement with experiment is observed relative to conventional TDDFT.

\bibliographystyle{apsrev4-2}
\bibliography{references}